\documentclass{aa}
\usepackage{graphicx}
\usepackage{txfonts}
\usepackage{xcolor}
\usepackage{natbib}
\usepackage{float}
\usepackage{makecell}
\usepackage{comment}
\usepackage{subcaption}
\usepackage{soul}
\usepackage{lscape}
\usepackage{placeins}
\usepackage[colorlinks=true, linkcolor=blue, citecolor=blue, urlcolor=blue]{hyperref}

\begin{document}

\title{New morpho-kinematic classification: The two-dimensional spatial distribution of stellar specific angular momentum in late-type galaxies}

\author{Juan Manuel Pacheco-Arias \inst{1}, Philippe Amram \inst{1}, Benoît Epinat\inst{1,2,3} \and Wilfried Mercier \inst{1}}

\institute{Aix-Marseille Univ., CNRS, CNES, LAM, 38 Rue Frédéric Joliot Curie, 13338 Marseille, France\\ 
\email{juan.pacheco@lam.fr}
\and Canada-France-Hawaii Telescope, 65-1238 Mamalahoa Highway, Kamuela, HI 96743, USA
\and French-Chilean Laboratory for Astronomy, IRL 3386, CNRS and Universidad de Concepción, Departamento de Astronomía, Barrio Universitario s/n, Concepción, Chile}

\date{Received 16 September 2025 / Accepted 16 January 2026}

\abstract
{}
{The two-dimensional spatial distribution of stellar specific angular momentum (sAM) within galaxies has never been analysed. We seek to identify the morpho-kinematics of this property and its correlation with the total stellar sAM ($j_\star$) and the total stellar mass ($M_\star$) of late-type galaxies in the local Universe.}
{We estimated the two-dimensional distribution of the stellar sAM of 30 spiral and irregular galaxies part of the Gassendi H$\alpha$ survey of SPirals (GHASP). The galaxies are mostly located in nearby low-density environments. To perform the estimations, we produced, for the first time, high-resolution stellar sAM surface density (sAMSD) maps by using a formalism that combines the 3.4 $\mu$m WISE photometry with H$\alpha$ velocity fields and \ion{H}{i} rotation curves. We quantified the features of these new maps using standard non-parametric morphological indicators such as concentration, asymmetry, and smoothness, augmented by two additional coefficients that respectively measure the degree of similarity to an axisymmetric Freeman disc and the prominence of bi-symmetric substructures within the stellar sAMSD space.}
{Each galaxy in our sample was classified into one of the five new morpho-kinematic categories proposed in this study: $j_{\star}$- ring, $j_{\star}$- spiral, $j_{\star}$- bar, $j_{\star}$- clump, and $j_{\star}$- irregular. These classes are named after the predominant galactic substructure of their stellar sAMSD map and form a new classification scheme that combines both morphological characteristics and dynamic properties. For 14 galaxies, the morphological classification does not coincide with the corresponding morpho-kinematic category. As expected, there is a high correlation between $j_\star$ and $M_\star$ for our sample. Similarly, our galaxies exhibit a correlation between $j_\star$ and their star formation rate, and between $j_\star$ and their total \ion{H}{i} mass. The mean $j_\star$ and $M_\star$ are distributed along different regions along the Fall relation for all $j_\star$ types, although there is a considerable scatter within each type.}
{There is a link between the two-dimensional distribution of stellar sAMSD  and $j_\star$. The location of the different $j_{\star}$ types in the $j_{\star}$-$M_{\star}$ diagram could indicate a morpho-kinematic evolution of late-type galaxies within the hierarchical paradigm of galaxy formation. The physical mechanisms capable of modifying the shape of galaxies in the stellar sAMSD space seem to be related with the stability of the galactic disc. The spatial redistribution of angular momentum for low-mass late-type galaxies may be driven by a combination of feedback, dynamical friction, shock waves, and resonances, while the quasi-stationary rotating density wave may be responsible for the aforementioned process in the massive spirals.}


\titlerunning{Two-dimensional angular momentum distribution in late-type galaxies}
\authorrunning{Pacheco-Arias, J. M., et al.}
\maketitle

\section{Introduction}

Angular momentum is a fundamental property of galaxies that plays a crucial role in their formation, evolution, and structural dynamics. However, a comprehensive picture of how galaxies acquire it and redistribute it across cosmic time is still under construction. \cite{hoyle1949problems} was the first to mention the importance of galactic angular momentum in his early theoretical models, proposing that proto-galaxies acquire rotation through tidal torques. This idea was further developed in the tidal torque theory (TTT) formalised by \cite{peebles1969origin} and \cite{doroshkevich1970space}, which describes how galaxies acquire their initial angular momentum through interactions with their surrounding density field. Subsequently, TTT was integrated within the hierarchical framework of galaxy formation in the Lambda cold dark matter ($\Lambda$CDM) paradigm, where galactic angular momentum evolves through a combination of accretion, mergers, and gravitational interactions \citep{white1984angular,barnes1987angular,navarro1994simulations,mo1998formation,zavala2008bulges}. In this formalism, the specific angular momentum (sAM) of a galaxy increases with the mass, implying a correlation between the total stellar sAM ($j_\star$) and total stellar mass ($M_\star$) of galaxies in our Universe, where $j_{\star} \propto M_{\star}^{\alpha}$ with the slope $\alpha = 2/3$ coming directly from dimensional arguments \citep{peebles1969origin,doroshkevich1970space,white1984angular}.

In addition to the theoretical considerations of TTT in the cosmological context, which leads to a scaling relation between halo mass and angular momentum, the development of galactic formation theories together with the refinement of numerical simulations have allowed for hypotheses of the physical mechanisms responsible for the angular momentum transference between the halo and the disc and its propagation within the disc itself. In this scenario, stars and gas exchange sAM with dark matter through dynamical friction \citep{white1978core,navarro1997effects} and resonances \citep{lynden1972generating,athanassoula2003determines}, while feedback \citep{governato2010bulgeless,el2018gas}, shock waves \citep{roberts1969large,athanassoula1992existence}, and semi-stationary density waves \citep{lynden1972generating,sellwood2014secular} propagate stellar and gaseous sAM throughout the disc. The evolution of clumps, bulges, bars, and spiral arms in late-type galaxies as well as the stability of their discs cannot be understood without considering their baryonic sAM redistribution \citep{toomre1964gravitational,binney2008,naab2017theoretical}.

From the observational side, \cite{fall1983} was the first to estimate $j_\star$ for a set of late- and early-type galaxies by measuring their characteristic galactic radius and velocity. When comparing $j_{\star}$ with $M_{\star}$, he confirmed observationally that both galaxy types follow the correlation predicted by the TTT, establishing what is now known as the Fall relation. \cite{fall1983} also found that both morphological types are $j_\star$ offset from each other by roughly 0.5–1 dex at a fixed $M_\star$, with early-type galaxies having less total stellar sAM. This result demonstrates that angular momentum retention is a key factor during galaxy formation, as disc-like galaxies conserve more angular momentum than ellipticals, which lose a significant fraction of it, likely due to dissipative processes and mergers. In contrast to the continuous theoretical development, observations of galactic stellar sAM were interrupted for about 30 years, reappearing in the early 2010s with \cite{romanowsky2012angular}. This hiatus can be attributed to the fact that $j_\star$ combines the mass distribution and stellar kinematics of galaxies and therefore requires the high-resolution photometric and kinematic data that only became widely available with the advent of modern integral-field spectroscopic surveys such as ATLAS$^{\rm 3D}$ \citep{cappellari2011atlas3d}, CALIFA \citep{sanchez2012califa}, and MaNGA \citep{bundy2014overview}, which extended this observations to much larger and more diverse samples. Subsequently, \cite{romanowsky2012angular} reaffirmed Fall's pioneering results for spiral discs by establishing a standard procedure to estimate $j_\star$. This method consists of radially integrating the one-dimensional surface brightness (SB) profile ($B_\star$) of each galaxy multiplied by its corresponding rotation curve (RC) and radial coordinate and then normalising by the galaxy’s total flux. From this point on, most observational studies of late-type galaxies in the local Universe focused on the one-dimensional estimation of $j_\star$ for increasingly larger samples while seeking to precisely determine the parameters of the Fall relation and to correlate $j_\star$ with other global properties \citep{fall2013angular,obreschkow2014fundamental,fall2018angular,posti2018angular,pina2021baryonic,hardwick2022xgass,sorgho2024amiga}.

So far, most of the observational contributions to the discussion on galactic angular momentum have been based on its integrated global definition $j_\star$, thus allowing only the validation of large-scale $\Lambda$CDM predictions. By combining simulations and observations, some studies have shown, from a statistical perspective, that the spatially resolved sAM distributions for galactic discs and dark matter halos differ significantly in the $\Lambda$CDM paradigm \citep{van2001angular,sharma2005angular,sharma2012origin}. Indeed, the stellar sAM distribution seems to be linked to galaxy morphology and can be used as a tool to separate the different kinematic components of galaxies \citep{sweet2020stellar}. Understanding the physical mechanisms behind the redistribution of stellar sAM in the morphological substructures within galactic discs remains an important challenge for our picture of galaxy formation. To address this issue, we employed a methodology that estimates the stellar sAM surface density (sAMSD) of discs for a testing sample of 30 nearby late-type galaxies. In contrast to the approach established by \cite{romanowsky2012angular}, this methodology is based on the discrete computation of the spin parameter developed by \cite{emsellem2007sauron}, which has also been applied in \cite{obreschkow2015,cortese2016sami,sweet2018revisiting,bouche2021muse}, and \cite{mercier2023stellar} to estimate $j_{\star}$. However, we are the first to extend its use for the generation of high-resolution stellar sAMSD maps. With this method, we aim to understand which substructures store a significant portion of the stellar angular momentum of galaxies in order to identify whether there are morpho-kinematic characteristics that group them together, and ultimately we seek to determine which mechanisms could be responsible for the propagation of angular momentum within the galactic disc.

This paper is organised as follows: We present the full dataset in Sect. \ref{sec:The sample}. We describe the methodology in Sect. \ref{Sec: Methodology}. The results on both the stellar sAMSD distribution and the Fall relation are presented in Sect. \ref{sec: Results}. We interpret and discuss our results in Sect. \ref{sec: Discussion} and finally conclude in Sect. \ref{sec: Conclusions}.

\section{Sample and dataset}\label{sec:The sample}

The galaxies analysed in this study are described in detail in \cite{korsaga2019ghasp} and constitute a subsample of the GHASP survey \citep{epinat2008ghaspb, epinat2008ghaspa}. This set of 30 spiral and irregular galaxies in the local Universe includes 3.4 $\mu$m photometric images, H$\alpha$ velocity fields, and \ion{H}{i} RCs. The high spectral and spatial quality of these data makes them ideally suited for the computation of their stellar sAMSD with high accuracy.

\subsection{Surface brightness maps}\label{sec: flux_maps}

We used the $W_1$ filter (3.4 $\mu$m) images of WISE, with a resolution of 6\arcsec\ \citep{jarrett2012extending}. The near-infrared band is ideal for tracing the stellar mass of each galaxy since it is less sensitive to star formation and dust \citep{korsaga2019ghasp}. For this study we used the discs SB maps obtained after bulge subtraction. According to the bulge–disc decompositions of \cite{korsaga2018ghasp}, 11 of the 30 galaxies have bulges that are negligible to be modelled as a distinct component. From the 19 modelled bulges, 18 are classified as pseudobulges (Sérsic index $n < 2$) while only one is a classical bulge ($n = 2.29$). In terms of bulge to total ratio (B/T), 17 out of 19 galaxies have B/T $< 0.30$, one has B/T $= 0.45$, and one has B/T $=0.72$. In addition, we employed the \textsc{MaskingStars}\footnote{An unpublished code developed by Orient-Dorey, which makes use of the photutils segmentation package to mask all sources exceeding a specified flux threshold relative to the main target.} tool to remove contaminating sources in some of the images. The flux in the masked regions was subsequently replaced with the values derived from the radially averaged disc SB profile of the galaxies. A further discussion on the impact of bulges is provided in Sect. \ref{sec: stellar sAM maps}.

\subsection{Rotation curves}\label{sec: rotation_curve}

The H$\alpha$ velocity field and the \ion{H}{i} RCs are presented in \cite{korsaga2019ghasp}. The H$\alpha$ observations were carried out with the scanning Fabry-Perot interferometer at the 1.93\,m telescope of the Observatoire de Haute Provence \citep{garrido2002ghasp,epinat2008ghaspb}. They are part of the GHASP survey, which contains a total of 203 galaxies, mostly located in nearby low-density environments, for which we observed 3D H$\alpha$ seeing limited data cubes ($\sim$2\arcsec ), at high resolution power \citep[$\sim$10 000; see][]{epinat2008ghaspa,korsaga2019ghasp}. The velocity along the line of sight ($v_{\rm los}$) is the first moment of the 3D data cube, obtained after sky emission subtraction and adaptive spatial binning. A targeted signal-to-noise ratio on the H$\alpha$ emission line of $\rm S/N_{H\alpha}>7$ \citep[see][]{epinat2008ghaspa} was achieved through a pixel accretion algorithm based on the Voronoi tessellations technique \citep[method described in detail in][]{daigle2006improved}.\\

The H$\alpha$ RCs were calculated from the $v_{\rm los}$ following the method described in \cite{epinat2008ghaspa}. In this methodology, the geometrical parameters; namely the centre ($\alpha,\delta$), inclination ($i$), and major axis position angle (PA) of the galaxy (see Sect. \ref{sec: geometrical_params}), in addition to the systemic velocity, are obtained using all the 2D information of the observed velocity field during the Levenberg-Marquardt $\chi^{2}$-minimisation (this routine is coded in \textsc{MocKinG}\footnote{\url{https://gitlab.lam.fr/bepinat/MocKinG}}). The parameters are then used to deproject the $v_{\rm los}$ and sample it using rings, excluding the sectors around 22.5\degr\ of the minor axis, and weighting each pixel with its corresponding |$\cos\theta$|, where $\theta$ is the angle from the galaxy major axis. The width of the rings is variable in this procedure; for the inner region it is set to match half the seeing, while for the outer part it must ensure that each ring contains the same amount of uncorrelated bins \citep[the number of bins is determined as a compromise between signal-to-noise ratio and spatial coverage in][]{epinat2008ghaspa}. The angular limit for the excluded sector is arbitrarily chose to balance the deprojection effects and the number of bins per ring. The uncertainties in the H$\alpha$ RCs are determined by the dispersion of the velocity measurements divided by the square root of bins in each ring \citep{korsaga2019ghasp}. For well-behaved galaxies, the iterative process is relatively easy and rapidly converges. For five galaxies with asymmetric RCs (namely UGC4499, UGC7323, UGC9649, UGC10359 and UGC10470), the geometrical parameters have larger uncertainties compared to the rest of the sample.

The \ion{H}{i} component was taken from \cite{van2011lopsidedness} and \cite{lelli2016sparc}. These RCs were constructed by fitting tilted-ring models to the observed \ion{H}{i} velocity fields, and their uncertainties correspond to the difference between the approaching and receding sides of the galaxies \citep[e.g.][]{begeman1987,lelli2016sparc}. The hybrid RCs were constructed by \cite{korsaga2019ghasp} accounting for resolution discrepancies and beam smearing effects, as well as correcting for the differences in inclination and distance for the H$\alpha$ and \ion{H}{i} observations. The last two differences can be explained by the greater extent and presence of warps in the \ion{H}{i} disc compared to the H$\alpha$ disc. Because the \ion{H}{i} RCs were collected from different sources, the inclinations and distances used as reference to homogenise the sample were the H$\alpha$ ones. The \ion{H}{i} RCs velocities were rescaled by multiplying them by the ratio between the sine of \ion{H}{i} and H$\alpha$ inclinations whereas the RCs galactic radii were rescaled by the ratio between H$\alpha$ and \ion{H}{i} distances. The corrections for the amplitude of the \ion{H}{i} RCs are above 10\% only for six galaxies of our sample, without systematic trends. Only one galaxy (UGC8490) shows a change of more than 10\% for the corrected galactic radii.\\

\section{Methodology}\label{Sec: Methodology}

To accurately analyse the 2D spatial distribution of the stellar disc sAMSD of galaxies, we computed this property pixel by pixel, considering the bulge-free SB maps and the velocity fields as well as the RCs described in Sect. \ref{sec:The sample}. The use of infrared photometry, under the assumption that the stellar disc kinematics is, at first order, traced by the H$\alpha$ + \ion{H}{i} kinematics, produced high-resolution stellar sAMSD maps for each individual galaxy.  In addition, this methodology allowed us to compute integrated quantities to estimate the $j_{\star}$-$M_{\star}$ Fall relation for our sample.

\subsection{Stellar discs sAMSD maps}\label{sec: stellar sAM maps}

By definition, the stellar sAM of a galaxy within a volume, $\mathcal{V}$, is its angular momentum, $\vec{J_{\star}}(\mathcal{V})$, normalised by its total stellar mass, $M_{\star}$, computed over the total volume the galaxy, $\mathcal{V}_{\rm max}$:

\begin{equation}
    \vec{j_{\star}}(\mathcal{V}) \equiv \frac{\vec{J_{\star}}(\mathcal{V})}{M_{\star}} = \frac{ \int_{\mathcal{V}} \hspace{0.1cm} \vec{r}\times\vec{v}(\vec{r}) \hspace{0.1cm} \rho_\star(\vec{r}) \hspace{0.1cm} \mathrm{d}V}{\int_{\mathcal{V}_{\rm max}}\rho_\star(\vec{r}) \hspace{0.1cm} \mathrm{d}V},
    \label{eq: Full_j}
\end{equation}

\noindent where $\vec{r}$ is the 3D position from the centre of the galaxy, $\rho_\star(\vec{r})$ is the 3D stellar mass density, and $\vec{v}(\vec{r})$ is the 3D stellar velocity field. Equation (\ref{eq: Full_j}) requires knowledge of the 6D phase space of stars, which can only be partially measured from observations, even for nearby galaxies. It is therefore necessary to make a couple of assumptions to transform Eq. (\ref{eq: Full_j}) into a computable expression from observables. $\rho_\star(\vec{r})$ is often approximated as an axisymmetric, infinitely thin disc modelled by a decreasing exponential surface density profile. The above, combined with a constant RC, leads to a very simple expression for $j_\star$ first proposed in \cite{fall1983} and justified by \cite{romanowsky2012angular}. Although this approach is still widely used in the literature, especially because of its simplicity, several authors have shown that variability in RCs and stellar surface density profiles introduces substantial changes in the stellar sAM \citep{posti2018angular,pina2021baryonic,bouche2021muse}, especially for galaxies with clumps or kinematic substructures \citep{sweet2019angular,espejo2022multiresolution}. To tackle this issue, the strong assumptions on $\rho_\star(\vec{r})$ and $\vec{v}(\vec{r})$ must be relaxed.

Using cylindrical coordinates $(R, \theta, z)$ with the galactic plane as a reference ($z=0$) and assuming that $\rho_\star(\vec{r})$ can be decoupled into the stellar surface mass density of the disc, $\Sigma_{\star}(R,\theta)$, and the stellar disc thickness profile, $f(z)$, Eq. (\ref{eq: Full_j}) can be rewritten as

\begin{equation}
    \vec{j_{\star}}(\mathcal{V}) = \frac{\int_{\mathcal{V}} \left[Rv_{\theta} \hspace{0.1cm} \mathbf{\hat{z}} + (zv_R - Rv_z) \hspace{0.1cm} \boldsymbol{\hat{\theta}} - zv_{\theta} \hspace{0.1cm} \mathbf{\hat{R}} \right] \hspace{0.1cm} \Sigma_{\star}(R,\theta)f(z) \hspace{0.1cm} \mathrm{d}V}{\int_{\mathcal{V}_{\rm max}}\Sigma_{\star}(R,\theta)f(z) \hspace{0.1cm} \mathrm{d}V },
    \label{eq: Expanded_cross_product}
\end{equation}

\noindent where $\mathbf{\hat{R}}$, $\boldsymbol{\hat{\theta}}$, and $\boldsymbol{\hat{z}}$ are respectively the radial, tangential, and vertical unitary vectors. Knowing that the predominant component of the velocity field in a disc-like distribution is $v_\theta$, the radial and vertical velocities can be neglected, implying $v_R = v_z = 0$\footnote{Assuming that the vertical component is symmetric with respect to the galactic plane also leads to $v_z = 0$.}. Further assuming that the thickness profile and the tangential velocity are symmetric with respect to the galactic plane, namely $f(z) = f(-z)$ and $v_{\theta}(R,\theta,z) = v_{\theta}(R,\theta,-z)$, the integration of Eq. (\ref{eq: Expanded_cross_product}) along the vertical axis leads to

\begin{equation}
    \vec{j_{\star}}(\mathcal{S})
    = \frac{\int_{\mathcal{S}} v_{\theta}  \hspace{0.1cm} \Sigma_{\star}R \mathrm{d}S}{\int_{\mathcal{S}_{\rm max}} \Sigma_{\star} \hspace{0.1cm} \mathrm{d}S}\hspace{0.1cm} \mathbf{\hat{z}}  = \frac{\int_{\mathcal{S}} v_{\theta}  \hspace{0.1cm} \Sigma_{\star}R \mathrm{d}S}{M_{\star} \hspace{0.1cm} }\hspace{0.1cm} \mathbf{\hat{z}},
    \label{eq: j_cartesian}
\end{equation}

\noindent where $\mathrm{d}S$ is the surface element, $\mathcal{S}$ represents the surface over which the integration is carried out, $\mathcal{S}_{\rm max}$ is the delimiting total disc surface defined by the maximum radius $R_{\rm max}$ and $M_{\star}$ is the total stellar mass of the disc. Since the stellar sAM in Eq. (\ref{eq: j_cartesian}) points only in the direction perpendicular to the galactic plane ($\mathbf{\hat{z}}$), there is no need to use its vector notation any further, so we refer to it as $j_{\star}$ from now on. In this scenario, the stellar bulge does not contribute to $j_{\star}$ since it is considered a pressure-supported system with no significant net rotation.

Following \cite{emsellem2007sauron}, \cite{cortese2016sami}, and \cite{mercier2023stellar}, we can discretize Eq. (\ref{eq: j_cartesian}) along the spatial dimensions of the stellar SB maps. Integrals are replaced by summations over the entire image surface, allowing the use of both 2D photometric and kinematic information while accounting for the non-axisymmetric features of the galaxies in the calculation. This approach contrasts with what is traditionally used in the literature, where $\Sigma_\star$ and $v_\theta$ are assumed to be independent of $\theta$, and therefore the integrals over $\mathcal{S}$ are transformed into a single integral over $R$ to build a cumulative profile \citep{romanowsky2012angular,posti2018angular,pina2021baryonic,hardwick2022xgass,sorgho2024amiga}.

For the discretization, we defined the stellar sAMSD as
\begin{equation}
    \iota_\star (R, \theta) = \frac{\mathrm{d}j_\star}{\mathrm{d}S} (R, \theta) = \frac{v_\theta(R, \theta)  \hspace{0.1cm} \Sigma_\star(R, \theta) R}{M_\star}\hspace{0.1cm} ,
    \label{eq: iota}
\end{equation}
which we abbreviate to $\iota_\star$ for simplicity. With this formalism, $\iota_\star$ is independent of the angular pixel size and the distance of the galaxy. In practise, $\iota_\star$ is determined in each pixel, assuming that $\Sigma_\star$, $v_{\theta}$, $R$, and $\theta$ are constant within the surface of each pixel of position ($x^{\prime},y^{\prime}$).
By imposing a constant mass-to-light ratio throughout the galaxy, it can be expressed as
\begin{equation}
    \iota_\star (x^{\prime},y^{\prime}) = \frac{v_\theta (x^{\prime},y^{\prime})  \hspace{0.1cm} B_\star (x^{\prime},y^{\prime}) R}{F_\star} \times \frac{\cos{(i)}}{\tan(\alpha)^2 D^2}  \hspace{0.1cm} ,
    \label{eq: iota_pix}
\end{equation}
\noindent where $B_\star$ corresponds to the stellar disc SB of each individual pixel and is computed from the bulge-free 3.4 $\mu$m WISE SB image, $F_{\star}$ is the bulge-free stellar disc flux within $R_{\rm max}$, $i$ is the disc inclination (see Sect. \ref{sec: geometrical_params}), $\alpha=\pi/648000$ corresponds to one arcsecond in radians, and $D$ is the distance to the galaxy in pc. The second term embeds both the inclination effect on SB of the disc and the conversion from arcsec$^2$ to pc$^2$. $R$ is computed at the position ($x,y$) in the galactic plane, while the conversion from the apparent position on the sky ($x^{\prime},y^{\prime}$) to its location ($x,y$) in the disc plane is done by assuming an infinitely flat disc following the prescriptions presented in Appendix C of \cite{adamczyk2023}, considering all the geometrical parameters (($\alpha,\delta$), $i$ and PA) of each galaxy. Equation (\ref{eq: iota_pix}) is the core of this work since it allows us to generate, for the first time, the stellar sAMSD maps, shown in Sect. \ref{sec: sAM maps results}. This is a new way of studying and visualising galaxies that has not been explored to date; the stellar sAM has been mostly considered as an integrated global quantity, while its spatial distribution has only been analysed from a statistical basis.

To compare the results of this methodology with the traditional cumulative profiles, the cumulative $j_{\star}$ within a circular aperture of radius $r$ in the galactic plane can be obtained from the sum of all the pixels within this region:

\begin{equation}
    j_{\star}(<r) = \mathcal{S}_{\textrm{pix}}\sum_{\{x^{\prime},y^{\prime}|R<r\}} \iota_{\star}(x^{\prime},y^{\prime}) = \mathcal{S}_{R_{\textrm{max}}} \times \overline{\iota_{\star}(R<r)} \hspace{0.1cm} ,
    \label{eq: j}
\end{equation}
where $\mathcal{S}_{\textrm{pix}}$ is the surface of a pixel in the galaxy plane in pc$^2$, $\mathcal{S}_{R_{\textrm{max}}}$ is the surface of the galaxy up to the maximum radius $R_{\textrm{max}}$ in pc$^2$, and $\overline{\iota_{\star}(R<r)}$ is the mean stellar sAMSD up to radius $r$.

We decided to express $\iota_\star$ in kiloparsecs times kilometres per second per square parsec ($\rm kpc~km~s^{-1}~pc^{-2}$). It is homogeneous to the inverse of a time. Indeed, from Eq. (\ref{eq: iota}), the stellar sAMSD can be expressed as
\begin{equation}
    \iota_\star = \omega_D \times \frac{\Sigma_{\star} R^2}{M_\star} = \nu_D \times \frac{\Sigma_\star \times 2\pi R^2}{M_\star}\hspace{0.1cm},
    \label{eq: iota_nu_omega}
\end{equation}

\noindent where $\omega_D=v_\theta / R$ is the dynamical angular frequency and $\nu_D$ is the dynamical frequency, i.e. the inverse of the dynamical time $T_D=2\pi R/v_\theta$ needed to achieve one revolution around the galaxy centre. The sAMSD can therefore be interpreted as the dynamical angular frequency weighted by the stellar inertia momentum element contained per surface unit ($\Sigma_{\star} R^2$) normalised by the total mass (enclosed within $R_{\rm max}$) or as the dynamical frequency weighted by the ratio of twice the mass included in a disc of radius $R$ ($\Sigma_\star \times 2\pi R^2$) over the total mass.
Finally, the inclusion of the bulge in the calculation of $\iota_\star$ and $j_\star$ has a minimal impact on late-type galaxies, as expected from Eq. (\ref{eq: iota}), since this is a small substructure located very close to the galactic centre (low $R$) and with a small $v_\theta$ even for rotating pseudo bulges. 
Indeed, the $j_\star$ mean value for our sample only differs by 0.04 dex when bulges are considered compared to our bulge-free approach.

\subsection{Geometrical parameters}\label{sec: geometrical_params}

Obtaining the galactic plane coordinates of each pixel is a non-trivial operation necessary to compute $\iota_\star (x^{\prime},y^{\prime})$. This process involves knowing precisely ($\alpha,\delta$), $i$ and PA for each galaxy. For our sample, these geometrical parameters were determined independently from spatially resolved photometric and kinematic observations. The parameters extracted from photometry, hereafter called photometric parameters, were taken from the HyperLeda\footnote{\url{http://atlas.obs-hp.fr/hyperleda/}.} database. The kinematical parameters were determined by \cite{epinat2008ghaspa,epinat2008ghaspb} following the method described in Sec. \ref{sec: rotation_curve}, used to build the H$\alpha$ RCs. 

The values used to compute the stellar sAMSD for our sample are shown from column three to six of Table \ref{table: geometrical_params_full}.
For the coordinates ($\alpha,\delta$), we adopt the photometric centre of the galaxy, defined as the brightest pixel in the $W_1$-band image. The inclination is taken from the kinematical value $i$, as is the PA in most cases.  However, for UGC7323, UGC10470, and UGC11012,\footnote{In these three galaxies, the photometric PA provides a better description of the major axis orientation compared to the kinematic estimate derived from the isophotes.} the photometric PA was used instead (flagged as PA$^{\mathrm{Mor}}$ in Table \ref{table: geometrical_params_full}).

Following \cite{epinat2008ghaspa,epinat2008ghaspb}, the $i$ and PA kinematical uncertainties were computed using a Monte Carlo method on the power spectrum of the residual velocity field to include the correlation scales of the gas turbulence as well as the non-random noise from non-circular motions. The uncertainties in the photometric PA were estimated by propagating the axial ratio and optical radius uncertainties. The uncertainties on $i$ and PA are taken into account to infer the uncertainties associated with $j_{\star}(<r)$ by changing the projected shape of the circular aperture over which the summation is applied (see Eq. \ref{eq: j}).

\subsection{Tangential velocity}

Once the coordinates in the galactic plane have been assigned to each pixel, the only missing ingredient to compute $\iota_\star (x^{\prime},y^{\prime})$ is the tangential velocity $v_\theta(x^{\prime},y^{\prime})$ (see Eq. \ref{eq: iota_pix}). Under the assumption that the stars and the gas are co-rotating, the stars are located in a dynamically cold disc. This scenario can be held if a galaxy is in equilibrium, meaning that the dynamics of gas and stars are governed by the total gravitational potential \citep{mercier2023stellar}. All galaxies in our sample exhibit undisturbed rotational disc kinematics and are located in low-density environments, so the dynamic equilibrium approximation is expected to hold. It is well-known that the stellar circular rotation needs to be corrected for pressure-supported motions \citep{binney2008}, and asymmetric drift effects \citep{posti2018angular,pina2021baryonic}. 
However, stellar asymmetric drift corrections strongly depend on the stellar velocity dispersion, and since we have no estimate of this quantity, we cannot apply such corrections. Moreover, previous studies have shown that these corrections are negligible for massive spiral galaxies with $M_{\star} \gtrsim 10^9 , M_{\odot}$ \citep[e.g.][]{de2008high,pina2021baryonic}. Given that most galaxies in our sample satisfy this mass criterion (see Table \ref{table: geometrical_params_full}), we derive their stellar tangential velocity map by attributing the value of the RC at the galactic disc radius of each pixel.

\subsection{Stellar disc mass}\label{sec: Mass}

By calculating $M_{\star}$ for each of our galaxies, we could locate our sample on the $j_{\star}$-$M_{\star}$ plane. To do so, we implemented the same method as in  \cite{posti2018angular}, where $M_{\star}$ is obtained by integrating the infrared SB profile of the disc ($B_{3.4\mu\mathrm{m}}$) up to $R_{\rm max}$, namely,

\begin{equation}
    M_{\star} = \Upsilon_{\mathrm{d}}\hspace{0.1cm}2\pi\int_0^{R_{\rm max}}R \hspace{0.1cm} B_{3.4\mu\mathrm{m}}(R)\mathrm{d}R,
    \label{eq: Mass}
\end{equation}

\noindent where $\Upsilon_{\mathrm{d}}$ is the disc mass-to-light ratio. In \cite{posti2018angular} $\Upsilon_{\mathrm{d}}$ is constant (0.7\footnote{This value was found to reduce the scatter in scaling relations that depend on $M_{\star}$, see \cite{lelli2016small}.}) for their entire dataset. Here we used the $\Upsilon_{\mathrm{d}}$ calculated for each galaxy in \cite{korsaga2019ghasp}, following \cite{cluver2014galaxy} relation. This value was derived from the $W1 - W2$ colour and was assumed to remain constant across the galactic disc, following the implementation of \citet{hardwick2022xgass}.
The stellar disc masses for our sample are reported in column nine of Table \ref{table: geometrical_params_full}.

\subsection{Total stellar sAM}

We note that $j_\star$ is an integrated global quantity
obtained from Eq. (\ref{eq: j}). It is defined as $j_{\star}(\leq R_{\rm max})$, where $R_{\rm max}$ is the radius used to normalise $\iota_\star (x^{\prime},y^{\prime})$ (defining $F_{\star,\mathrm{tot}}$ in Eq. \ref{eq: iota_pix}) and the upper limit of the integral in Eq. (\ref{eq: Mass}). This ensures consistency between the values of $j_{\star}$ and $M_{\star}$ that are compared in the Fall relation (see Sect. \ref{sec: our Fall relation}).

Determining $R_{\rm max}$ for our sample is a critical step, as it directly impacts all the derived quantities presented in this study. To ensure consistency, we adopt a standard criterion whereby $R_{\rm max}$ corresponds to the galactocentric radius (in kpc) at which the cumulative radial profile of $j_{\star}$ reaches a stable value. Mathematically, we defined a condition on the variation of cumulative stellar sAM as

\begin{equation}
    \left. \frac{\mathrm{d}j_{\star}(<r)}{\mathrm{d}r} \right|_{r=R_{\rm max}} \leq {\gamma},
    \label{eq: R_max}
\end{equation}

\noindent where we find that $\gamma = 0.5$ km s$^{-1}$ provides the most robust and stable convergence for our dataset. This approach ensures representative $j_{\star}$ and $M_{\star}$ values. However, for 15 galaxies, it has been necessary to extrapolate the tangential velocity beyond the limits of the hybrid RC to achieve this condition\footnote{These galaxies are marked as $\ast$ in column eight of Table \ref{table: geometrical_params_full}.}. In these cases, we performed individual inspections in which we verified that their RC is in the flat regime before being extrapolated. The values of $R_{\rm max}$ and $j_{\star}$ are displayed in columns eight and ten of Table \ref{table: geometrical_params_full}. The cumulative stellar sAM profiles for all the sample are shown in Fig. \ref{fig: app_j_prof}.

\section{Results}\label{sec: Results}

The analysis of the stellar sAMSD maps computed in Sect. \ref{Sec: Methodology} is twofold. We describe the spatial distribution diversity of the stellar sAM and use it to propose a new morpho-kinematic classification of galaxies in Sect. \ref{sec: sAM maps results}, while the study of the Fall relation for our sample is presented in Sect. \ref{sec: our Fall relation}.

\subsection{The angular momentum diversity}\label{sec: sAM maps results}

\begin{figure*}[ht!]
    \centering
    \includegraphics[width=\textwidth]{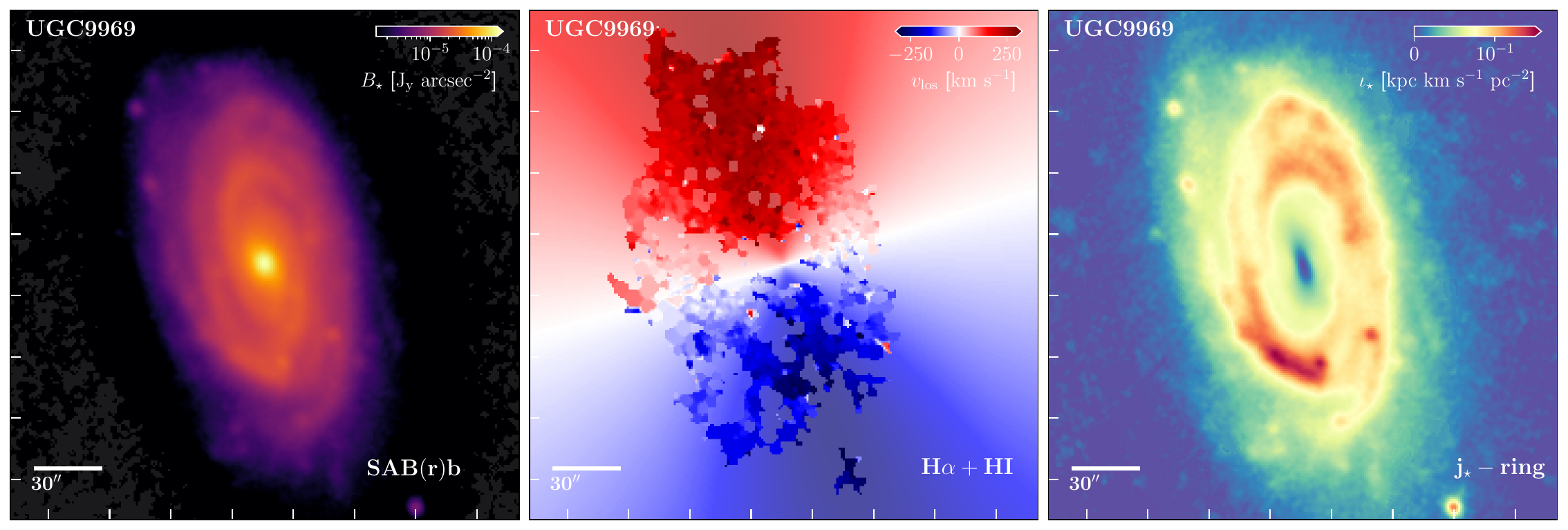}
    \caption{Maps of $B_\star$, $v_{\rm los}$, and stellar sAMSD for UGC9969. Left panel: SB in the 3.4 $\mu$m WISE $W_1$ band. The transparent pixels correspond to the regions where the SB profile was used to fill the map after the foreground star removal process. Middle panel: H$\alpha$ velocity map (solid colour $v_{\rm los}$) on top of the H$\alpha$ + \ion{H}{i} RC projected on the sky (transparent $v_{\rm los}$). Right panel: Stellar sAMSD per pixel computed using Eq. (\ref{eq: iota_pix}).}
    \label{fig: Total_UGC9969}
\end{figure*}

\begin{figure*}[ht!]
    \centering
    \begin{subfigure}[b]{0.245\textwidth}
        \includegraphics[width=\textwidth]{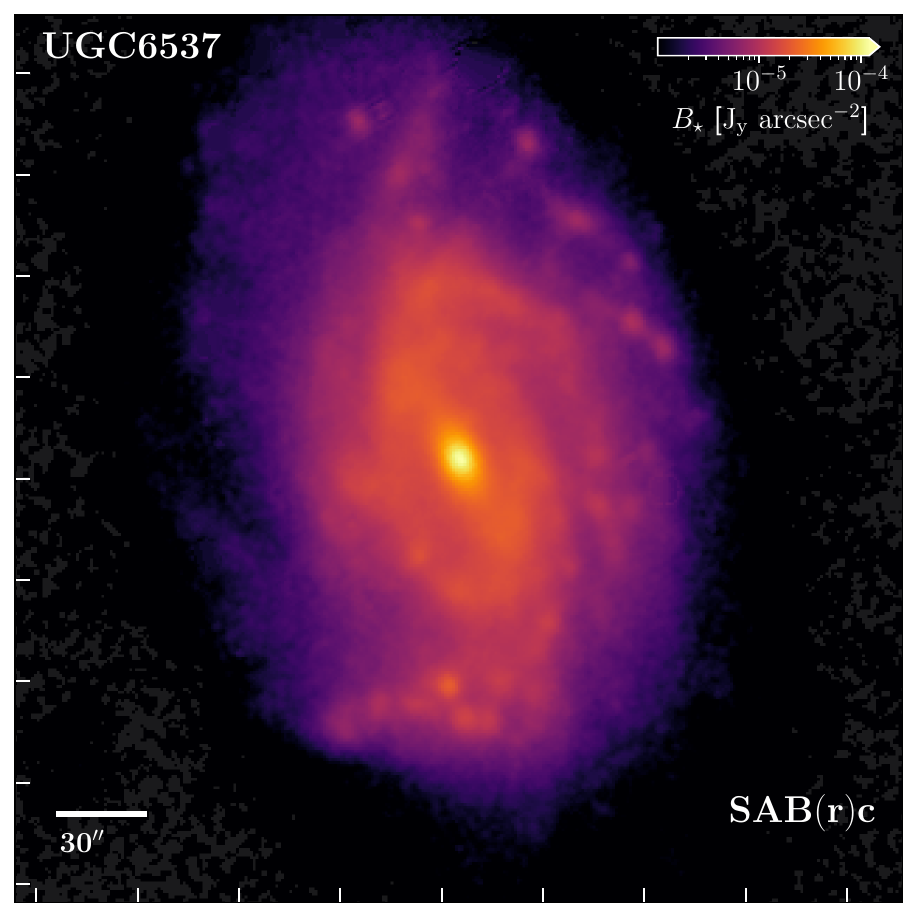}
    \end{subfigure}
    \begin{subfigure}[b]{0.245\textwidth}
        \includegraphics[width=\textwidth]{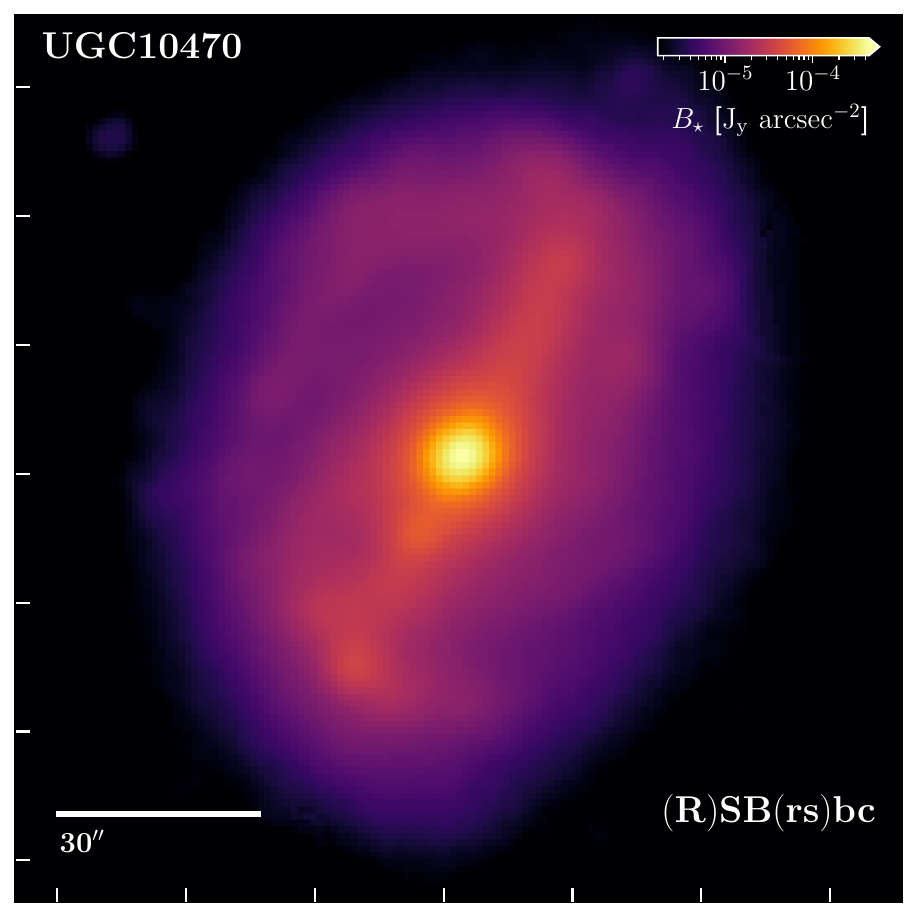}
    \end{subfigure}
    \begin{subfigure}[b]{0.245\textwidth}
        \includegraphics[width=\textwidth]{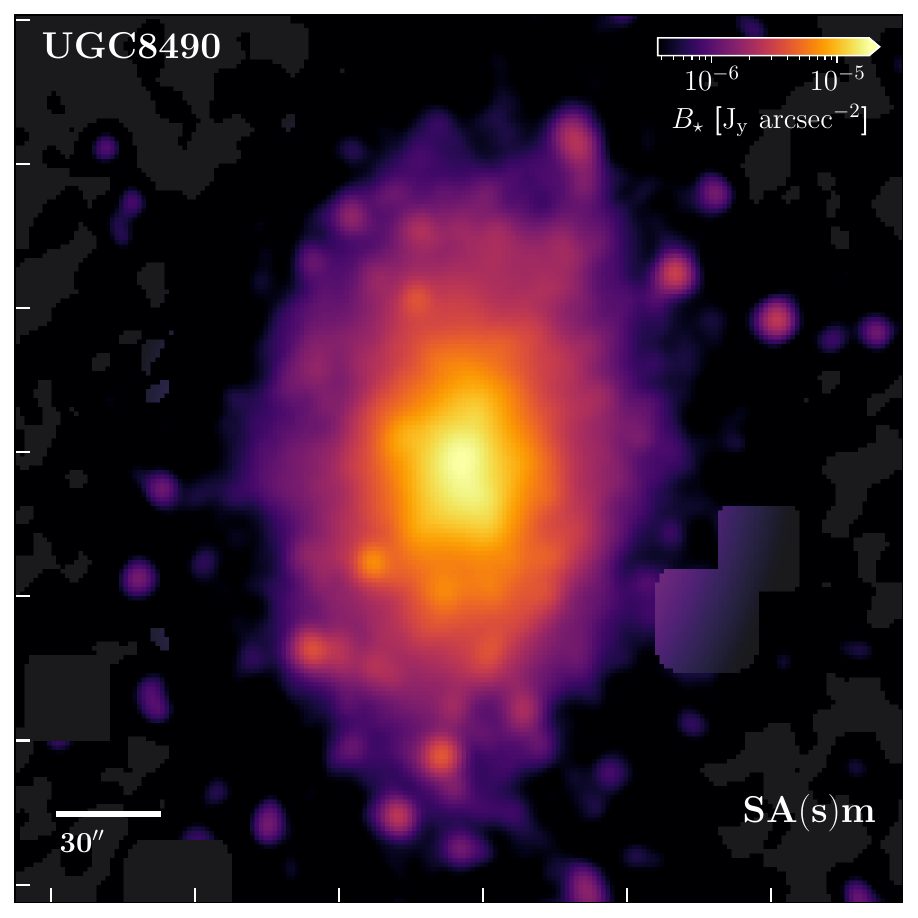}
    \end{subfigure}
    \begin{subfigure}[b]{0.245\textwidth}
        \includegraphics[width=\textwidth]{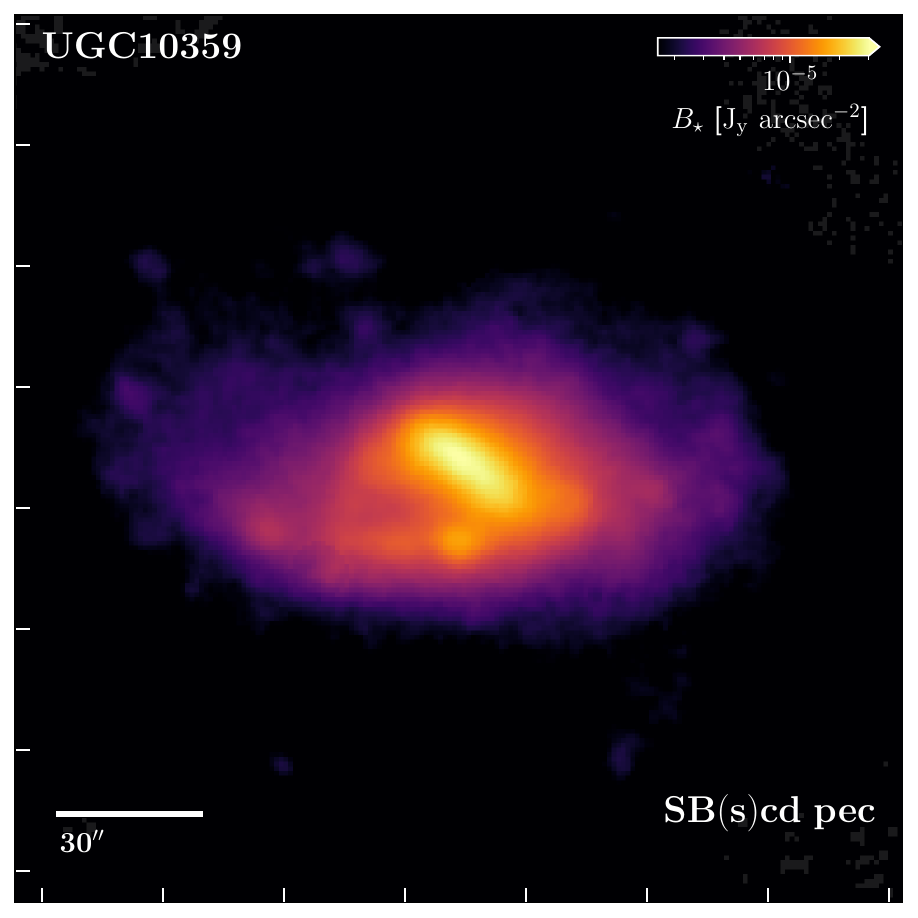}
    \end{subfigure}
        \\
    \begin{subfigure}[b]{0.245\textwidth}
        \includegraphics[width=\textwidth]{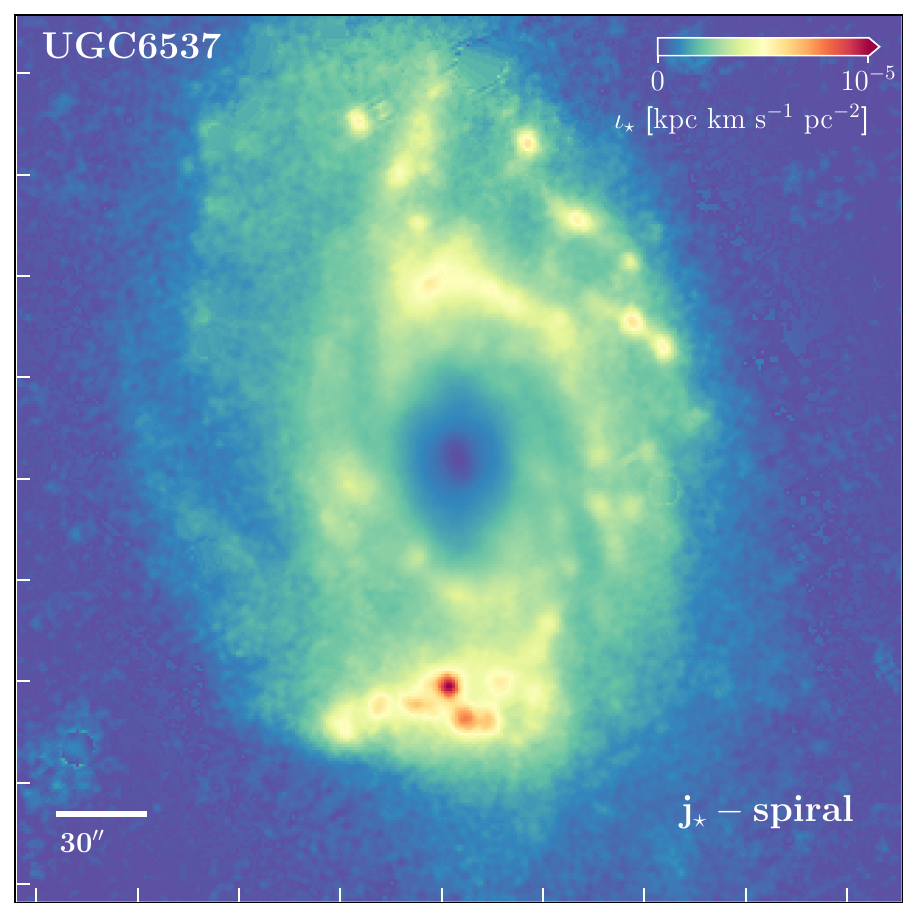}
    \end{subfigure}
    \begin{subfigure}[b]{0.245\textwidth}
        \includegraphics[width=\textwidth]{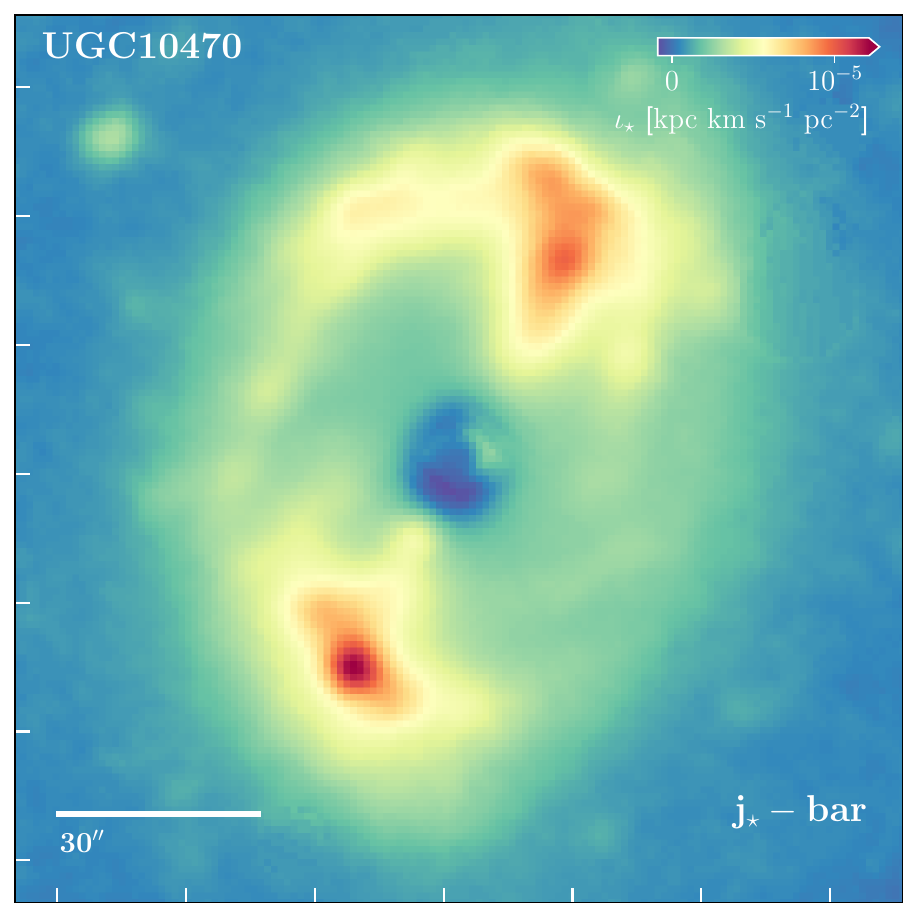}
    \end{subfigure}
    \begin{subfigure}[b]{0.245\textwidth}
        \includegraphics[width=\textwidth]{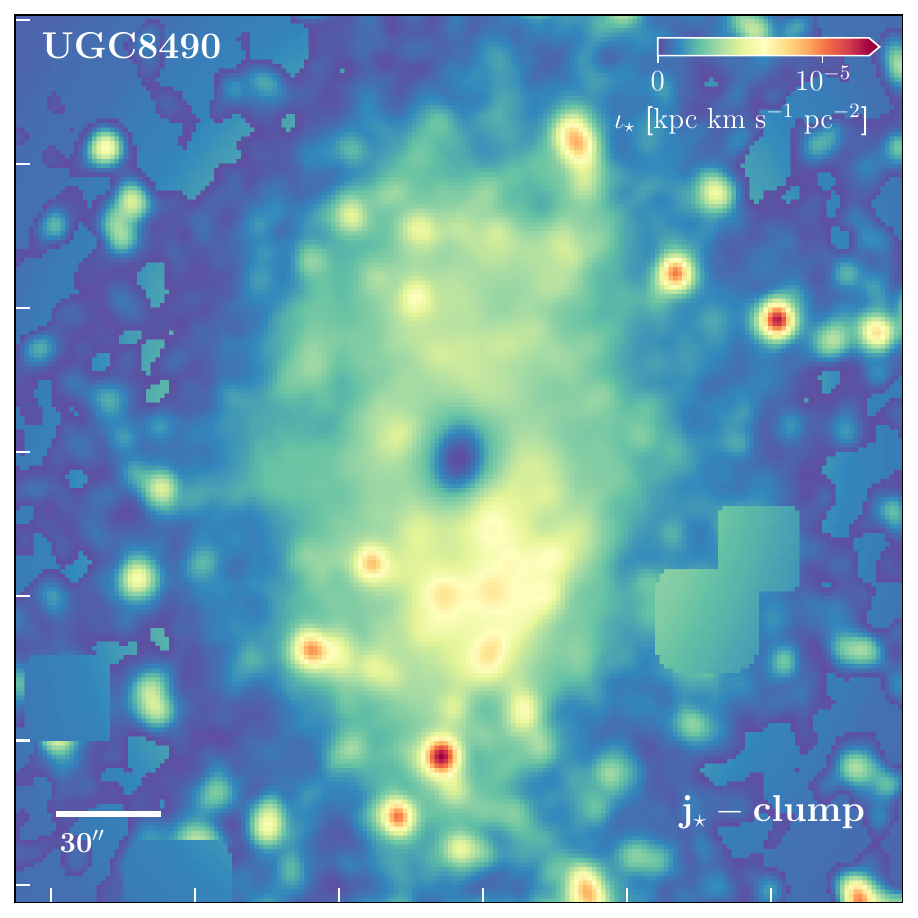}
    \end{subfigure}
    \begin{subfigure}[b]{0.245\textwidth}
        \includegraphics[width=\textwidth]{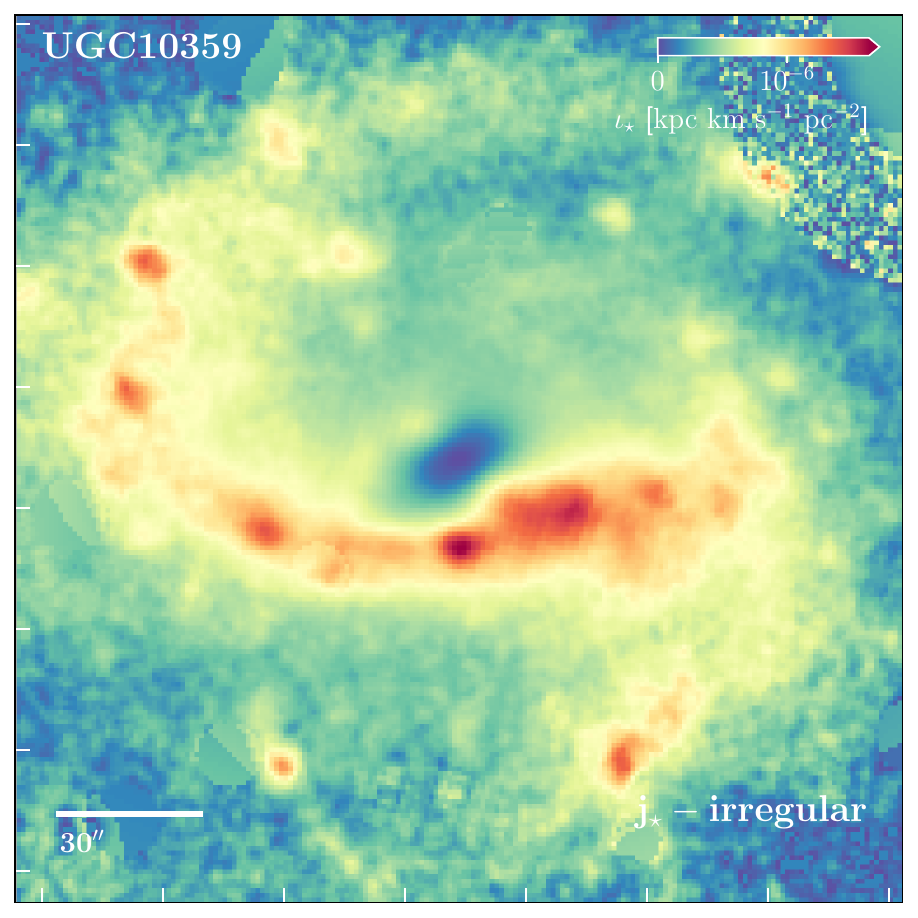}
    \end{subfigure}
    \caption{Maps of $B_\star$ and stellar sAMSD for UGC6537, UGC10470, UGC8490, and UGC10359 (from left to right). The first row is the SB in the 3.4 $\mu$m WISE $W_1$ band for each galaxy. The second row shows the stellar sAMSD per pixel for each galaxy, computed using Eq. (\ref{eq: iota_pix}).}
    \label{fig: stellar_sAM_maps}
\end{figure*}

The right panel of Fig. \ref{fig: Total_UGC9969} shows the stellar sAMSD map for UGC9969 constructed following the methodology described in Sect. \ref{sec: stellar sAM maps}. In this example, it is clear how the spatial distribution of the stellar sAM along the galactic disc forms substructures that are not obvious in the infrared image (left panel) nor in the H$\alpha$ + \ion{H}{i} kinematic (middle panel). This effect is enhanced by the high contrast at the intermediate scales of the galactic disc in the stellar sAMSD map, produced by the combination of two factors: the low $v_{\theta}$ and small $R$ close to the galactic centre that suppress any flux excess in the inner region of the galaxy, and the exponential decay of the average light profile that attenuates the stellar sAMSD on the large scales (see Eq. \ref{eq: iota_pix}). The difference between the infrared morphology and the stellar sAMSD map is not only observed in UGC9969 but in all our galaxies (see Fig. \ref{fig: stellar_sAM_maps} and Appendix \ref{app: stelar sAM maps}). We visually identified five main $j_{\star}$ types\footnote{We decided to use the prefix $j_{\star}$ whenever we refer to the morpho-kinematic of the stellar sAMSD maps to avoid confusion with the traditional morphological classification.} illustrated in Figs. \ref{fig: Total_UGC9969} and \ref{fig: stellar_sAM_maps}, and named after the substructures that stores the bulk of their angular momentum: $j_{\star}$- ring, $j_{\star}$- spiral, $j_{\star}$- bar, $j_{\star}$- clump and $j_{\star}$- irregular.

We used five metrics in order to quantify this diversity in the stellar sAMSD space. In the case of a disc with an axisymmetric mass distribution and a physical RC (either flat or decreasing at a large radius), the sAMSD map forms a torus structure where the sAM peaks (see Appendix \ref{app: Freeman disc} for the Freeman disc case). We therefore defined our first metric as the coefficient of determination ($\rm R^2$) calculated as the difference in pixel-to-pixel values between the stellar sAMSD map and the distribution of sAMSD in a Freeman disc (see Appendix \ref{app: R^2} for the detailed description). In morpho-kinematic terms, $\rm R^2$ determines how $j_{\star}$- ringed our galaxies are (see Sect. \ref{sec:j-ring}). The second metric is the amplitude of the $m=2$ Fourier mode ($\rm A_2$) in our stellar sAMSD maps (see Appendix \ref{app: A_2} for the detailed description). This number quantifies the predominance of bi-symmetric substructures, mainly $j_{\star}$- bars (see Sect. \ref{sec:j-bar}). Finally, we measure the concentration, asymmetry, and smoothness \citep[CAS; see][]{conselice2003relationship} of the stellar sAM spatial distribution. The CAS triplet aims at performing a model-independent non-parametric morphological classification of galaxies, allowing us to extend its use to the morpho-kinematic classification in the stellar sAMSD space. These three parameters, calculated using the \textsc{statmorph} Python package developed by \cite{rodriguez2019optical}, quantify the prevalence of $j_{\star}$- clumps (see Sect. \ref{sec:j-clump}) and $j_{\star}$- irregularities (see Sect. \ref{sec:j-irregular}). The values for each of these five metrics are tabulated per galaxy in Table \ref{table: morphokinematic_params}. We calculated the $\rm R^2$ and CAS parameters by considering all pixels in the stellar sAMSD map with $r \leq R_{25}$, where $R_{25}$ is the isophotal radius at the limiting SB of 25 mag arcsec$^{-2}$, taken from \cite{korsaga2018ghasp}. The methodology we used to determine the various $j_{\star}$ types is provided as a decision tree in Fig. \ref{fig: decision_tree} and is described thoroughly in the following sections.

\begin{table*}
\caption{Average parameters by morpho-kinematic category.}
\label{table: mean_values}      
\centering
\setlength{\tabcolsep}{2.2pt}
\setlength\extrarowheight{1.5mm}
\begin{tabular}{c c c c c c c c}
\hline\hline\\[-2ex]
\makecell{} & \makecell{Number\\ of galaxies \\ (1)} & \makecell{Morphological \\ type code \\ (2)} & \makecell{$\mathrm{log_{10}}M_{\star}$ \\ {$[\mathrm{M_{\odot}}]$} \\ (3)} & \makecell{$\mathrm{log_{10}}j_{\star}$ \\ {$[\mathrm{kpc}$ $\mathrm{km}$ $\mathrm{s^{-1}}]$} \\ (4)} & \makecell{$\rm R^2$ \\  \\ (5)} & \makecell{$\rm A_2$ \\ \\ (6)} & \makecell{CAS \\  \\ (7)}\\[.5ex]
\hline
Full sample & 30 & $5.43 \pm 2.30$ & $10.53 \pm 0.59$ & $2.94 \pm 0.33$ & $0.37 \pm 0.40$ & $0.11 \pm 0.07$ & $(2.15,0.41,0.16) \pm (0.48,0.16,0.08)$\\[.5ex]
\hline
$j_{\star}$- ring & 7 & $3.81 \pm 1.60$ & $10.90 \pm 0.40$ & $3.08 \pm 0.32$ & $0.73 \pm 0.17$ & $0.07 \pm 0.04$ & $(2.44,0.25,0.10) \pm (0.38,0.12,0.05)$\\
$j_{\star}$- spiral & 3 & $5.07 \pm 0.05$ & $10.46 \pm 0.14$ & $3.03 \pm 0.13$ & $0.63 \pm 0.04$ & $0.07 \pm 0.02$ & $(1.90,0.37,0.10) \pm (0.16,0.02,0.01)$\\
$j_{\star}$- bar & 4 & $3.43 \pm 1.94$ & $10.18 \pm 0.51$ & $2.72 \pm 0.22$ & $0.51 \pm 0.10$ & $0.22 \pm 0.03$ & $(2.04,0.33,0.09) \pm (0.73,0.11,0.02)$\\
$j_{\star}$- clump & 9 & $7.72 \pm 1.31$ & $9.99 \pm 0.51$ & $2.63 \pm 0.32$ & $0.10 \pm 0.23$ & $0.13 \pm 0.06$ & $(1.96,0.56,0.24) \pm (0.16,0.12,0.07)$\\
$j_{\star}$- irregular & 7 & $5.41 \pm 2.02$ & $10.51 \pm 0.30$ & $3.10 \pm 0.19$ & $0.15 \pm 0.53$ & $0.08 \pm 0.05$ & $(2.27,0.46,0.18) \pm (0.56,0.12,0.06)$\\[1ex]
\hline           
\end{tabular}
\tablefoot{(1) Number of galaxies in each $j_{\star}$ type, (2) mean morphological type code, and its 1$\sigma$ standard deviation, (3) mean stellar mass and its 1$\sigma$ standard deviation, and (4) mean total stellar sAM and its 1$\sigma$ standard deviation, (5) mean ringness and its 1$\sigma$ standard deviation, (6) mean bar predominance and its 1$\sigma$ standard deviation, (7) mean concentration, asymmetry and smoothness and its 1$\sigma$ standard deviation. The detailed definition of parameters (5), (6), and (7) can be found in Sect. \ref{sec: sAM maps results} and Appendix \ref{app: metrics}.}
\end{table*}

\subsubsection{The $j_{\star}$- ring galaxies}
\label{sec:j-ring}

The first morpho-kinematic category comprises all galaxies that store the bulk of their angular momentum in a ring-like structure. We labelled such galaxies as $j_{\star}$- ring, and they were selected based on two metric criteria: either their stellar sAMSD map reproduces at least 70\% of what is expected by their equivalent Freeman mass distribution ($\mathrm{R}^2>0.7$ ), or it is symmetrical by more than 80\% (A $<0.2$). The second condition is necessary to identify $j_\star$- rings that are radially displaced with respect to their Freeman counterpart. Within this class, UGC9969 is the most representative example (see right panel of Fig. \ref{fig: Total_UGC9969}). This galaxy has $\mathrm{R}^2 = 0.85$ and $A = 0.03$, the second highest and lowest values, respectively, in our sample. Furthermore, UGC9969 is the galaxy with the highest $j_\star$ (see Table \ref{table: geometrical_params_full}). This category is completed by UGC11852, UGC10075, UGC8334, UGC3734, UGC11012, and UGC5253, sorted by descending $\rm R^2$, and their stellar sAMSD maps are shown in the first three rows of Fig. \ref{fig: app_j_maps}. Of the seven galaxies presented above, six meet both metric conditions; only UGC5253 has $\mathrm{R}^2 = 0.32 < 0.7$ but A $<0.2$. The stellar sAMSD radial distribution for this galaxy peaks far from its disc scale length, penalising its $\mathrm{R}^2$. However, the high azimuthal symmetry exhibit in its stellar sAMSD map favours its classification as a $j_\star$- ring.

Following the traditional morphological classification of the de Vaucouleurs system \citep{Vaucouleurs1959}, 40\% of the full sample (12 galaxies) reports the presence of some type of ring in the optical wavelength range. This contrasts with our morpho-kinematic classification, for which only 23\% (7 galaxies) are $j_{\star}$- ring. Among these seven galaxies, four have a ring also visible in their optical image: UGC9969 exhibits a well-defined inner ring, UGC8334 and UGC3734 report a partial inner ring, and UGC5253 shows a morphological outer pseudo-ring. The remaining three galaxies are $j_\star$- ringed but do not host a ring in their morphology: UGC11852, UGC10075 and UGC11012. Finally, there are eight galaxies that host a morphological ring in their optical image but none in their stellar sAMSD map: UGC11914 has both a well-defined outer and inner ring, UGC10470 has a well-defined outer ring and a partial inner ring, UGC6537 and UGC11670 only have a well-defined inner ring, and finally UGC6778, UGC7766, UGC9649 and UGC11597 complete the list with only a partial inner ring.

Within the $j_{\star}$- ring galaxies, five are SA (UGC10075, UGC8334, UGC3734, UGC11012 and UGC5253), one is SAB (UGC9969), and one is SB (UGC11852). With regard to the predominance and distribution of their spiral arms, categories a, ab, b, bc, and c have one representative each (UGC11852, UGC5253, UGC9969, UGC8334, and UGC3734, respectively), while there are two galaxies categorised as cd (UGC10075 and UGC 11012). In general, this subset has an average morphological type code of $3.81$ (Sb/Sbc in the de Vaucouleurs system) with a standard deviation of $1.60$ (both values are below the sample average, see Table \ref{table: mean_values}). Regarding the average morpho-kinematic metrics, this category has the highest C values and the second-lowest S values of all $j_{\star}$ types. As expected from a 2D ring, this subset has a high ringness, concentration, and symmetry combined with low azimuthal variation in their stellar sAM spatial distribution. Finally, $j_{\star}$- ring galaxies tend to have a higher $M_{\star}$ and $j_{\star}$ (0.37 dex and 0.13 dex, respectively) compared to the sample average (see Table \ref{table: mean_values}).

\subsubsection{The $j_{\star}$- spiral galaxies}
\label{sec:j-spiral}

The second morpho-kinematic category gathers all galaxies whose stellar sAMSD is accumulated in the overdensities corresponding to their spiral arms. These galaxies are called $j_{\star}$- spiral and are identified by combining three metric criteria in a logical conjunction: their ringness must be between 50\% and 70\% ($0.5 < \mathrm{R}^2 < 0.7$), they cannot have a predominant bar in their stellar sAMSD map ($\rm A_2 < 0.1$), and their asymmetry cannot exceed 50\% (A $< 0.5$). The mix of parameters $\rm R^2$, $\rm A_2$, and A quantifies the radial and azimuthal variation of the spiral geometry.

Only three galaxies make up this category, making it the $j_{\star}$ type with the fewest members in our study. The most representative example is UGC6537, whose infrared photometry and stellar sAMSD can be seen in the leftmost column of Fig. \ref{fig: stellar_sAM_maps}. This subset is completed by UGC2855 and UGC6778, ordered in descending $\rm R^2$ (their stellar sAMSD maps can be seen in the fourth row of Fig. \ref{fig: app_j_maps}). Traditional morphology classifies these three galaxies as SABc, with the presence of a defined and a partial inner ring in UGC6537 and UGC6778, respectively. These galaxies are the only c-type among the 11 SABs of the sample. Contrary to the $j_{\star}$- ring classification, this morpho-kinematic category fully coincides with the de Vaucouleurs system. This is expected given the composition of our data, in which only two galaxies out of a total of 30 are not spirals of any type. It is important to note that no SA galaxy classifies as a $j_{\star}$- spiral, even though they account for about 37\% (11 galaxies) of the full set.

The average values for this category indicate a morphological type code of 5.07 (Sc for the de Vaucouleurs system) with a standard deviation close to zero (0.05), as would be expected. The morpho-kinematic metrics indicate that $j_{\star}$- spirals have the lowest C of all categories and the same S as $j_\star$- rings. Their $M_\star$ is lower than the sample average by 0.07 dex, while their $j_\star$ remains 0.09 dex above that reference (see Table \ref{table: mean_values}).

\subsubsection{The $j_{\star}$- bar galaxies}
\label{sec:j-bar}

The third morpho-kinematic category includes all galaxies whose stellar sAMSD peaks along their bar. These galaxies are identified as $j_{\star}$- bar based on two metric criteria: the first, and most important, is the predominance of $\rm A_2$ ($\rm A_2 \geq 0.18$); the second, which must be satisfied in parallel, is to exhibit a smooth stellar sAMSD (S $<0.1$). The last condition ensures that the azimuthal variations are not the result of small-scale overdensities (clumps, for example). Among the four galaxies that make up this class, UGC10470 is the clearest example of angular momentum accumulation along the bar. The stellar sAMSD map of this galaxy shows how this substructure is highlighted, especially at its edges (see the second column, from left to right, in Fig. \ref{fig: stellar_sAM_maps}). UGC11670, UGC9649, and UGC12754, sorted by descending $\rm A_2$, complete this category, and their stellar sAMSD maps are shown in the fifth and sixth rows of Fig. \ref{fig: app_j_maps}.

Three of the four $j_{\star}$- bar galaxies are classified as SB (UGC10470, UGC9649, and UGC12754), and only one of them is SA (UGC11670). The absence of a bar in the morphology of UGC11670 is surprising since this galaxy has the highest $\rm A_2$ in this category. In addition, UGC11670 is the only one in our sample whose classification in the de Vaucouleurs system is 0/a, indicating that it is in a transitional state between lenticular and spiral type a. The other galaxies in this subset are type b, b/c, and c/d (UGC9649, UGC10470, and UGC12754, respectively). It is also surprising that UGC10470 belongs to this category since it is classified as a galaxy with a well-defined outer ring and a partial inner ring. However, these rings are not predominate in its stellar sAMSD map. This is also the case for UGC11670 and UGC9649, whose inner morphological rings are eclipsed by their $j_{\star}$- bar. 

Statistically, our sample consists of 18 morphologically barred galaxies (60\%), of which 11 are SAB (around 37\%), six are SB (20\%), and one is IAB. Meanwhile, the  $j_{\star}$- bar morpho-kinematic category represents only 13\% of the full dataset. The mean morphological type code for this category is 3.43 (Sb in the de Vaucouleurs system) with a standard deviation of 1.94. This is the lowest mean and the second highest standard deviation of all $j_{\star}$ types (see Table \ref{table: mean_values}). $j_{\star}$- bar galaxies have, on average, a C, A, $M_{\star}$ and $j_{\star}$ below the sample average (by 0.11, 0.8, 0.35 dex, 0.23 dex, respectively).

\subsubsection{The $j_{\star}$- clump galaxies}
\label{sec:j-clump}

The fourth morpho-kinematic category groups together all galaxies that concentrate their stellar sAMSD in small-scale overdensities (clumps) distributed across their galactic plane. Following this definition, a galaxy is classified as $j_{\star}$- clump if it meets two metric conditions: its stellar sAMSD map must be inhomogeneous (S $>0.14$) and radially diffused (C $<2.3$). The last condition prevents galaxies that have inhomogeneities only near their galactic centre (characteristic of a deficient bulge removal and/or inaccuracies in the geometrical parameters) from being included in this category. The galaxy that best exhibits these features is UGC8490 (see the third column, from left to right, in Fig. \ref{fig: stellar_sAM_maps}), and is accompanied by UGC5414, UGC4325, UGC4499, UGC1913, UGC9179, UGC11597, UGC4284, and UGC7323, organised in descending S (see rows six, seven, eight, nine and ten in Fig. \ref{fig: app_j_maps}). UGC5414 has a high $\rm A_2$ (greater than some $j_\star$- bar galaxies) that can be explained by the presence of diametrically opposed concentrations of stellar sAMSD close to its centre. However, because this galaxy exhibits inhomogeneities throughout its galactic disc, its S is high enough (the second highest in our sample) to be classified as a $j_\star$-clump.

This is the largest morpho-kinematic category of all, with a total of nine members. It does not contain any galaxies classified as SB, has three SA (UGC4325, UGC8490, and UGC4284), five SAB (UGC4499, UGC1913, UGC9179, UGC11597, and UGC7323), and one IAB (UGC5414). The latter is the most $j_{\star}$- clumpy galaxy in the entire sample with an S $= 0.39$. Within this subset, there are no defined morphological rings; only UGC11597 reports a partial inner ring. In terms of stages in the de Vaucouleurs system, these galaxies are distributed as follows: there are two cd (UGC11597 and UGC4284), two d (UGC1913 and UGC9179), two dm (UGC4499 and UGC7323), and three m (UGC5414, UGC4325, and UGC8490). This tendency towards late stages (from c to m) is not surprising, given that clumpy/irregular galaxies are located in this region of the classification scheme. The average morphological type code in this class is 7.73 (Sd/Sdm for the de Vaucouleurs system), by far the highest of all $j_{\star}$ types, with a standard deviation of 1.31. The $j_{\star}$- clumps are, on average, the galaxies with the highest A and lowest $\rm R^2$, $M_{\star}$, and $j_{\star}$ in our entire sample (see Table \ref{table: mean_values}).

\subsubsection{The $j_{\star}$- irregular galaxies}
\label{sec:j-irregular}

The last morpho-kinematic category brings together galaxies that gather their angular momentum in irregular substructures. To classify a galaxy as $j_{\star}$- irregular, there are two sets of metric conditions: the first one identifies asymmetrical galaxies (A $>0.5$) with a high concentration (C $\geq 2.3$); while the second one targets galaxies with non-clumpy stellar sAMSD maps (S $<0.14$) that differ from the Freeman disc stellar sAM spatial distribution ($\mathrm{R}^2<0.3$). UGC10359 is the archetype of this category, and its 3.4 $\mu$m $B_\star$ and stellar sAMSD maps are shown in the rightmost column of Fig. \ref{fig: stellar_sAM_maps}. In this galaxy, the elongated, arched, and non-axisymmetric high stellar sAMSD region stands out clearly in the lower left part near the galactic centre. The other $j_{\star}$- irregular galaxies are UGC9858, UGC2800, UGC5251, UGC7766, UGC3574 and UGC11914, ordered in descending A (see the last panels in Fig. \ref{fig: app_j_maps}.

This morpho-kinematic category represents around 23\% (7 galaxies) of the total sample, while irregular galaxies, from a morphological point of view, are only $\sim$7\% (2 galaxies). UGC2800 is the only one that shares both labels, being a $j_{\star}$- irregular and an Im. The other members of this $j_{\star}$ type represent all the spiral classes of the de Vaucouleurs system equally: there are two SA (UGC3574 and UGC11914), two SAB (UGC9858 and UGC7766), and two SB (UGC5251 and UGC10359). Within this class, there are galaxies such as UGC11914 that reports a well-defined outer and inner ring, while UGC7766 is classified as having a partial inner ring. Additionally, UGC10359 is the only one in our entire dataset designated as peculiar in its morphological classification. Regarding the de Vaucouleurs stages, we have one ab (UGC11914), two bc (UGC9858 and UGC5251), three cd (UGC7766, UGC3574, and UGC10359), and one m (UGC2800), covering almost the entire spectrum.

In terms of average properties, $j_{\star}$- irregulars have the most similar morphological type code to the full sample, with a value of $5.41 \pm 2.02$ (Sc in de Vaucouleurs system). Concerning $M_\star$ and $j_{\star}$, this subset is 0.02 dex below and 0.16 dex above the sample average. Finally, the morpho-kinematic metrics show that this category has the lowest $\rm A_2$ among the $j_{\star}$ types (see Table \ref{table: mean_values}).

\subsection{Fall relation}\label{sec: our Fall relation}

\begin{figure}
\centering
\includegraphics[width=\hsize]{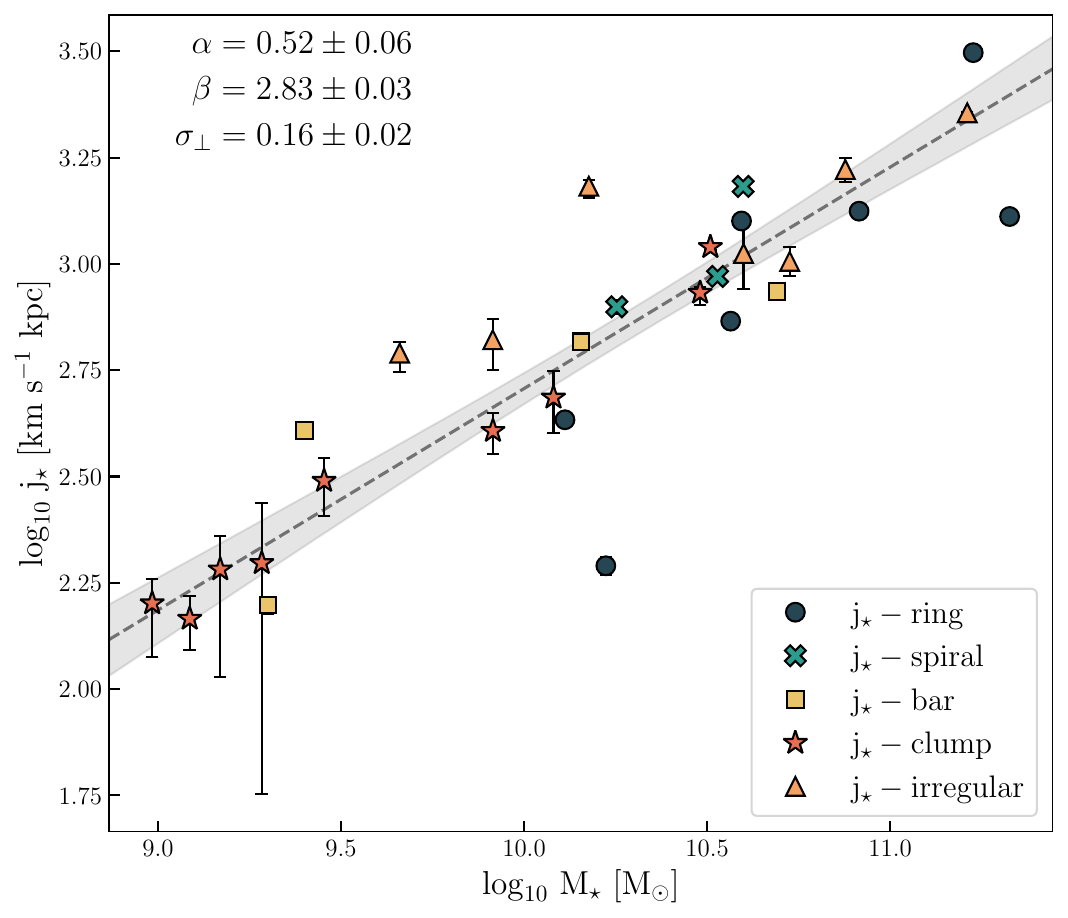}
\caption{Fall relation for our sample. The x-axis is the stellar mass, in logarithmic scale, calculated following the method described in Sect. \ref{sec: Mass}. The y-axis is the total stellar sAM, in logarithmic scale, computed using Eq. (\ref{eq: j}). The error bars represent the variation in $j_\star$ when uncertainties in inclination and PA are considered. Each of the five $j_{\star}$ types described in Sect. \ref{sec: sAM maps results} is represented by a different colour and marker, as can be seen in the legend box. The dotted grey line and the shaded region around it are the best fit of the Fall relation for our sample, with its respective $1\sigma$ uncertainty. The alpha coefficient, the zero point, and the intrinsic scatter for the relation are displayed in the upper-left corner.}
\label{fig: Fall_relation}
\end{figure}

Figure \ref{fig: Fall_relation} shows the $M_\star$ and $j_\star$ for our sample. The error bars have been derived from the uncertainties in the geometrical parameters of each galaxy, and they show how the variability in $j_{\star}$ is higher in the low-mass end of our dataset. In order to visualise the distribution of the morpho-kinematic categories (see Sect. \ref{sec: sAM maps results}) along the Fall relation, each galaxy is represented on the $j_{\star}$-$M_{\star}$ plane with a different set of markers and colours depending on its $j_{\star}$ type. In the lower left part of Fig. \ref{fig: Fall_relation}, most $j_{\star}$- clump galaxies are clustered together. As we move diagonally across the plane towards higher stellar masses, $j_{\star}$- bar and $j_{\star}$- irregular galaxies begin to appear, spreading across the middle range of masses and stellar sAM. Continuing in this direction, we find $j_{\star}$- spirals highly concentrated in a small range of $\mathrm{log}_{10} M_{\star}$ and $\mathrm{log}_{10}j_{\star}$ (from 10.19 to 10.58 and from 2.90 to 3.18, respectively). Finally, galaxies whose stellar sAMSD map resembles the spatial distribution of sAM for a Freeman disc ($j_{\star}$- ring) are mostly located at the high-$M_\star$ and high-$j_\star$ end of our sample.

As expected, there is a strong correlation between $j_{\star}$ and $M_{\star}$ for our 30 galaxies, characterised by a Spearman coefficient $\rho_s = 0.87$ and a p-value $ < 10^{-10}$. The linear fit for both properties in the log-log scale was performed following \cite{posti2018angular} and \cite{pina2021baryonic}. The idea is to assume uninformative priors for the parameters of the fit and explore their posterior distributions with a Monte Carlo Markov Chain method implemented by \cite{foreman2013emcee}. For the Fall relation stated as
\begin{equation}
    \log_{10}{j_{\star}} = \alpha \left(\log_{10} {M_{\star}} - \log_{10} {\widetilde{M_{\star}}}\right) + \beta ~,
\end{equation}
where $\widetilde{M_{\star}}$ is the median stellar mass of the sample and $j_{\star}$ is expressed in kiloparsecs times kilometres per second  ($\rm kpc~km~s^{-1}$), we found $\alpha = 0.52 \pm 0.06$, $\beta = 2.83 \pm 0.03$, and an orthogonal intrinsic scatter of $\sigma_{\perp} = 0.16 \pm 0.02$. These values correspond to the median of their posterior distribution (constructed by maximising the log-likelihood), and their ranges are determined from their 16th and 84th percentiles in the symmetric approximation, building the Bayesian analogue of a $1\sigma$ error. The uncertainties in the fitting parameters are represented as the shaded grey region around the best-fitting line (dashed grey line) of Fig. \ref{fig: Fall_relation}. The lower and upper edges of the grey area correspond to the 16th and 84th percentiles of the model predictions obtained from the posterior samples of $\alpha$ and $\beta$.

It is important to note that the slope we determined is consistent, within the uncertainties, with the values reported by other papers using spiral discs in the local Universe, although it is lower by 0.08 \citep{romanowsky2012angular}, 0.05 \citep{posti2018angular}, and 0.07 \citep{hardwick2022xgass}. As a reference, our slope is 0.14 smaller than the $\alpha$ predicted by $\Lambda$CDM. Due to the different fitting strategy implemented in \cite{hardwick2022xgass}, our $\beta$ parameter can only be compared with \cite{romanowsky2012angular} and \cite{posti2018angular}, falling 0.47 dex and 0.59 dex below the established values in both cases. Regarding $\sigma_{\perp}$, we have the same value as \cite{romanowsky2012angular} that is 0.02 above \cite{posti2018angular} and 0.06 below \cite{hardwick2022xgass}. It should be noted that the methods for computing $j_{\star}$ and $M_{\star}$ vary in each of these studies, as well as their integration limits and mass-to-light ratios.

Finally, we identified a high degree of correlation between $j_{\star}$ and the disc scale length of our galaxies ($\rho_s = 0.91$ and p-value $<10^{-12}$), expected as a consequence of the mass-size relation. Similarly, we found that the star formation rate, obtained from the calibrated H$\alpha$ fluxes presented in \cite{epinat2008ghaspb}, and the total \ion{H}{i} mass ($M_{\ion{H}{i}}$), computed from the \ion{H}{i} line flux reported in HyperLeda, are global properties correlated with $j_{\star}$ in our sample ($\rho_s = 0.73$ and $\rho_s = 0.71 $, respectively).

\section{Discussion}\label{sec: Discussion}

For the first time, we introduce a set of five morpho-kinematic classes to characterise the distribution of stellar sAMSD within galactic discs. Although the galaxy sample under consideration is relatively modest in size, the proposed five-class scheme provides a meaningful classification that differs from the traditional morphological one, with each $j_\star$ type representing a subset of comparable relative importance. Morpho-kinematic classes are not mutually exclusive; some galaxies can be described by a combination of multiple $j_\star$ types. Examples include UGC11670, whose $j_\star$- bar gives rise to small spiral arms in $j_\star$ space; UGC11597, which can be identified as a combination of $j_\star$- clump and $j_\star$- spiral; UGC11914, classified as $j_\star$- irregular but exhibiting a faint outer ring in its stellar sAMSD map; and UGC3574, also classified as $j_\star$- irregular but showing an incipient appearance of asymmetric spiral arms in $j_\star$ space. Nevertheless, we define specific classification criteria for each metric in order to identify the dominant substructure in the sAMSD maps, although those maps are not used to reveal galactic morphological components such as classical rings or bars.

Within the limitations of our results, we emphasise the importance of correctly determining the geometrical parameters for each galaxy. Small discrepancies in the centre, inclination, and PA can lead to different stellar sAMSD maps. This impacts significantly the redistribution of the stellar sAM along the galaxy semi-major and semi-minor axes. Azimuthal variations in the stellar sAMSD are therefore expected considering that geometrical parameters are critical to compute the galactic plane coordinates used in Eq. (\ref{eq: iota_pix}). Moreover, some authors have shown that uncertainties in these values also affect the global integrated $j_\star$ computed from one-dimensional radial profiles \citep{romeo2023specific}. In general, adopting the photometric centre, with a kinematic inclination and PA, generates symmetric stellar sAMSD maps. We recommend using the photometric PA only in cases where a prominent bar is observed in the infrared photometry.

\begin{figure}
\centering
\includegraphics[width=\hsize]{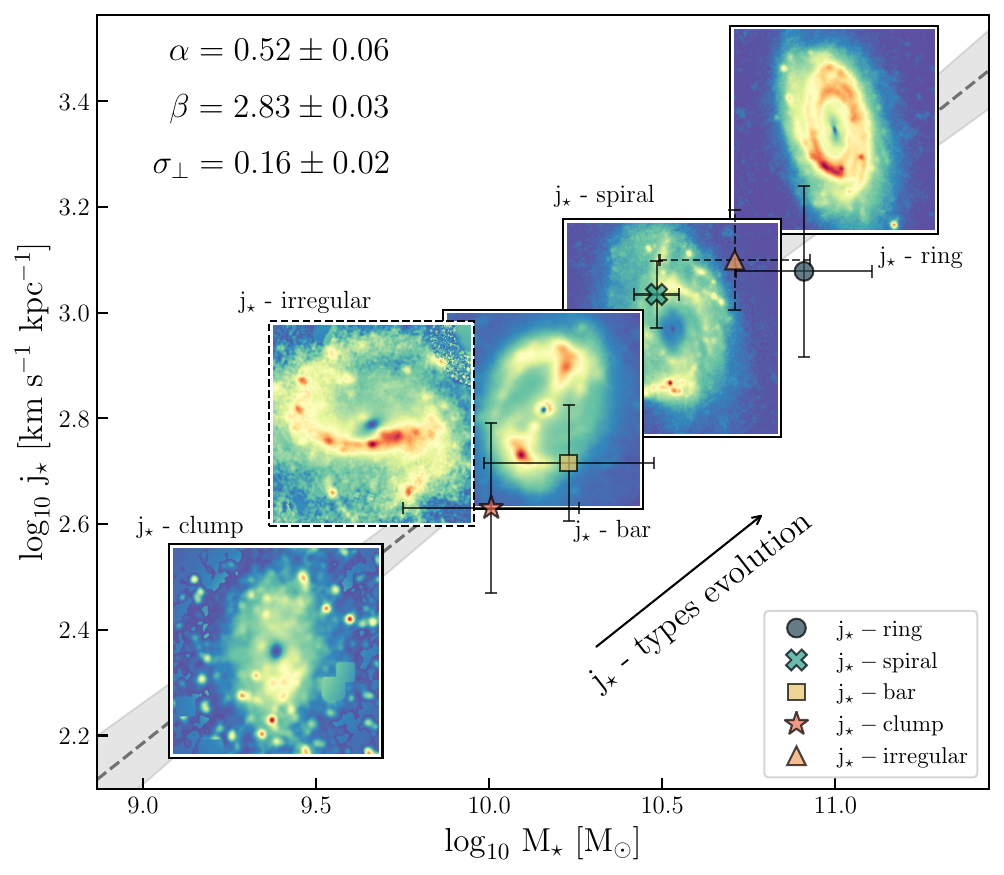}
\caption{Morpho-kinematic evolution diagram along the Fall relation. The average $M_\star$ and $j_\star$ for each morpho-kinematic class, and their 1$\sigma$ standard deviations, are represented following the colour and marker scheme introduced in Fig. \ref{fig: Fall_relation}. The stellar sAMSD maps for the representatives of the five $j_{\star}$ types are arranged on the $j_{\star}$-$M_{\star}$ plane in the location corresponding to their stellar disc mass and total stellar sAM. Each galaxy is labelled according to its morpho-kinematics. Starting from the lower-left corner and moving up along the best-fit line for our sample, we have UGC8490, UGC10359, UGC10470, UGC6537, and UGC9969. The axes and the parameters in the upper-left corner are the same as in Fig. \ref{fig: Fall_relation}.}
\label{fig: morphokinematic_evo}
\end{figure}

\begin{figure}
\centering
\includegraphics[width=\hsize]{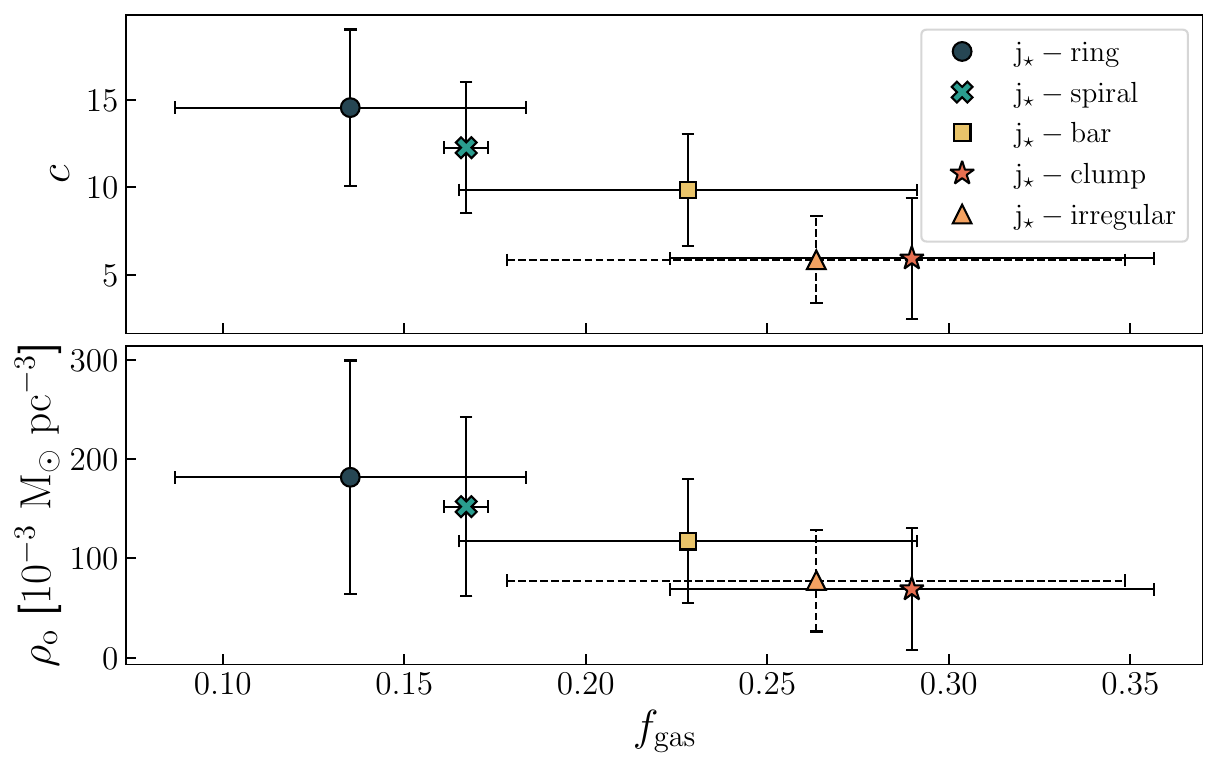}
\caption{Average $f_{\rm gas}$, $c$, and $\rho_{\rm o}$ for each $j_{\star}$ type. The values for each property and their 1$\sigma$ standard deviation are represented following the colour and marker scheme introduced in Fig. \ref{fig: Fall_relation}.}
\label{fig: fgas_c_rho}
\end{figure}

This study connects the $j_\star$ and $M_\star$ of late-type galaxies with their morpho-kinematics in the stellar sAMSD space. Figure \ref{fig: Fall_relation} shows how the galaxies in our sample are grouped into different overlapping regions distributed along the Fall relation, depending on the substructure of their disc that stores most of their stellar sAMSD. To make this clearer, in Fig. \ref{fig: morphokinematic_evo} we plot the average $j_\star$ and $M_\star$ for each morpho-kinematic category shown in Table \ref{table: mean_values}, as well as the representative stellar sAMSD map for each $j_{\star}$ type at its corresponding $j_{\star}$-$M_{\star}$ position. Following the hierarchical paradigm of galaxy formation, we suggest that there is a morpho-kinematic evolution in our dataset as we move along the best-fitting line (black arrow in Fig. \ref{fig: morphokinematic_evo}). Searching for a possible link between disc stability and angular momentum distribution \citep[see][]{Obreschkow2016} we plot in Fig. \ref{fig: fgas_c_rho} the average concentration ($c$) and central density ($\rho_{\rm o}$) for each $j_\star$ type, as a function of their average gas fraction ($f_{\rm gas}$). The gas fraction is computed as $f_{\rm gas} = 1.35M_{\ion{H}{i}}/(M_\star + 1.35M_{\ion{H}{i}})$, while the dark halo matter properties $c$ and $\rho_{\rm o}$ were taken from the mass modelling performed by \cite{korsaga2019ghasp} using the Navarro–Frenk–White (NFW) and Pseudo-isothermal (ISO) halo profiles, respectively. If we read Fig. \ref{fig: fgas_c_rho} from right to left, we see that $f_{\rm gas}$ decreases, while $c$ and $\rho_{\rm o}$ increase. This result suggests that the physical mechanisms responsible for the redistribution of sAMSD along the morpho-kinematic evolution may be related to the galactic disc stability.

We know that low-mass late-type galaxies tend to have a shallower gravitational potential, a low central stellar mass density, and a high $f_{\rm gas}$, which makes them prone to star formation instabilities that generate clumps, bars, and flocculent spiral arms. In contrast, massive late-type galaxies inhabit massive halos, with high central stellar mass density and low $f_{\rm gas}$, which gives them greater disc stability, uniform mass distribution, and tighter spiral arms (grand-design spiral morphology; \citealt{elmegreen1982flocculent,elmegreen1987arm,kauffmann2003stellar,catinella2010galex,dobbs2014dawes,lin1987spiral,gerola1978stochastic}). The transition from loose flocculent morphology to grand-design spirals fits with what we identified in the sAMSD spatial redistribution along the Fall relation, despite the fact that the various morpho-kinematic categories overlap in $M_\star$. At the beginning of their evolution, low mass galaxies, located at the lower left end of the $j_{\star}$-$M_{\star}$ plane, would have a clumpy and unstable disc that gives rise to the $j_\star$- clump category.
Over time, these galaxies would continue to accrete mass, leading to a redistribution of their stellar sAMSD. Initially, this redistribution would concentrate the stellar sAMSD towards the outer edges of the newly formed bars, giving rise to the $j_\star$- bar structures. As the galaxies evolve and their discs become dynamically more stable, the stellar sAMSD would progressively extend along the growing and consolidated spiral arms, ultimately producing the $j_\star$- spiral configurations.
Finally, when the $f_{\rm gas}$ is sufficiently low and the potential is deep enough, the most massive galaxies would exhibit uniform discs, which combined with low-pitch angle spiral arms, give rise to $j_\star$- rings. Under this description, moving along Fall relation, from left to right, implies that a spiral galaxy approaches the morpho-kinematics of an unperturbed galactic disc. This picture excludes the $j_{\star}$- irregulars (stellar sAMSD map marked with the dotted line in Fig. \ref{fig: morphokinematic_evo} and orange triangles with dotted error bars in Figs. \ref{fig: morphokinematic_evo} and \ref{fig: fgas_c_rho}) from the secular evolutionary line, since the appearance of asymmetric structures may be related to interactions with other galaxies. This would explain the variations in its position relative to the other $j_\star$ types in Figs. \ref{fig: morphokinematic_evo} and \ref{fig: fgas_c_rho}. In fact, $j_\star$- irregulars display elongated non-axisymmetric stellar substructures far from their galactic centre. This mass excess located in the high $v_\theta$ region tends to increase their $j_\star$, explaining why they are the $j_\star$ type with the highest average $j_\star$ without being the most massive or the most stable.

There are many physical processes responsible for the redistribution of angular momentum within the galactic disc in the context of secular evolution. Morphological clumps are usually linked to angular momentum transfer through feedback \citep{navarro1993simulations,navarro2000dark,governato2010bulgeless,brook2011hierarchical} and dynamical friction \citep{noguchi1998clumpy,noguchi1999early,immeli2004gas}, indicating that these may be the mechanisms responsible for the shape of the sAMSD in $j_\star$- clump galaxies. In the case of $j_\star$- bars, we know that the evolution and strengthening of morphological bars is driven by their emission of angular momentum in inner disc resonances \citep{athanassoula2003determines,athanassoula2005nature}, so we expect this process to lead the redistribution of angular momentum in this morpho-kinematic category in combination with dynamical friction \citep{tremaine1984kinematic,weinberg1985evolution} and shock propagation \citep{athanassoula1992existence,athanassoula2002bar}. Finally, the angular momentum propagation in stable, massive late-type galaxies, namely $j_\star$- spirals and $j_\star$- rings, would be carried out mainly by the quasi-stationary density wave that slowly rotates within their discs \citep{lynden1972generating} in conjunction with shocks propagation along their spiral arms \citep{roberts1969large,roberts1972application}. It is important to note that the physical processes mentioned above are not mutually exclusive, so the final shape of galaxies in sAMSD space must be generated by a combination of them which can only be understood and disentangled through hydrodynamic simulations.

\section{Conclusions}\label{sec: Conclusions}

We have studied the two-dimensional distribution of the stellar sAMSD for a set of 30 spiral and irregular galaxies in the local Universe that are part of the GHASP sample. For this purpose, we developed and implemented a new formalism that allows for the computation of the stellar sAM pixel by pixel using the 3.4 $\mu$m photometry of WISE and the H$\alpha$ + \ion{H}{i} RC as a proxy for the stellar kinematics. Following this approach, we were able to produce high-resolution stellar sAMSD maps of galaxies for the first time.

We have shown that a diversity of shapes exists in the stellar sAMSD space that do not always correspond to what is seen in the morphology or kinematics of our galaxies. The accumulation of stellar sAM in different substructures of the galactic disc allowed us to group our sample into a new set of five morpho-kinematic categories proposed by this study: $j_{\star}$- ring (seven galaxies), $j_{\star}$- spiral (three galaxies), $j_{\star}$- bar (four galaxies), $j_{\star}$- clump (nine galaxies), and $j_{\star}$- irregular (seven galaxies). We quantified the morpho-kinematic variability by combining the $\rm R^2$ coefficient of determination with traditional non-parametric morphological metrics such as the $\rm A_2$ Fourier mode dominance and the CAS triplet. In this way, we established quantitative criteria that outline all the $j_{\star}$ types. Each of the five morpho-kinematic classes encompasses a broad range of de Vaucouleurs morphological types. Nonetheless, these classes respectively correspond (on average) to Sb/Sbc, Sc, Sb, Sd/Sdm, and Sc galaxies within the de Vaucouleurs classification scheme, following the order listed above. After comparison with their equivalent photometric substructures, we found that eight galaxies have a morphological ring that does not correspond to a $j_{\star}$-ring, three galaxies are in the opposite situation, and four galaxies have both a morphological ring and a $j_{\star}$- ring. Spiral arms are the only substructure for their morpho-kinematic category, which agrees with their morphological classification since all $j_{\star}$- spirals are SABc galaxies. The situation for bars is more complex since only three of the 15 galaxies with a photometric bar are classified as $j_{\star}$- bars, and curiously, one galaxy has a $j_{\star}$- bar that is not reported in its photometry. All $j_{\star}$- clumps correspond to small-scale overdensities present in the $B_\star$ maps. Only one of the two photometrically irregular galaxies is classified as $j_{\star}$- irregular, while the remaining eight $j_{\star}$- irregulars are reported as spirals of some type. The sAMSD maps naturally reveal new information about our galaxies that is difficult to capture using traditional photometric techniques.

In terms of global integrated properties, we confirmed the expected high linear correlation between $M_\star$ and $j_\star$ for our sample ($\rho_s = 0.87$). Using a Monte Carlo Markov chain method implemented by \cite{foreman2013emcee}, we constructed a linear fit in log-log space that determined our parameters for the Fall relation ($\alpha = 0.52 \pm 0.06$, $\beta = 2.83 \pm 0.03$ and $\sigma_{\perp} = 0.16 \pm 0.02$). The values for $\alpha$, $\beta$, and $\sigma_{\perp}$ reported here are in agreement with those stated by other studies of spiral discs in the local Universe \citep{romanowsky2012angular,posti2018angular,hardwick2022xgass}. We also identified correlations between $j_{\star}$ and the star formation rate, and between $j_{\star}$ and $M_{\ion{H}{i}}$ in our sample ($\rho_s = 0.73$ and $\rho_s = 0.71$, respectively). Regarding the link between the two-dimensional distribution of the stellar sAMSD and $j_\star$, we identified how the $j_\star$ types are grouped into different overlapping sectors along the Fall relation. This result suggests that there may be a morpho-kinematic evolution of late-type galaxies within the hierarchical paradigm of galaxy formation. We found that the average values of $f_{\rm gas}$, $c$, and $\rho_{\rm o}$ for each $j_\star$ type validate this hypothesis and may relate the redistribution of stellar sAMSD to the stability of the galactic disc. This picture excludes the $j_{\star}$- irregulars from the secular evolutionary line.

We have discussed how our morpho-kinematic classification groups galaxies in the same way as the spiral galaxy classification system proposed by \cite{elmegreen1987arm}. In this scenario, the low-mass $j_\star$ types ($j_\star$- clump and $j_\star$- bar) share the same characteristics as flocculent spiral galaxies. Meanwhile, the massive $j_\star$ types ($j_\star$- spirals and $j_\star$- ring) share the same properties of grand-design spiral galaxies. We propose that the $j_\star$- evolution is related to the various physical mechanisms of angular momentum propagation in the absence of tidal interactions. Following this framework, angular momentum is primarily redistributed in $j_\star$- clump galaxies through feedback processes and dynamical friction; in $j_\star$- bar galaxies through a combination of inner disc resonances, dynamical friction, and shocks; and in $j_\star$- spiral and $j_\star$- ring galaxies through shocks and the propagation of their quasi-stationary rotating density waves. It should be noted that these redistribution mechanisms are not mutually exclusive, and it is likely that all of them form part of the substructures that we identified in the sAMSD maps without being restricted to just one $j_\star$ type.

Finally, it is necessary to extend the applicability of our methodology to a larger sample of observations in parallel with its implementation for the computation of stellar sAMSD maps from synthetic observations of hydrodynamic simulations. This approach would not only validate our observational results but also deepen our understanding of the physical mechanisms driving galaxy evolution within the stellar $j_\star$–$M_\star$ parameter space. Expanding observational samples and incorporating analysis of numerical simulations may lead to further refinements of the $j_\star$- type classification, ultimately strengthening foundations of galaxy formation and evolution.

\section*{Data availability}

The sAMSD maps for our entire sample and the Tables \ref{table: mean_values}, \ref{table: geometrical_params_full}, and \ref{table: morphokinematic_params} are only available in electronic form at the CDS via anonymous ftp to cdsarc.u-strasbg.fr (130.79.128.5) or via \url{http://cdsweb.u-strasbg.fr/cgi-bin/qcat?J/A+A/}.

\begin{acknowledgements} We gratefully acknowledge the anonymous referee for insightful comments and constructive suggestions. We warmly thank Orient Dorey for making the unpublished \textsc{MaskingStars} software available (private communication), which was used in this analysis. W.M. acknowledges the funding of the French Agence Nationale de la Recherche for the project iMAGE (grant ANR-22-CE31-0007).
\end{acknowledgements}

\bibliographystyle{aa}
\bibliography{References}

\begin{appendix}

\clearpage

\onecolumn

\section{Table of general properties}

\FloatBarrier

\begin{table*}[h!]
\caption{Galaxy parameters of all galaxies in our sample.}
\label{table: geometrical_params_full}
\centering
\setlength\extrarowheight{1.5mm}
\begin{tabular}{c c c c c c c c c c}
\hline\hline\\[-2ex]
\makecell{N$^{\circ}$ \\ UGC \\ (1)} & \makecell{Morphological \\ type \\ (2)} & \makecell{$\alpha$ \\ {[J2000]} \\ (3)} & \makecell{$\delta$ \\ {[J2000]} \\ (4)} & \makecell{$i$ \\ {[\degr]} \\ (5)} & \makecell{PA \\ {[\degr]} \\ (6)} & \makecell{$D$ \\ {[Mpc]} \\ (7)}  & \makecell{$R_{\rm max}$ \\ {[kpc]} \\ (8)} & \makecell{$\mathrm{log_{10}}M_{\star}$ \\ {$[\mathrm{M_{\odot}}]$} \\ (9)} & \makecell{$\mathrm{log_{10}}j_{\star}$ \\ {$[\mathrm{kpc}$ $\mathrm{km}$ $\mathrm{s^{-1}}]$} \\ (10)}\\
\hline
   1913 & SAB(s)d & $02^{\mathrm{h}}27^{\mathrm{m}}16^{\mathrm{s}}.7$ & 33\degr34\arcmin 44\arcsec & $48 \pm 9$ & $108 \pm 3$ & $9.30$ & $23.91^{\ast}$ & $10.48$ & $2.93$ $^{0.01}_{0.03}$\\
   2800 & Im & $03^{\mathrm{h}}40^{\mathrm{m}}02^{\mathrm{s}}.4$ & $71^{\circ}24^{\prime}22^{\prime\prime}$ & $52 \pm 13$ & $11 \pm 4$ & $20.60$ & $12.05$ & $9.91$ & $2.82$ $^{0.05}_{0.07}$\\
   2855 & SABc & $03^{\mathrm{h}}48^{\mathrm{m}}20^{\mathrm{s}}.6$ & $70^{\circ}07^{\prime}58^{\prime\prime}$ & $68 \pm 2$ & $100 \pm 2$ & $17.50$ & $15.33^{\ast}$ & $10.58$ & $3.18$ $^{0.00}_{0.01}$\\
   3574 & SA(s)cd & $06^{\mathrm{h}}53^{\mathrm{m}}10^{\mathrm{s}}.3$ & $57^{\circ}10^{\prime}41^{\prime\prime}$ & $19 \pm 10$ & $99 \pm 3$ & $21.80$ & $20.11^{\ast}$ & $10.06$ & $3.18$ $^{0.01}_{0.03}$ \\
   3734 & SA(rs)c & $07^{\mathrm{h}}12^{\mathrm{m}}28^{\mathrm{s}}.7$ & $47^{\circ}10^{\prime}02^{\prime\prime}$ & $43 \pm 7$ & $139 \pm 2$ & $15.90$ & $11.23$ & $10.05$ & $2.63$ $^{0.01}_{0.02}$\\
   4284 & SA(s)cd & $08^{\mathrm{h}}14^{\mathrm{m}}40^{\mathrm{s}}.1$ & $49^{\circ}03^{\prime}43^{\prime\prime}$ & $59 \pm 9$ & $176 \pm 3$ & $9.80$ & $15.06^{\ast}$ & $10.05$ & $2.69$ $^{0.06}_{0.08}$\\
   4325 & SA(s)m & $08^{\mathrm{h}}19^{\mathrm{m}}20^{\mathrm{s}}.5$ & $50^{\circ}00^{\prime}36^{\prime\prime}$ & $63 \pm 14$ & $57 \pm 3$ & $10.90$ & $8.49^{\ast}$ & $9.17$ & $2.28$ $^{0.08}_{0.25}$\\
   4499 & SABdm & $08^{\mathrm{h}}37^{\mathrm{m}}41^{\mathrm{s}}.3$ & $51^{\circ}39^{\prime}12^{\prime\prime}$ & $50 \pm 14$ & $141 \pm 4$ & $12.20$ & $6.83$ & $8.93$ & $2.20$ $^{0.06}_{0.13}$ \\
   5251 & SBbc & $09^{\mathrm{h}}48^{\mathrm{m}}35^{\mathrm{s}}.8$ & $33^{\circ}25^{\prime}18^{\prime\prime}$ & $73 \pm 6$ & $80 \pm 3$ & $21.50$ & $35.88^{\ast}$ & $10.57$ & $3.02$ $^{0.07}_{0.08}$ \\
   5253 & (R$^\prime$)SA(rs)ab & $09^{\mathrm{h}}50^{\mathrm{m}}21^{\mathrm{s}}.9$ & $72^{\circ}16^{\prime}46^{\prime\prime}$ & $40 \pm 4$ & $176 \pm 2$ & $21.10$ & $23.59$ & $10.92$ & $3.12$ $^{0.01}_{0.01}$\\
   5414 & IAB(s)m & $10^{\mathrm{h}}03^{\mathrm{m}}57^{\mathrm{s}}.3$ & $40^{\circ}45^{\prime}26^{\prime\prime}$ & $71 \pm 13$ & $39 \pm 4$ & $10.00$ & $7.30^{\ast}$ & $9.28$ & $2.30$ $^{0.14}_{0.54}$ \\
   6537 & SAB(r)c & $11^{\mathrm{h}}33^{\mathrm{m}}21^{\mathrm{s}}.2$ & $47^{\circ}01^{\prime}47^{\prime\prime}$ & $47 \pm 5$ & $20 \pm 2$ & $14.30$ & $18.07$ & $10.51$ & $2.97$ $^{0.01}_{0.01}$\\
   6778 & SAB(rs)c & $11^{\mathrm{h}}48^{\mathrm{m}}38^{\mathrm{s}}.0$ & $48^{\circ}42^{\prime}39^{\prime\prime}$ & $49 \pm 4$ & $163 \pm 2$ & $15.50$ & $18.08^{\ast}$ & $10.19$ & $2.90$ $^{0.02}_{0.02}$\\
   7323 & SAB(s)dm & $12^{\mathrm{h}}17^{\mathrm{m}}30^{\mathrm{s}}.2$ & $45^{\circ}37^{\prime}11^{\prime\prime}$ & $51 \pm 11$ & $19 \pm 21^{\mathrm{Mor}}$ & $8.10$ & $16.77^{\ast}$ & $9.92$ & $2.61$ $^{0.04}_{0.05}$\\
   7766 & SAB(rs)cd & $12^{\mathrm{h}}35^{\mathrm{m}}57^{\mathrm{s}}.7$ & $27^{\circ}57^{\prime}35^{\prime\prime}$ & $69 \pm 3$ & $143 \pm 2$ & $13.00$ & $37.12^{\ast}$ & $10.72$ & $3.01$ $^{0.03}_{0.03}$\\
   8334 & SA(rs)bc & $13^{\mathrm{h}}15^{\mathrm{m}}49^{\mathrm{s}}.3$ & $42^{\circ}01^{\prime}45^{\prime\prime}$ & $66 \pm 1$ & $100 \pm 1$ & $9.80$ & $28.78$ & $11.33$ & $3.11$ $^{0.00}_{0.00}$\\
   8490 & SA(s)m & $13^{\mathrm{h}}29^{\mathrm{m}}36^{\mathrm{s}}.6$ & $58^{\circ}25^{\prime}13^{\prime\prime}$ & $40 \pm 15$ & $167 \pm 3$ & $4.70$ & $5.66$ & $9.09$ & $2.17$ $^{0.05}_{0.07}$\\
   9179 & SAB(s)d & $14^{\mathrm{h}}19^{\mathrm{m}}48^{\mathrm{s}}.2$ & $56^{\circ}43^{\prime}46^{\prime\prime}$ & $36 \pm 14$ & $49 \pm 4$ & $5.70$ & $8.57$ & $9.42$ & $2.49$ $^{0.05}_{0.08}$\\
   9649 & SB(rs)b & $14^{\mathrm{h}}57^{\mathrm{m}}45^{\mathrm{s}}.4$ & $71^{\circ}40^{\prime}55^{\prime\prime}$ & $54 \pm 6$ & $55 \pm 3$ & $7.70$ & $5.44$ & $9.30$ & $2.20$ $^{0.02}_{0.02}$\\
   9858 & SABbc & $15^{\mathrm{h}}26^{\mathrm{m}}41^{\mathrm{s}}.6$ & $40^{\circ}33^{\prime}53^{\prime\prime}$ & $75 \pm 2$ & $70 \pm 2$ & $38.20$ & $37.14^{\ast}$ & $10.75$ & $2.22$ $^{0.03}_{0.03}$\\
   9969 & SAB(r)b & $15^{\mathrm{h}}39^{\mathrm{m}}36^{\mathrm{s}}.9$ & $59^{\circ}19^{\prime}55^{\prime\prime}$ & $61 \pm 1$ & $16 \pm 1$ & $36.00$ & $32.40^{\ast}$ & $11.21$ & $3.50$ $^{0.00}_{0.00}$\\
   10075 & SA(s)cd & $15^{\mathrm{h}}51^{\mathrm{m}}25^{\mathrm{s}}.3$ & $62^{\circ}18^{\prime}36^{\prime\prime}$ & $62 \pm 2$ & $30 \pm 1$ & $14.70$ & $19.66$ & $10.56$ & $2.87$ $^{0.01}_{0.01}$\\
   10359 & SB(s)cd pec & $16^{\mathrm{h}}20^{\mathrm{m}}57^{\mathrm{s}}.9$ & $65^{\circ}23^{\prime}26^{\prime\prime}$ & $44 \pm 12$ & $104 \pm 3$ & $16.00$ & $17.11$ & $9.63$ & $2.79$ $^{0.03}_{0.04}$ \\
   10470 & (R)SB(rs)bc & $16^{\mathrm{h}}32^{\mathrm{m}}39^{\mathrm{s}}.3$ & $78^{\circ}11^{\prime}53^{\prime\prime}$ & $34 \pm 9$ & $155 \pm 26^{\mathrm{Mor}}$ & $21.20$ & $17.54$ & $10.04$ & $2.82$ $^{0.01}_{0.01}$ \\
   11012 & SA(s)cd & $17^{\mathrm{h}}49^{\mathrm{m}}26^{\mathrm{s}}.3$ & $70^{\circ}08^{\prime}40^{\prime\prime}$ & $72 \pm 2$ & $123 \pm 7^{\mathrm{Mor}}$ & $5.30$ & $7.97$ & $10.22$ & $2.29$ $^{0.02}_{0.02}$\\
   11597 & SAB(rs)cd & $20^{\mathrm{h}}34^{\mathrm{m}}52^{\mathrm{s}}.4$ & $60^{\circ}09^{\prime}14^{\prime\prime}$ & $40 \pm 10$ & $61 \pm 3$ & $5.90$ & $20.08$ & $10.48$ & $3.04$ $^{0.00}_{0.01}$ \\
   11670 & SA(r)0/a & $21^{\mathrm{h}}03^{\mathrm{m}}33^{\mathrm{s}}.5$ & $29^{\circ}53^{\prime}52^{\prime\prime}$ & $65 \pm 2$ & $153 \pm 2$ & $12.80$ & $18.65^{\ast}$ & $10.65$ & $2.93$ $^{0.02}_{0.02}$ \\
   11852 & SBa & $21^{\mathrm{h}}55^{\mathrm{m}}59^{\mathrm{s}}.3$ & $27^{\circ}53^{\prime}55^{\prime\prime}$ & $47 \pm 7$ & $9 \pm 3$ & $80.10$ & $21.59$ & $10.49$ & $3.10$ $^{0.01}_{0.02}$\\
   11914 & (R)SA(r)ab & $22^{\mathrm{h}}07^{\mathrm{m}}52^{\mathrm{s}}.4$ & $31^{\circ}21^{\prime}34^{\prime\prime}$ & $33 \pm 4$ & $86 \pm 2$ & $15.00$ & $33.06^{\ast}$ & $10.76$ & $3.35$ $^{0.00}_{0.01}$ \\
   12754 & SB(s)cd & $23^{\mathrm{h}}43^{\mathrm{m}}54^{\mathrm{s}}.2$ & $26^{\circ}04^{\prime}35^{\prime\prime}$ & $53 \pm 5$ & $162 \pm 2$ & $8.90$ & $10.80^{\ast}$ & $9.37$ & $2.61$ $^{0.01}_{0.02}$\\[1ex]
\hline
\end{tabular}
\tablefoot{(1) Name of the galaxy in the UGC catalogue, (2) morphological type extracted from NASA/IPAC Extragalactic Database (NED), (3) and (4) right ascension and declination of the photometric centre determined as the brightest pixel of the WISE $W_1$ band image, (5) and (6) kinematic inclination and PA taken from the velocity field analysis carried out in \cite{epinat2008ghaspb}. The Mor superindex indicates that the photometric PA taken from HyperLeda replaces the kinematic PA to compute the stellar sAMSD. (7) distance to the galaxy taken from \cite{epinat2008ghaspb}. (8) maximum radius determined by the criterion presented in Eq. (\ref{eq: R_max}). The $\ast$ superindex flags the galaxies for which the RC was extrapolated to reach $R_{\rm max}$, (9) stellar mass calculated following Eq. (\ref{eq: Mass}), and (10) total stellar sAM computed using Eq. (\ref{eq: j}).}
\end{table*}

\clearpage

\section{Cumulative stellar sAM profiles}\label{app: cumulative stellar sAM}

\FloatBarrier

\begin{figure*}[ht!]
    \centering
    \begin{subfigure}[b]{0.245\textwidth}
        \includegraphics[width=\textwidth]{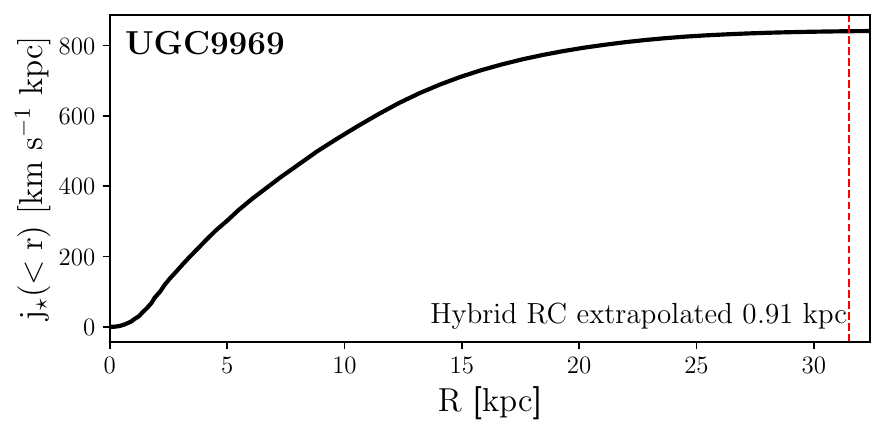}
    \end{subfigure}
    \begin{subfigure}[b]{0.245\textwidth}
        \includegraphics[width=\textwidth]{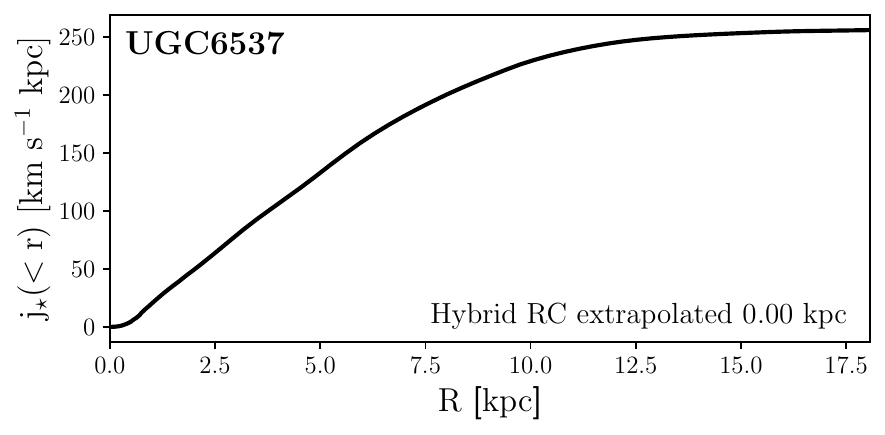}
    \end{subfigure}
    \begin{subfigure}[b]{0.245\textwidth}
        \includegraphics[width=\textwidth]{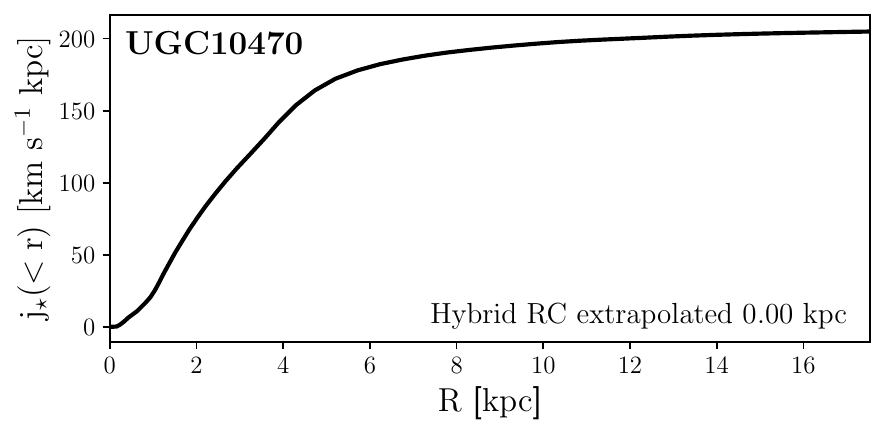}
    \end{subfigure}
    \begin{subfigure}[b]{0.245\textwidth}
        \includegraphics[width=\textwidth]{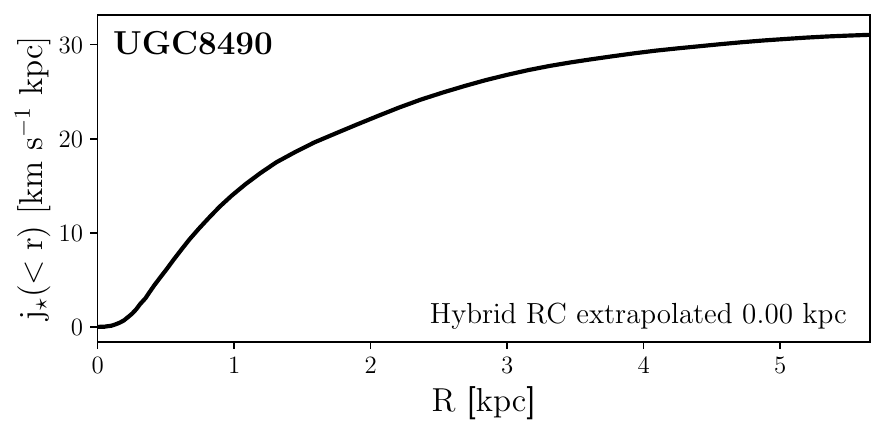}
    \end{subfigure}
    \\
    \begin{subfigure}[b]{0.245\textwidth}
        \includegraphics[width=\textwidth]{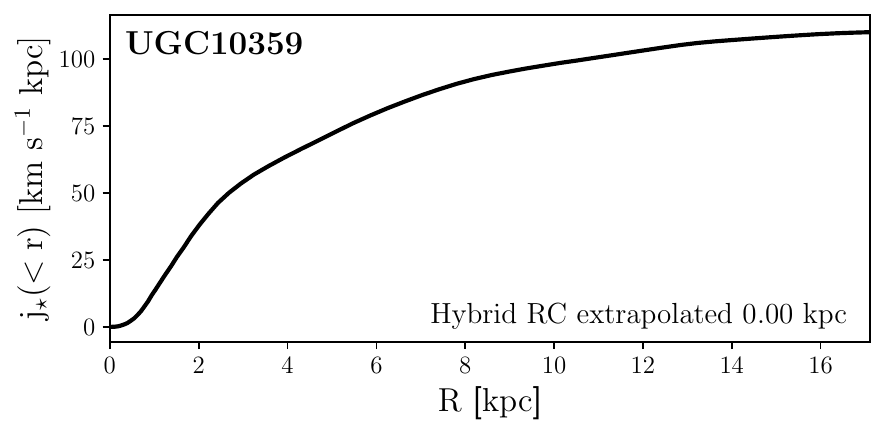}
    \end{subfigure}
    \begin{subfigure}[b]{0.245\textwidth}
        \includegraphics[width=\textwidth]{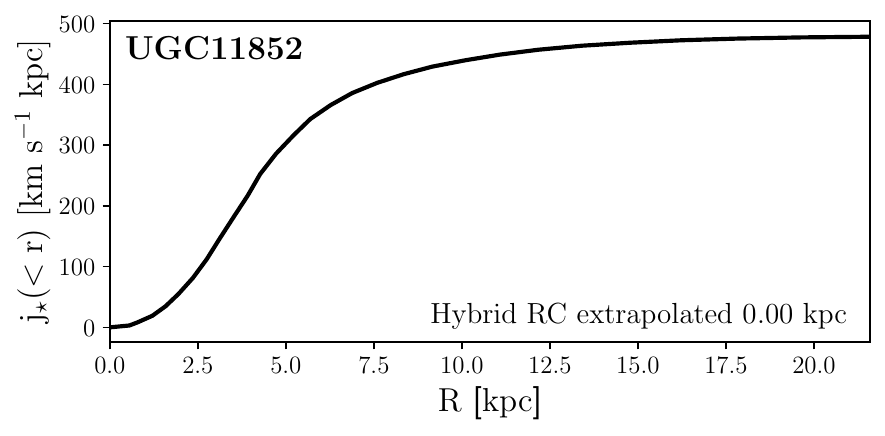}
    \end{subfigure}
    \begin{subfigure}[b]{0.245\textwidth}
        \includegraphics[width=\textwidth]{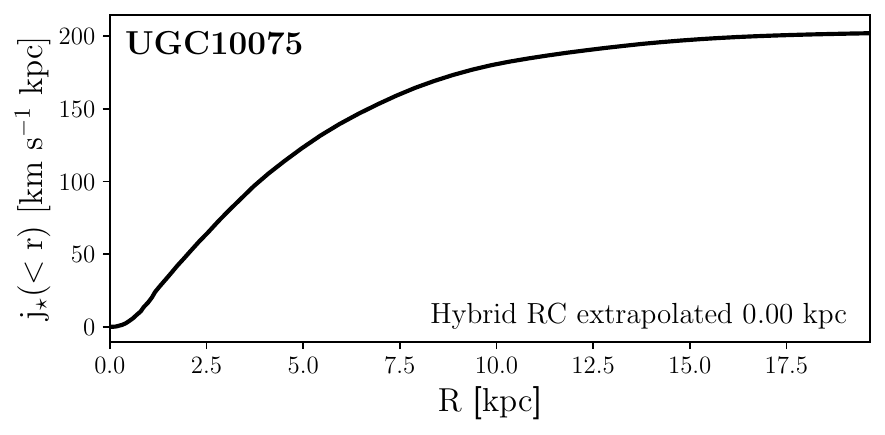}
    \end{subfigure}
    \begin{subfigure}[b]{0.245\textwidth}
        \includegraphics[width=\textwidth]{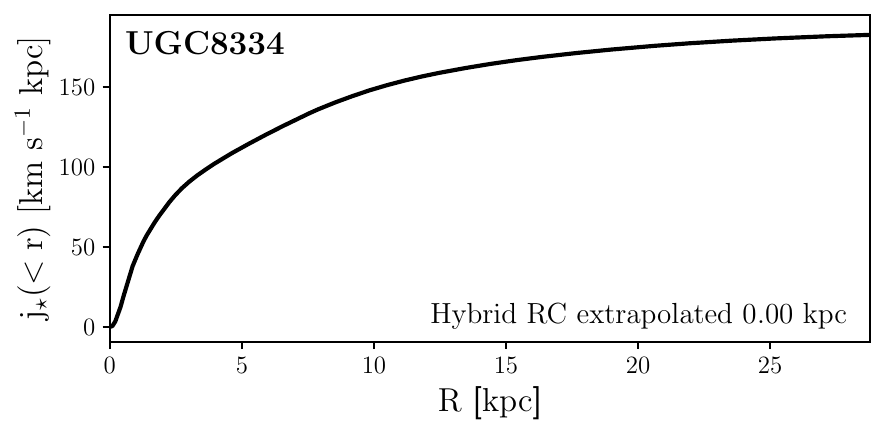}
    \end{subfigure}
    \\
    \begin{subfigure}[b]{0.245\textwidth}
        \includegraphics[width=\textwidth]{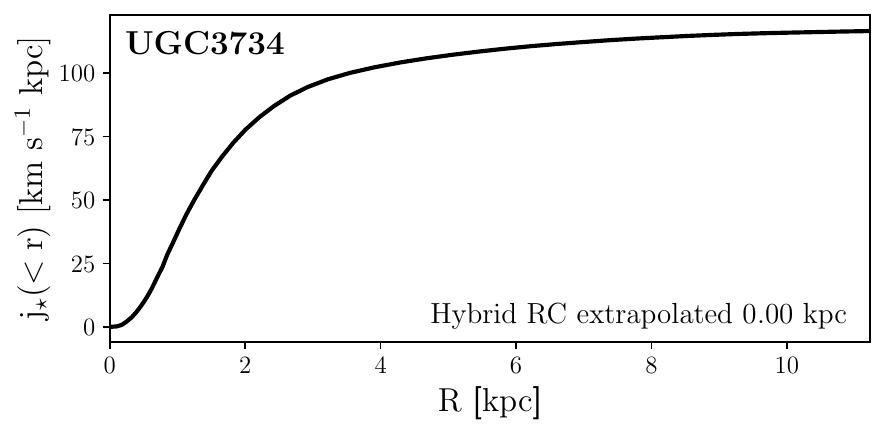}
    \end{subfigure}
    \begin{subfigure}[b]{0.245\textwidth}
        \includegraphics[width=\textwidth]{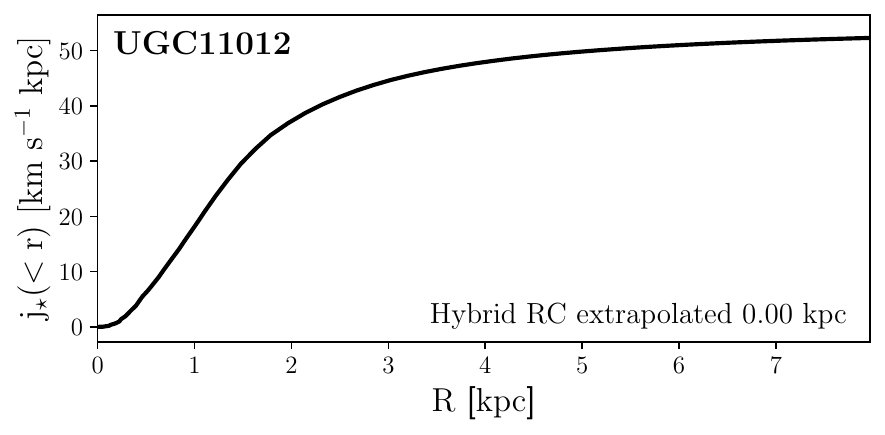}
    \end{subfigure}
    \begin{subfigure}[b]{0.245\textwidth}
        \includegraphics[width=\textwidth]{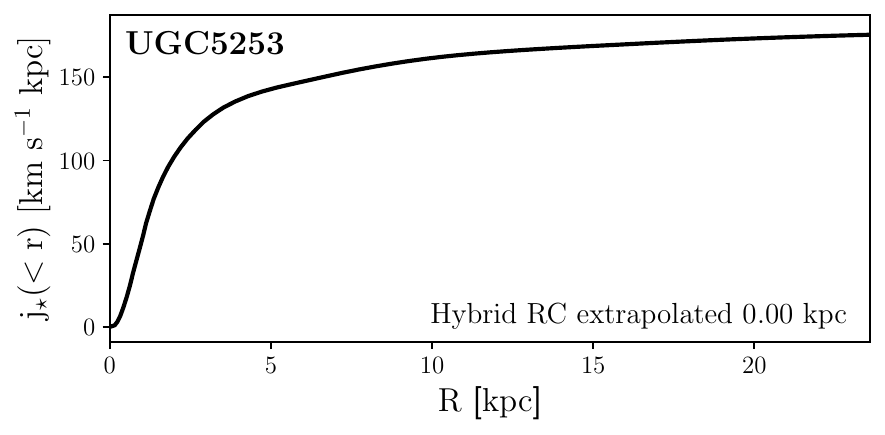}
    \end{subfigure}
    \begin{subfigure}[b]{0.245\textwidth}
        \includegraphics[width=\textwidth]{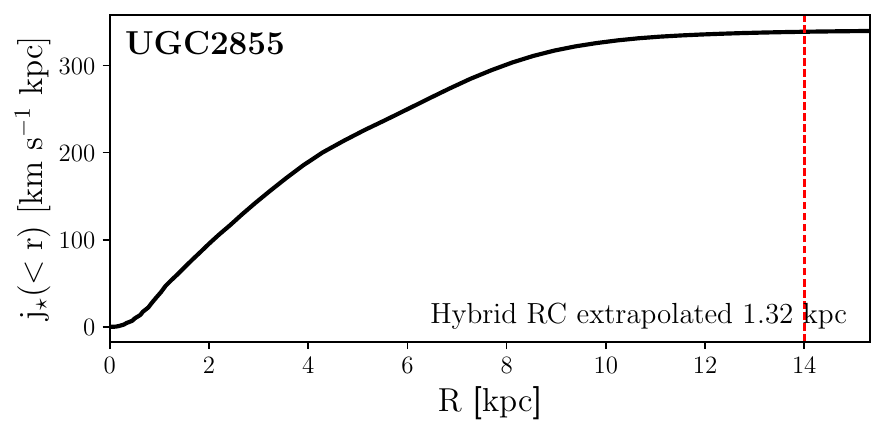}
    \end{subfigure}
    \\
    \begin{subfigure}[b]{0.245\textwidth}
        \includegraphics[width=\textwidth]{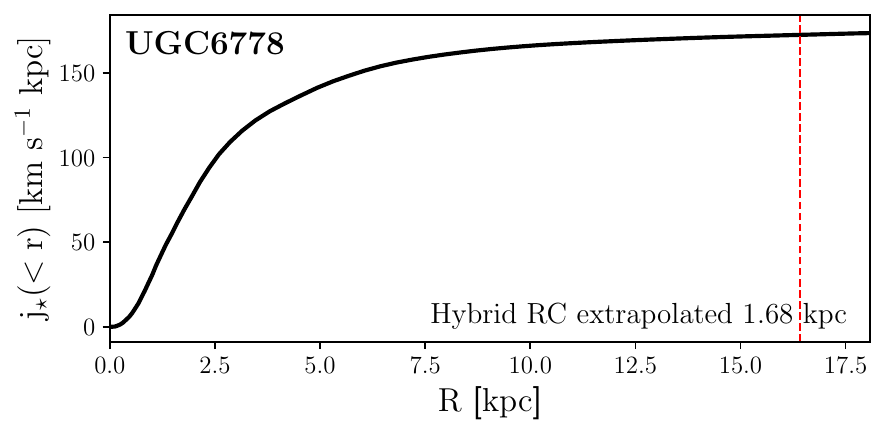}
    \end{subfigure}
    \begin{subfigure}[b]{0.245\textwidth}
        \includegraphics[width=\textwidth]{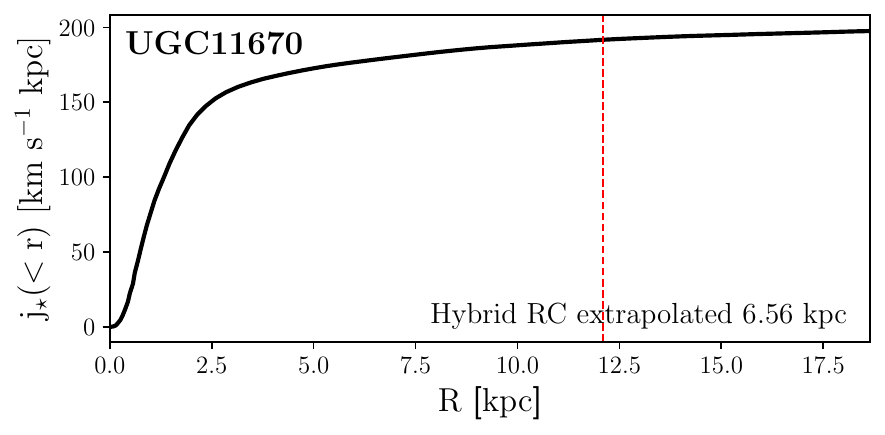}
    \end{subfigure}
    \begin{subfigure}[b]{0.245\textwidth}
        \includegraphics[width=\textwidth]{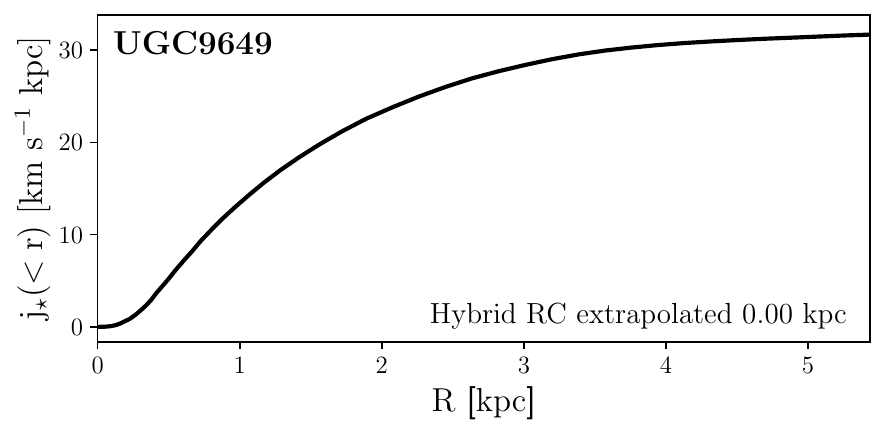}
    \end{subfigure}
    \begin{subfigure}[b]{0.245\textwidth}
        \includegraphics[width=\textwidth]{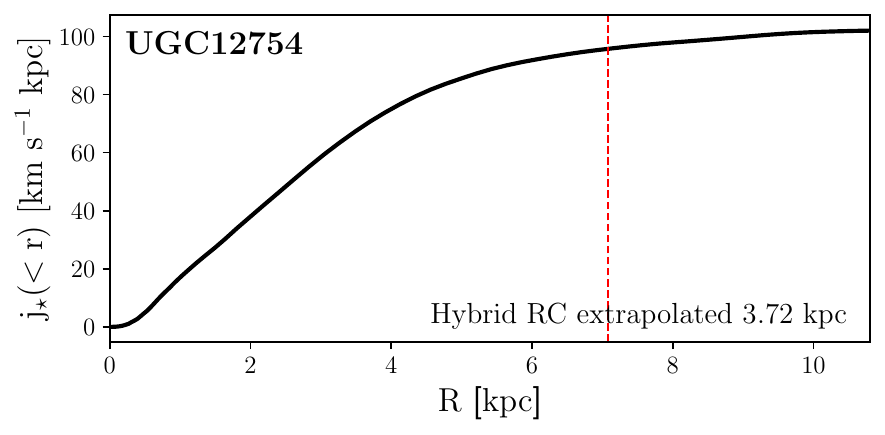}
    \end{subfigure}
    \\
    \begin{subfigure}[b]{0.245\textwidth}
        \includegraphics[width=\textwidth]{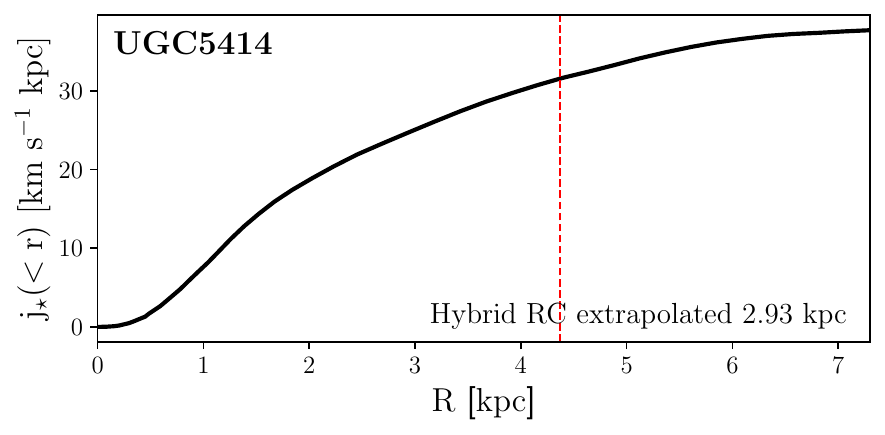}
    \end{subfigure}
    \begin{subfigure}[b]{0.245\textwidth}
        \includegraphics[width=\textwidth]{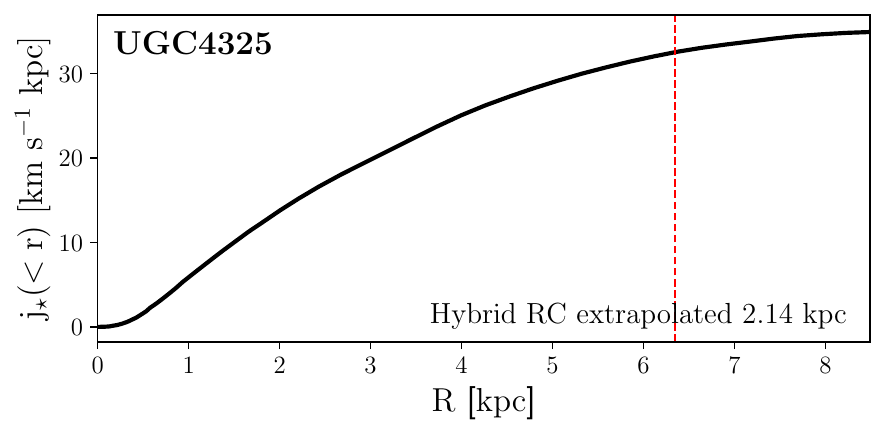}
    \end{subfigure}
    \begin{subfigure}[b]{0.245\textwidth}
        \includegraphics[width=\textwidth]{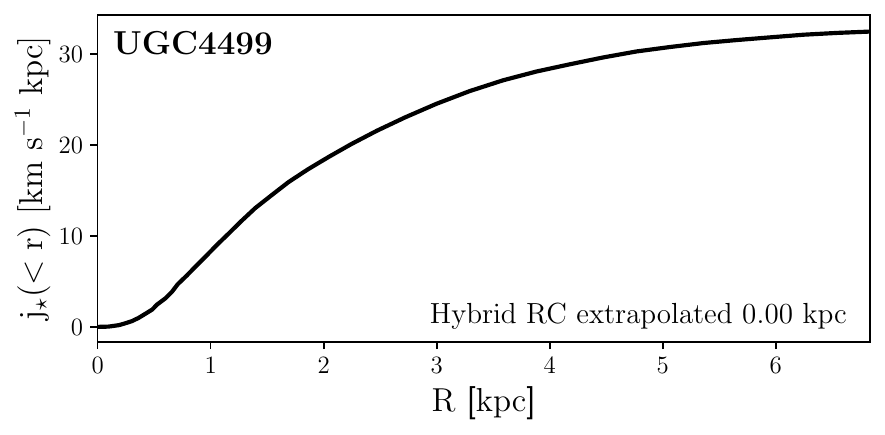}
    \end{subfigure}
    \begin{subfigure}[b]{0.245\textwidth}
        \includegraphics[width=\textwidth]{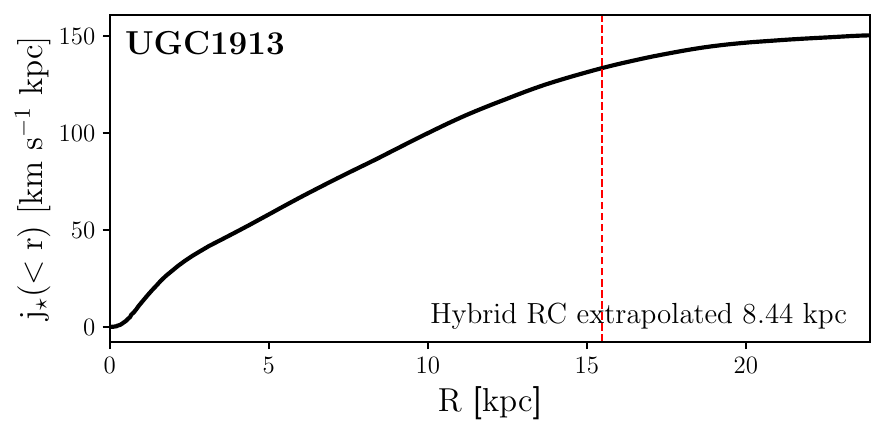}
    \end{subfigure}
    \\
    \begin{subfigure}[b]{0.245\textwidth}
        \includegraphics[width=\textwidth]{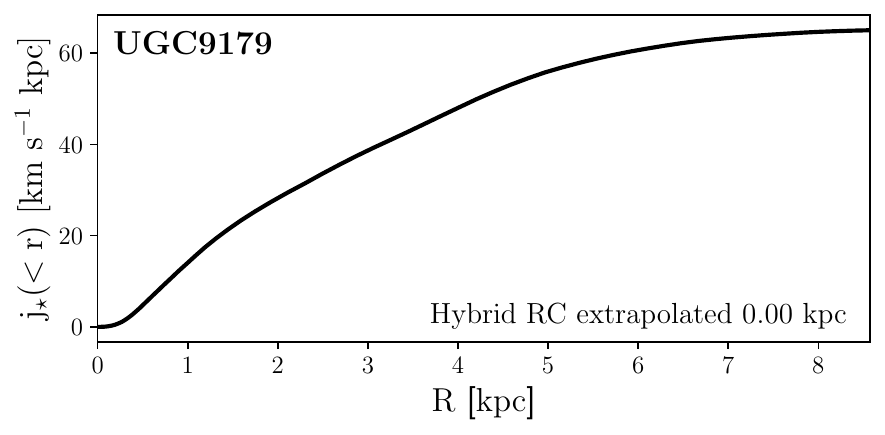}
    \end{subfigure}
    \begin{subfigure}[b]{0.245\textwidth}
        \includegraphics[width=\textwidth]{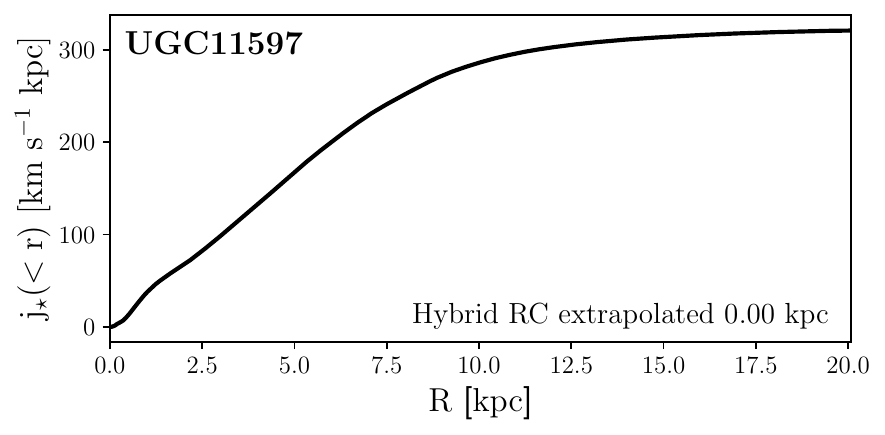}
    \end{subfigure}
    \begin{subfigure}[b]{0.245\textwidth}
        \includegraphics[width=\textwidth]{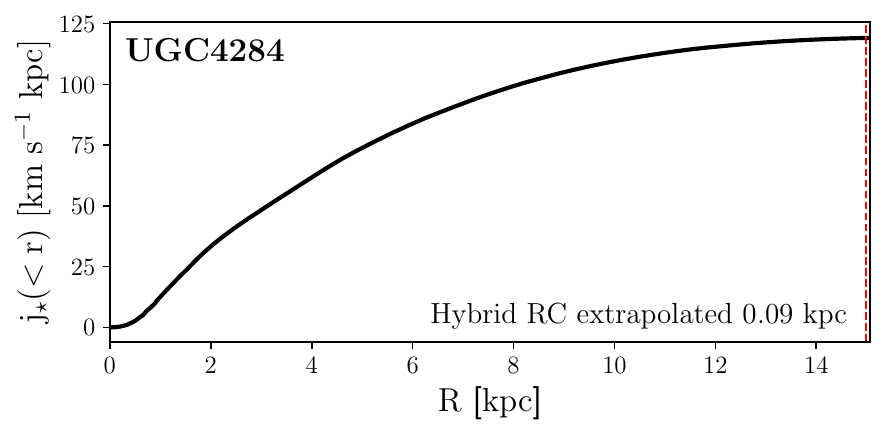}
    \end{subfigure}
    \begin{subfigure}[b]{0.245\textwidth}
        \includegraphics[width=\textwidth]{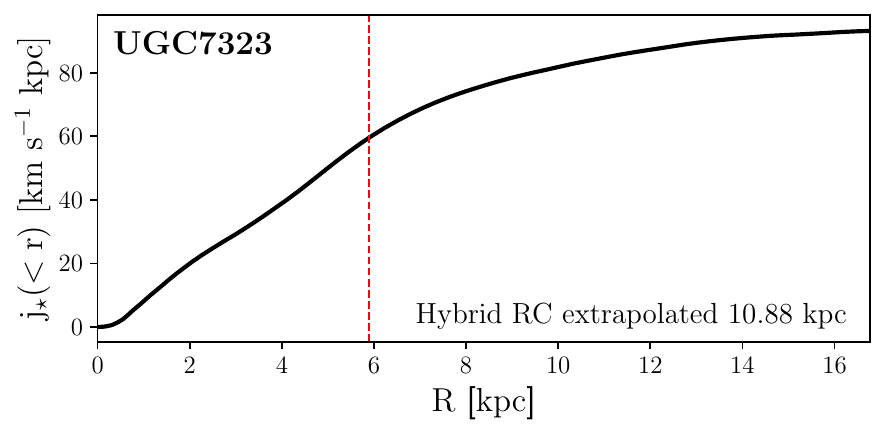}
    \end{subfigure}
    \\
    \begin{subfigure}[b]{0.245\textwidth}
        \includegraphics[width=\textwidth]{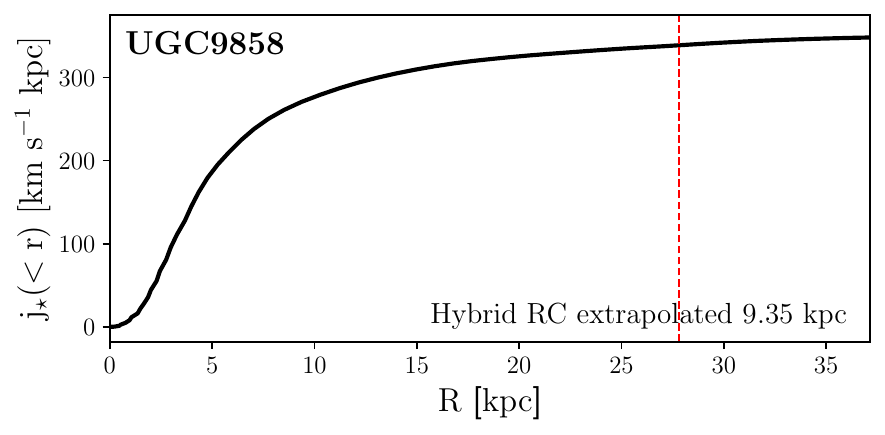}
    \end{subfigure}
    \begin{subfigure}[b]{0.245\textwidth}
        \includegraphics[width=\textwidth]{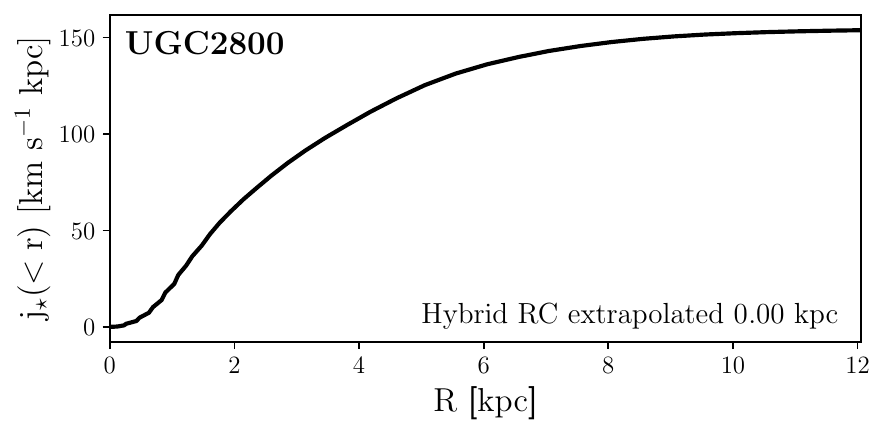}
    \end{subfigure}
    \begin{subfigure}[b]{0.245\textwidth}
        \includegraphics[width=\textwidth]{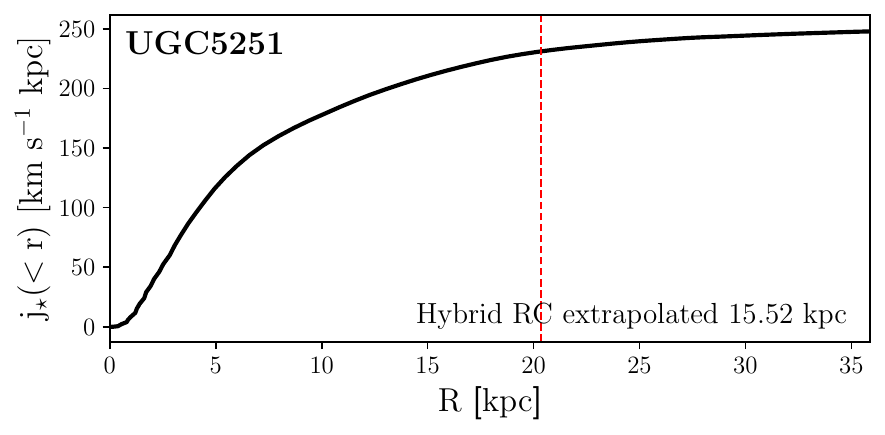}
    \end{subfigure}
    \begin{subfigure}[b]{0.245\textwidth}
        \includegraphics[width=\textwidth]{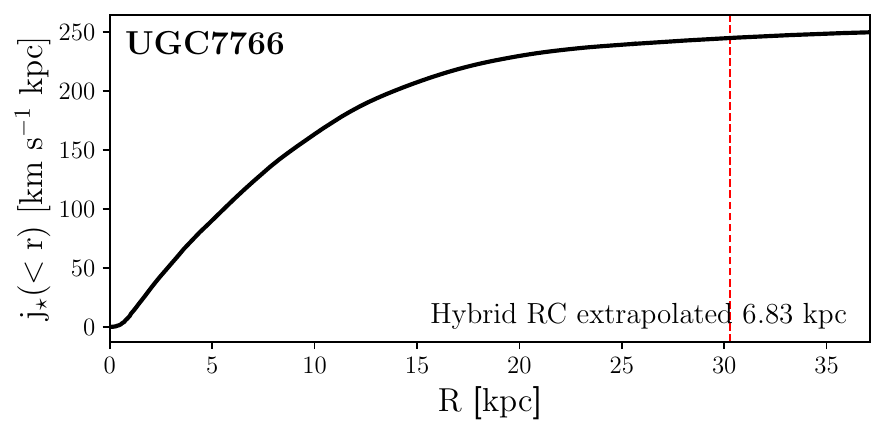}
    \end{subfigure}
    \\
    \begin{subfigure}[b]{0.245\textwidth}
        \includegraphics[width=\textwidth]{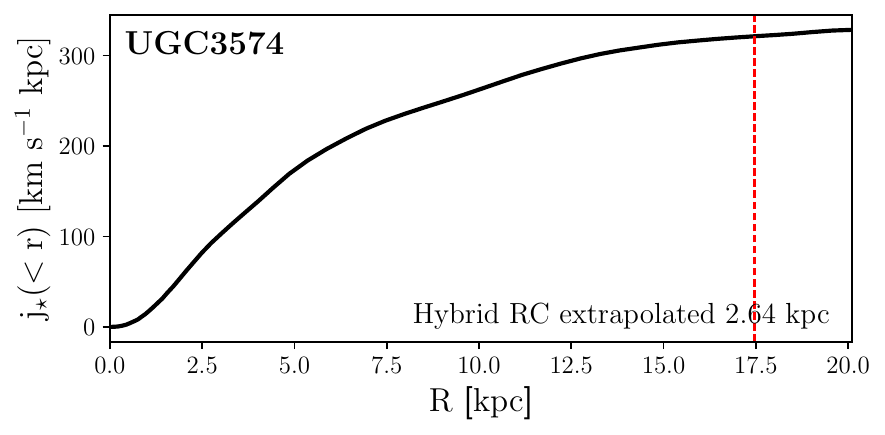}
    \end{subfigure}
    \begin{subfigure}[b]{0.245\textwidth}
        \includegraphics[width=\textwidth]{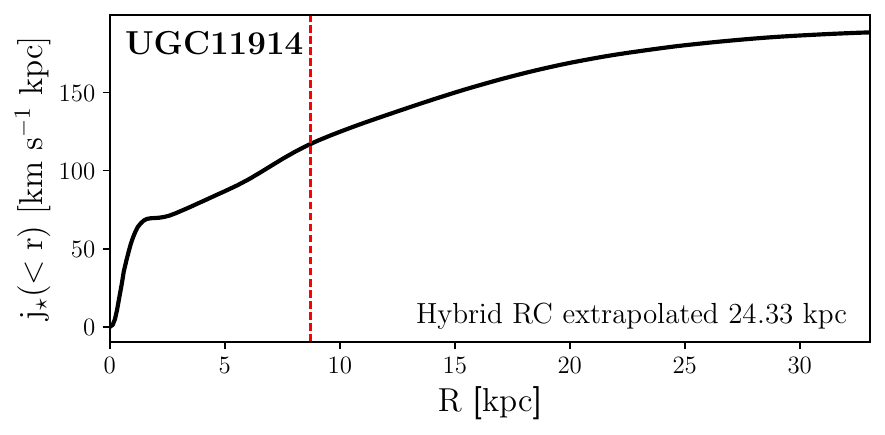}
    \end{subfigure}
    \caption{Cumulative stellar sAM profiles for UGC9969, UGC6537, UGC10470, UGC8490, UGC10359, UGC11852, UGC10075, UGC8334, UGC3734, UGC11012, UGC5253, UGC2855, UGC6778, UGC11670, UGC9649, UGC12754, UGC5414, UGC4325, UGC4499, UGC1913, UGC9179, UGC11597, UGC4284, UGC7323, UGC9858, UGC2800, UGC5251, UGC7766, UGC3574 and UGC11914. All the profiles are shown until they reach their $R_{\rm max}$. The dashed red line shows the last available radius in the hybrid RC of galaxies where extrapolation was necessary. The extrapolation remains below 50\% of $R_{\rm max}$ for 13 out of 15 galaxies. The impact on $j_{\star}(<r)$ is limited, as these extrapolations only induce a median rise of 3\%.
    The rise in $j_{\star}(<r)$ is between 36\% and 38\% for only two galaxies, between 11\% and 17\% for two other galaxies, and below 7\% for all other galaxies.
    }
    \label{fig: app_j_prof}
\end{figure*}

\clearpage

\section{Surface brightness and stellar sAMSD maps}\label{app: stelar sAM maps}

Figure \ref{fig: app_j_maps} shows the SB and the stellar sAMSD maps of the galaxies that complete our sample, excluding those representing each morpho-kinematic class described in Sect. \ref{sec: sAM maps results}. All are labelled with their corresponding morphological type and $j_{\star}$ type, and they are presented in the same order used for the description of the categories in Sect. \ref{sec: sAM maps results}.

\begin{figure*}[ht!]
    \centering
    \begin{subfigure}[b]{0.245\textwidth}
        \includegraphics[width=\textwidth]{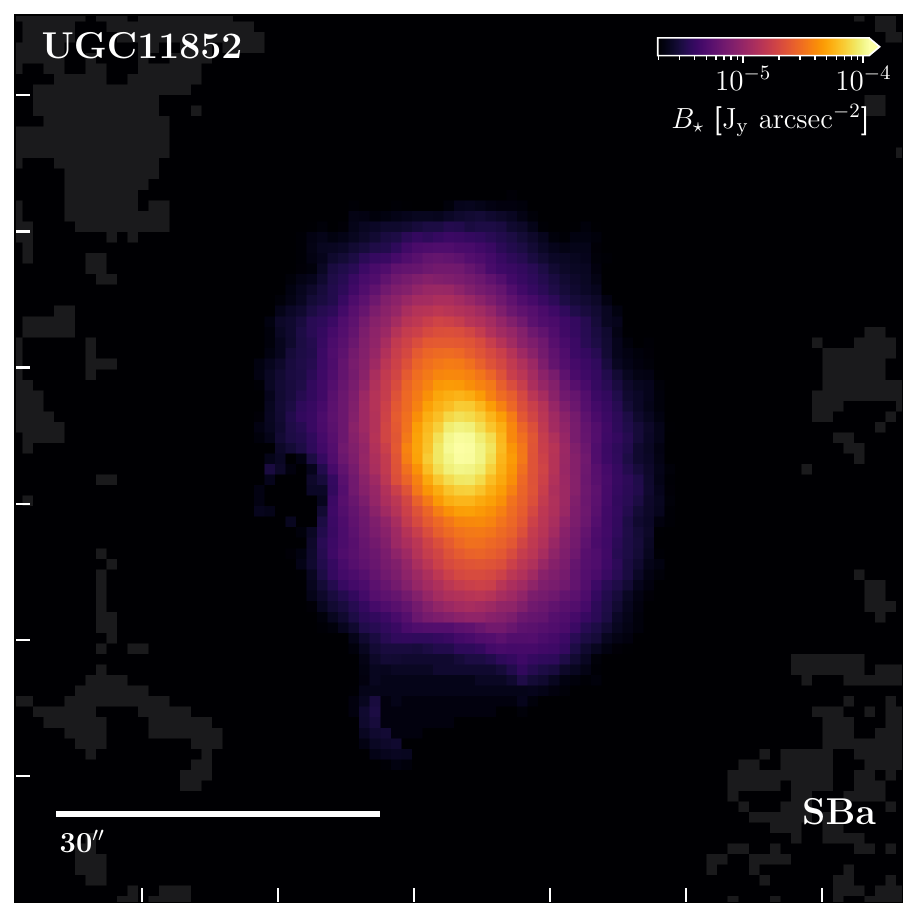}
    \end{subfigure}
    \begin{subfigure}[b]{0.245\textwidth}
        \includegraphics[width=\textwidth]{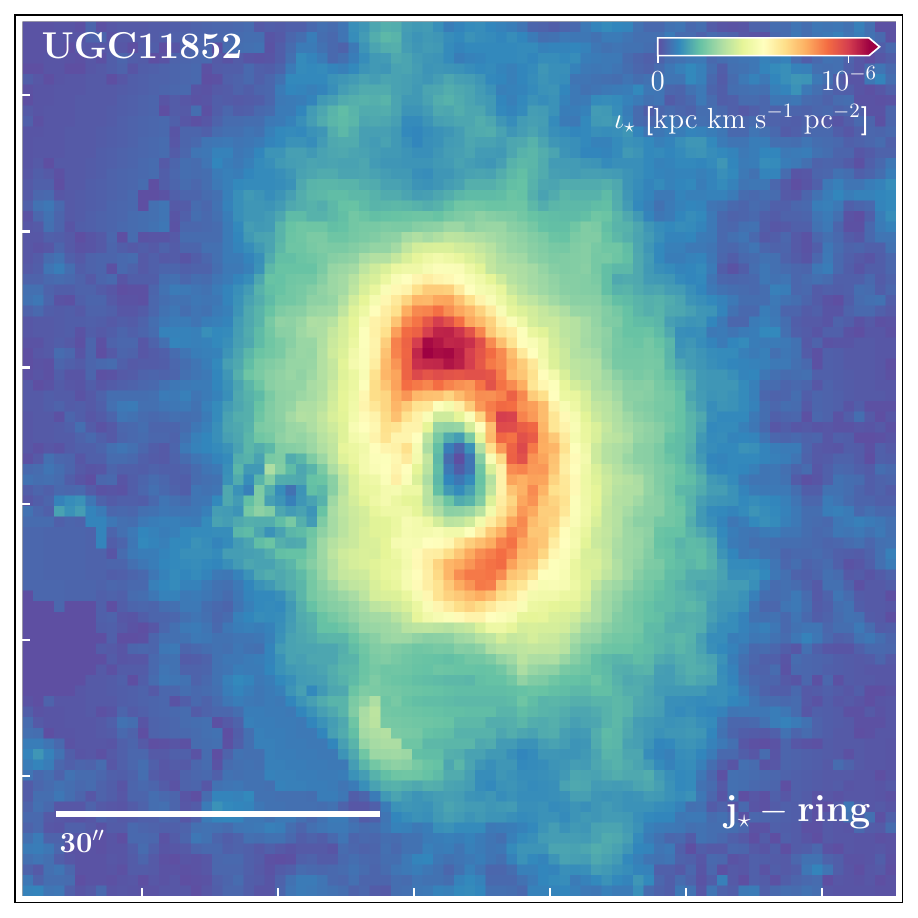}
    \end{subfigure}
    \begin{subfigure}[b]{0.245\textwidth}
        \includegraphics[width=\textwidth]{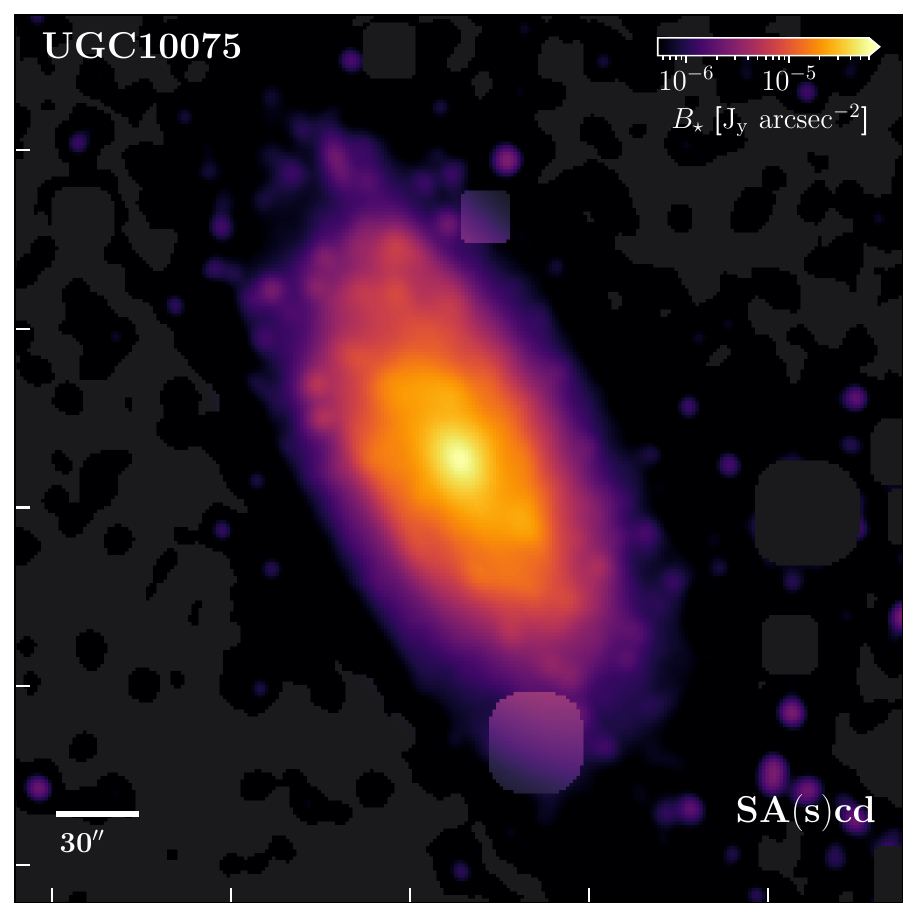}
    \end{subfigure}
    \begin{subfigure}[b]{0.245\textwidth}
        \includegraphics[width=\textwidth]{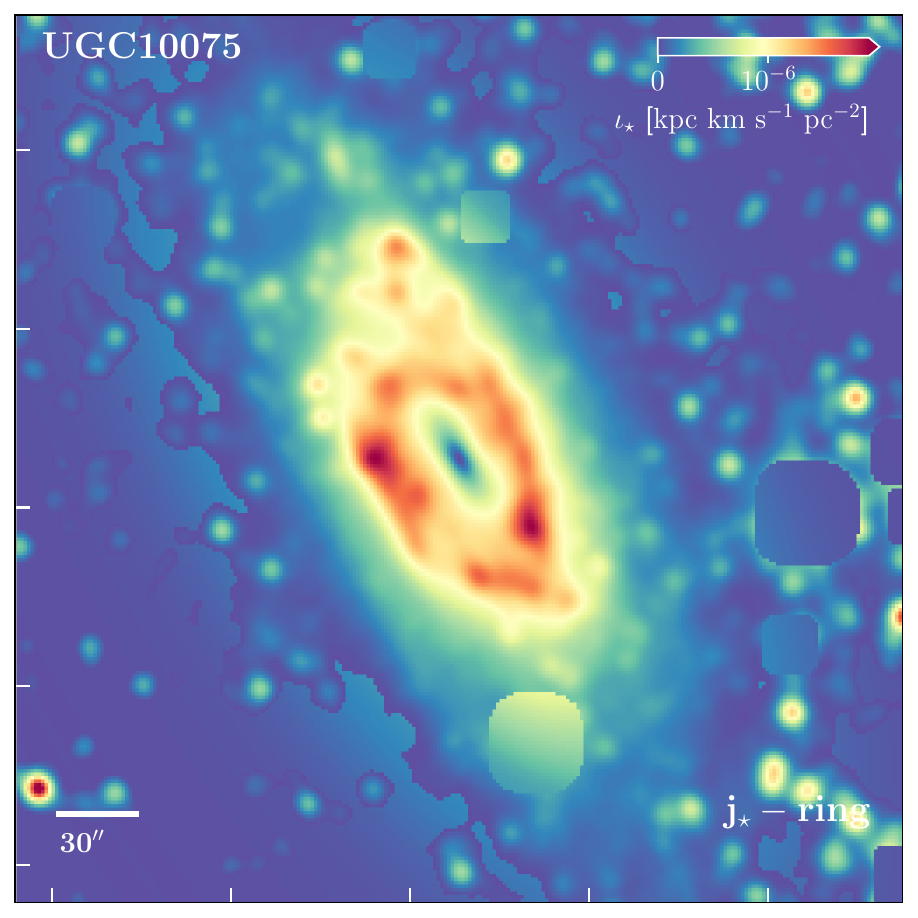}
    \end{subfigure}
    \\
    \begin{subfigure}[b]{0.245\textwidth}
        \includegraphics[width=\textwidth]{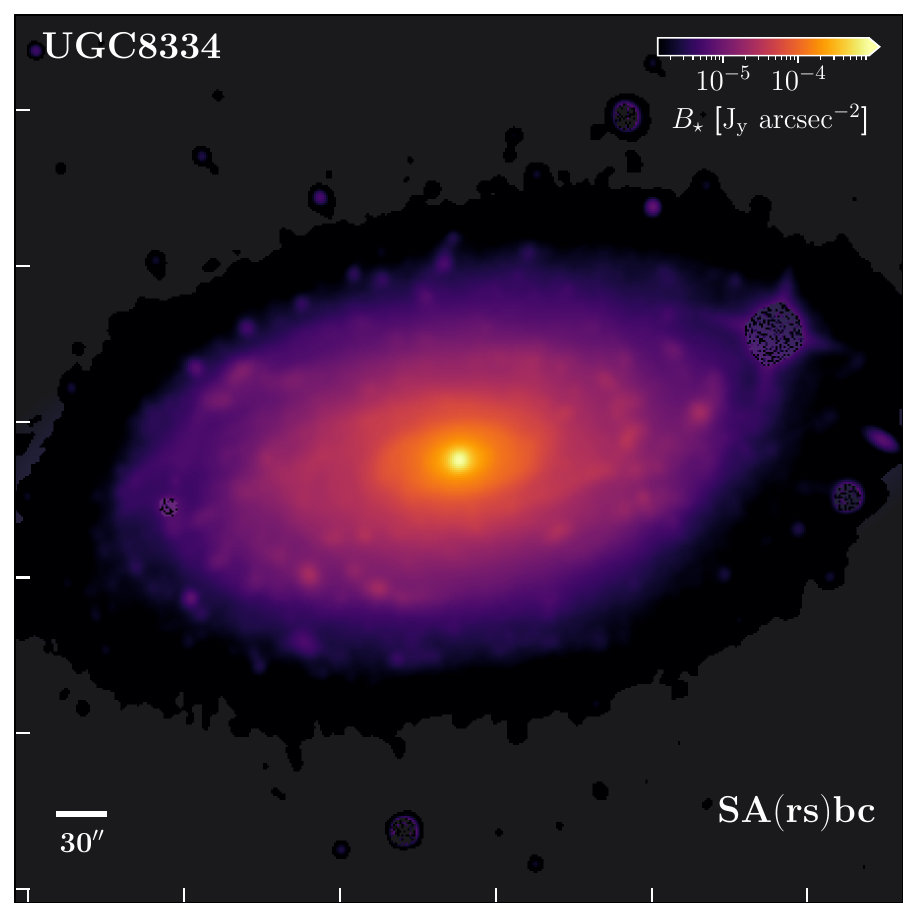}
    \end{subfigure}
    \begin{subfigure}[b]{0.245\textwidth}
        \includegraphics[width=\textwidth]{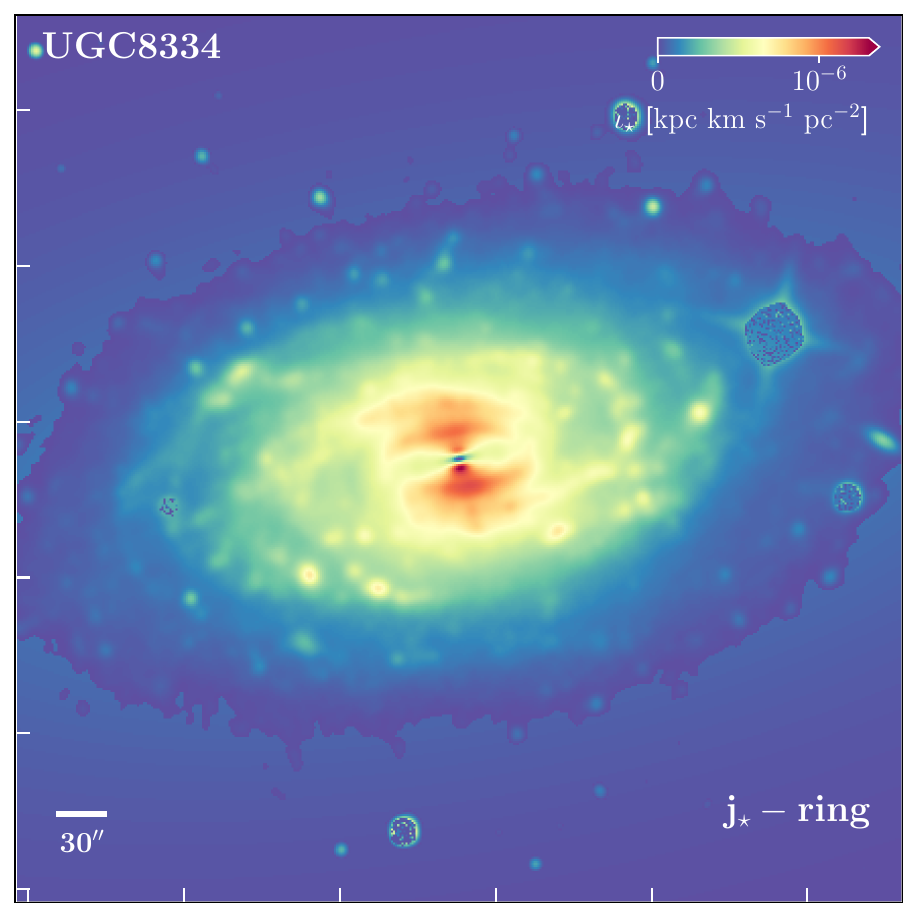}
    \end{subfigure}
    \begin{subfigure}[b]{0.245\textwidth}
        \includegraphics[width=\textwidth]{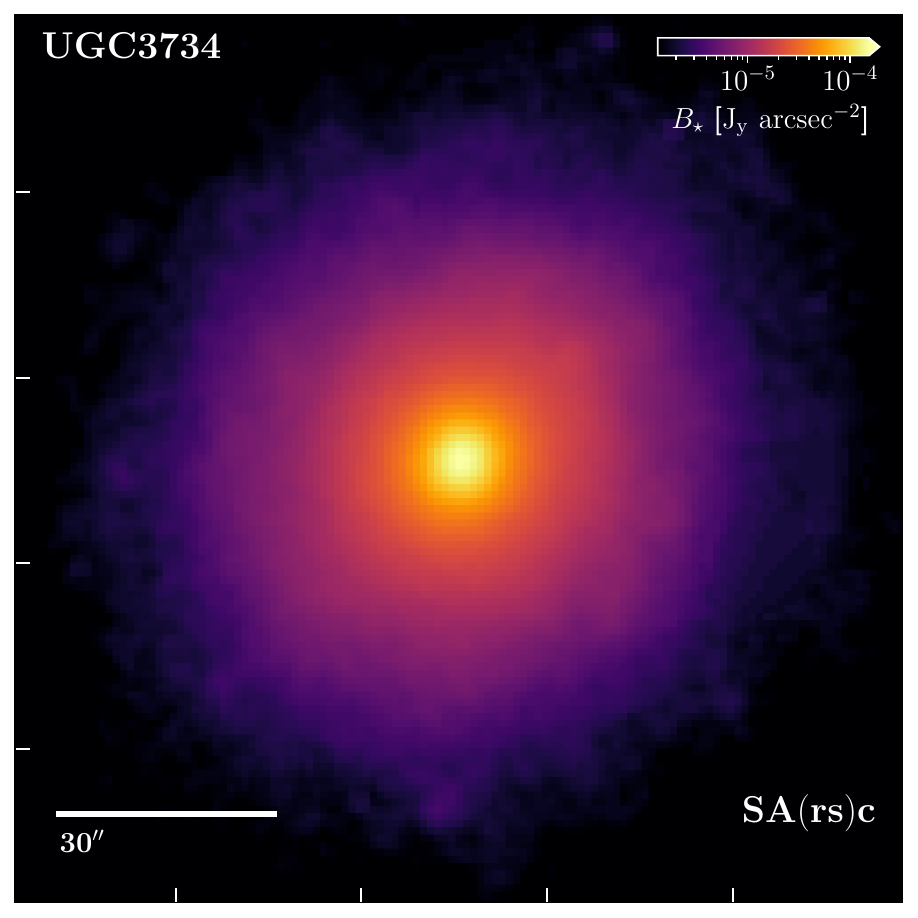}
    \end{subfigure}
    \begin{subfigure}[b]{0.245\textwidth}
        \includegraphics[width=\textwidth]{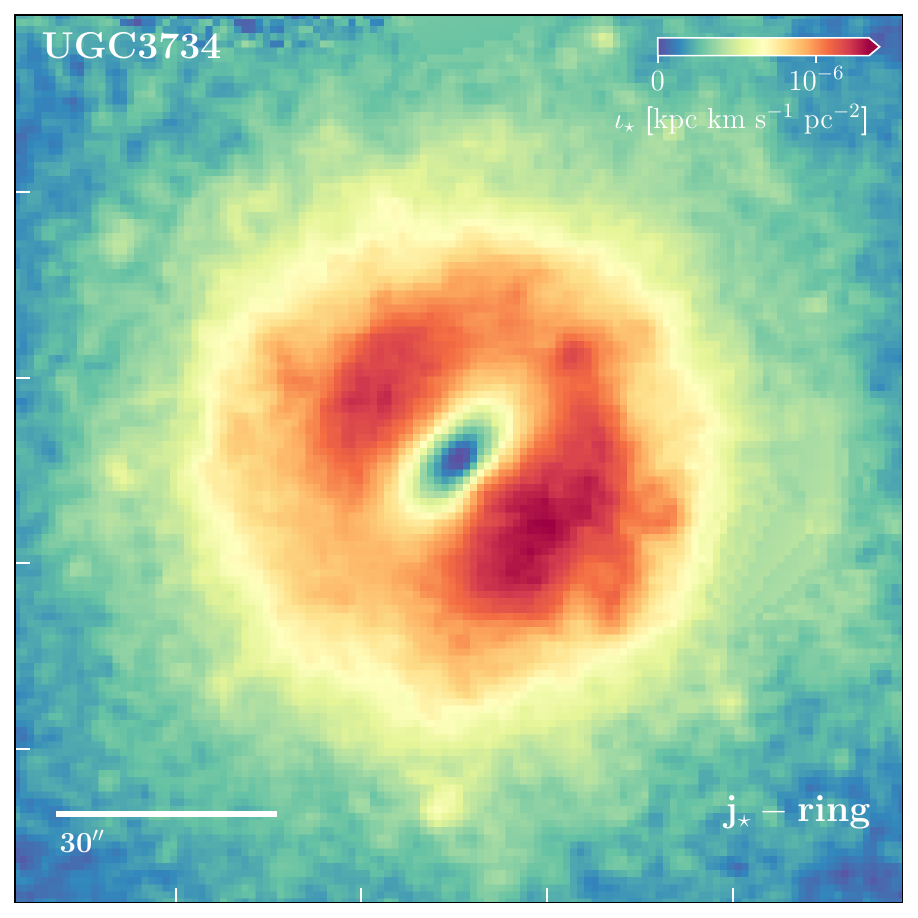}
    \end{subfigure}
    \\
    \begin{subfigure}[b]{0.245\textwidth}
        \includegraphics[width=\textwidth]{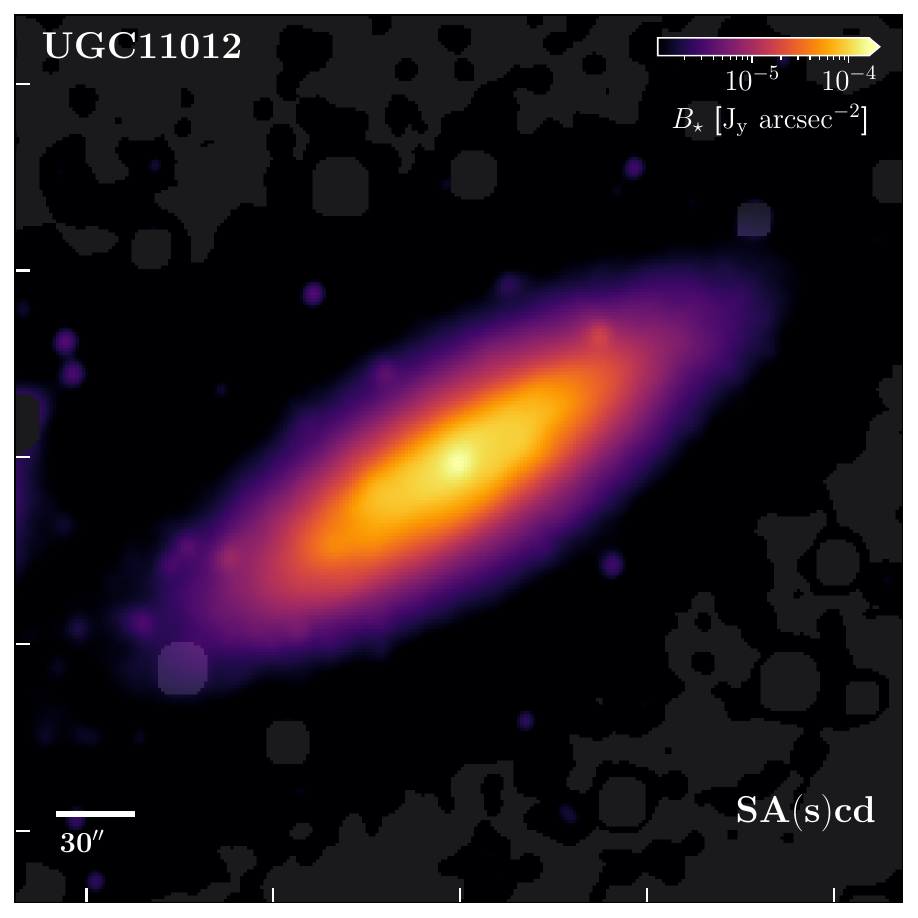}
    \end{subfigure}
    \begin{subfigure}[b]{0.245\textwidth}
        \includegraphics[width=\textwidth]{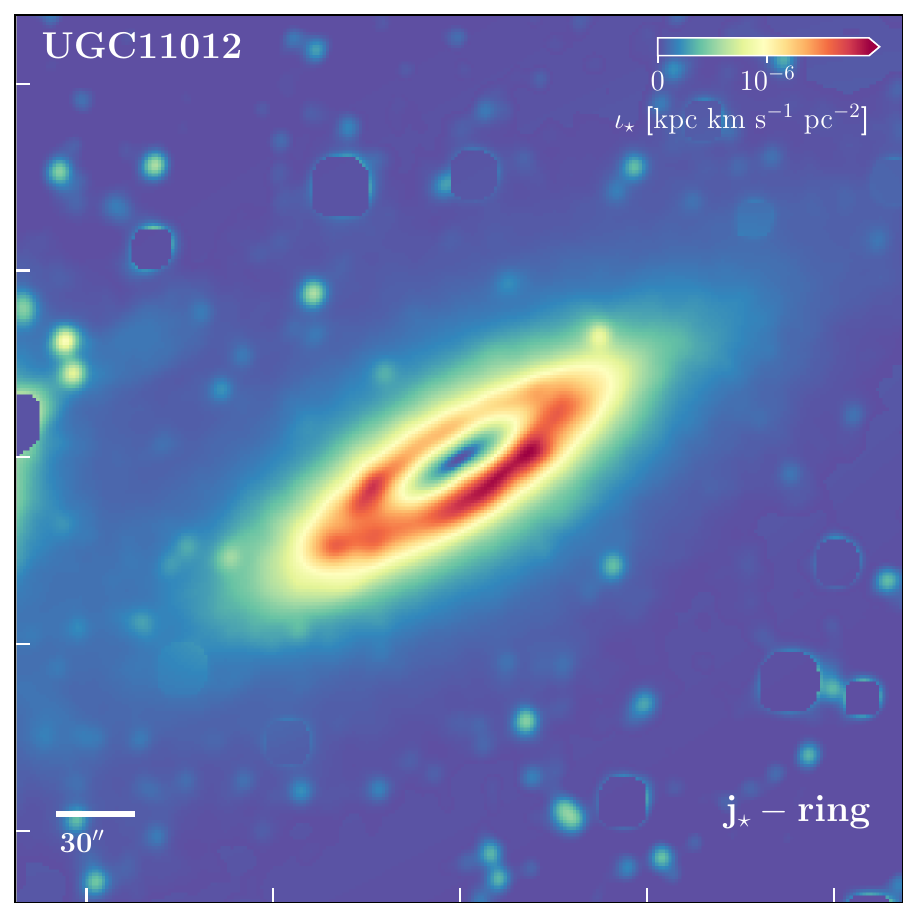}
    \end{subfigure}
    \begin{subfigure}[b]{0.245\textwidth}
        \includegraphics[width=\textwidth]{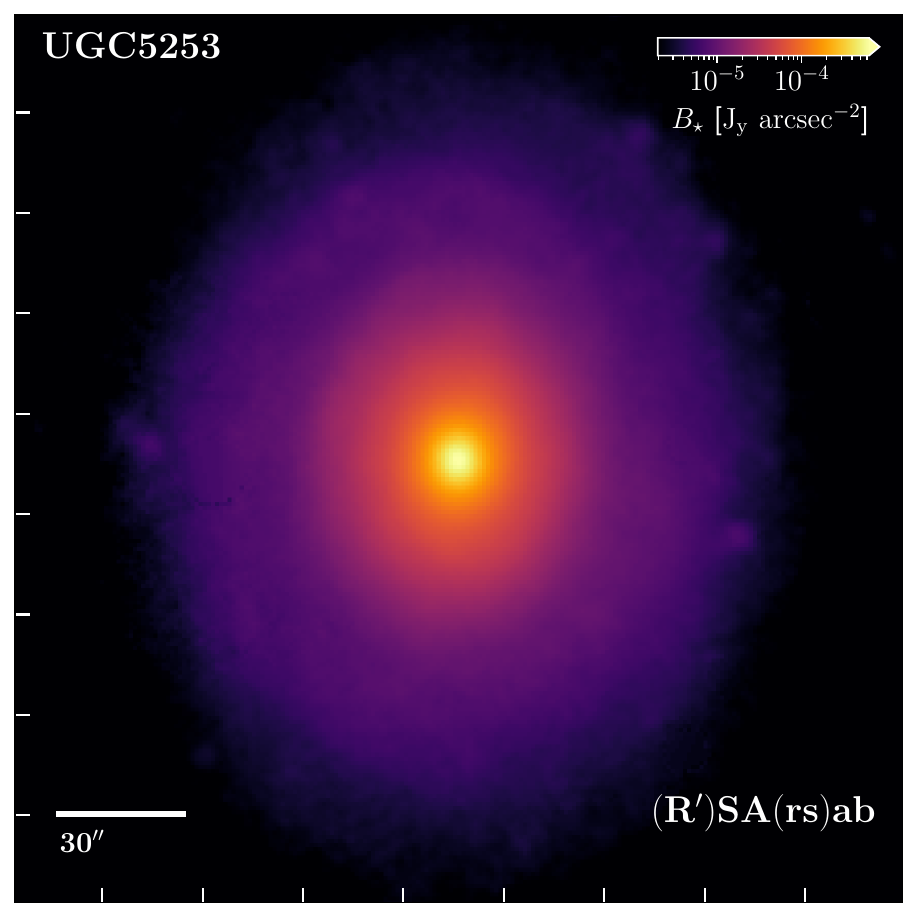}
    \end{subfigure}
    \begin{subfigure}[b]{0.245\textwidth}
        \includegraphics[width=\textwidth]{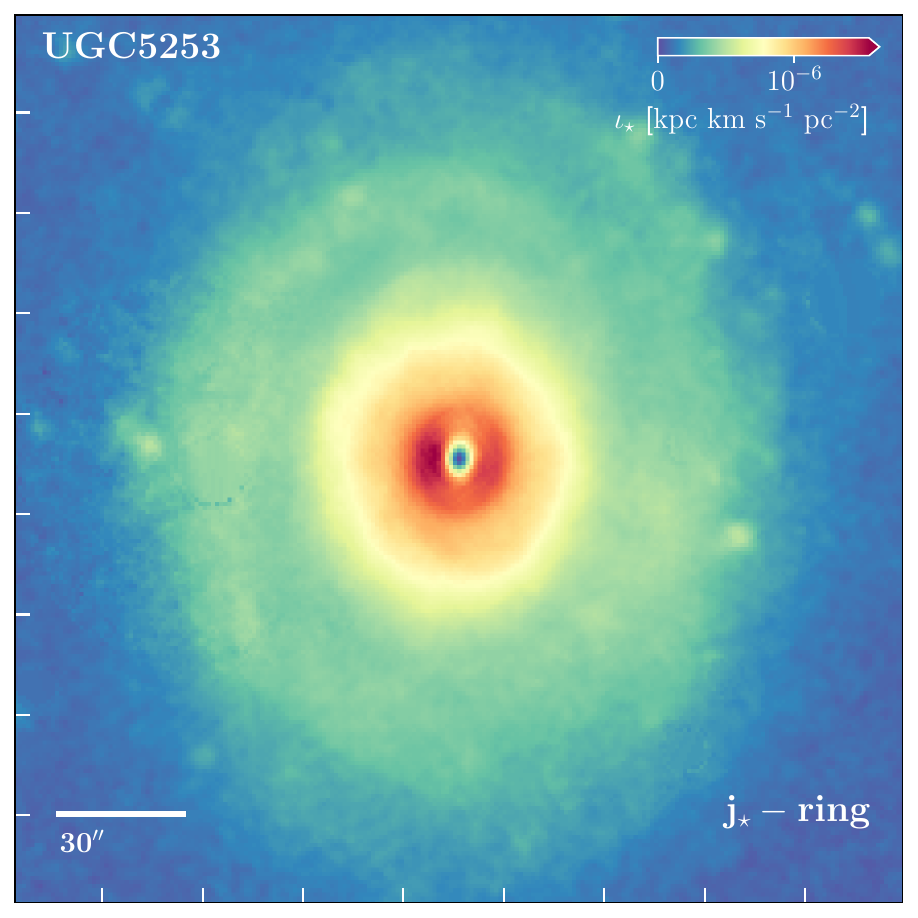}
    \end{subfigure}
    \\
    \begin{subfigure}[b]{0.245\textwidth}
        \includegraphics[width=\textwidth]{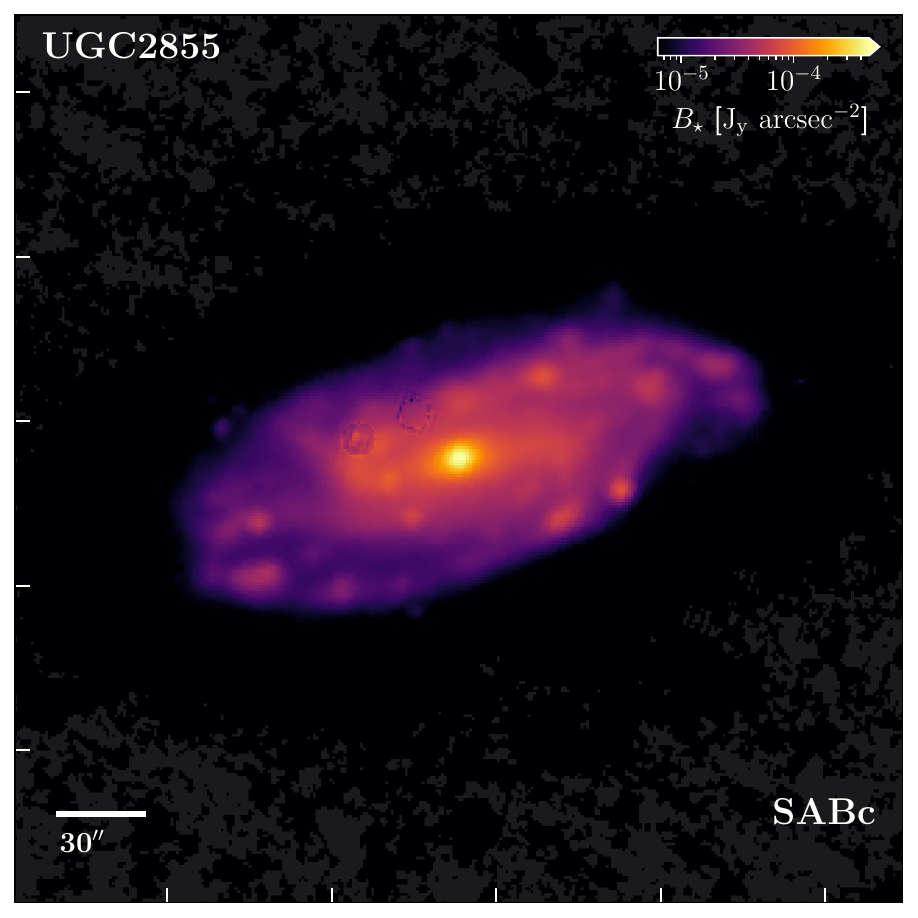}
    \end{subfigure}
    \begin{subfigure}[b]{0.245\textwidth}
        \includegraphics[width=\textwidth]{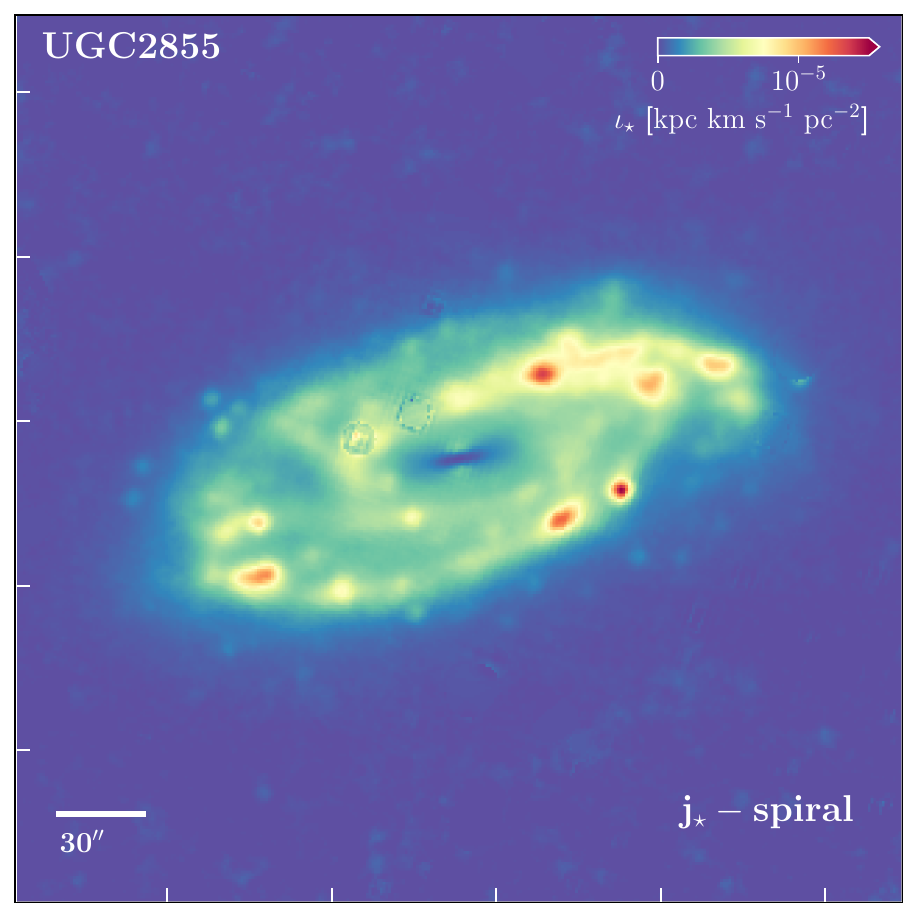}
    \end{subfigure}
    \begin{subfigure}[b]{0.245\textwidth}
        \includegraphics[width=\textwidth]{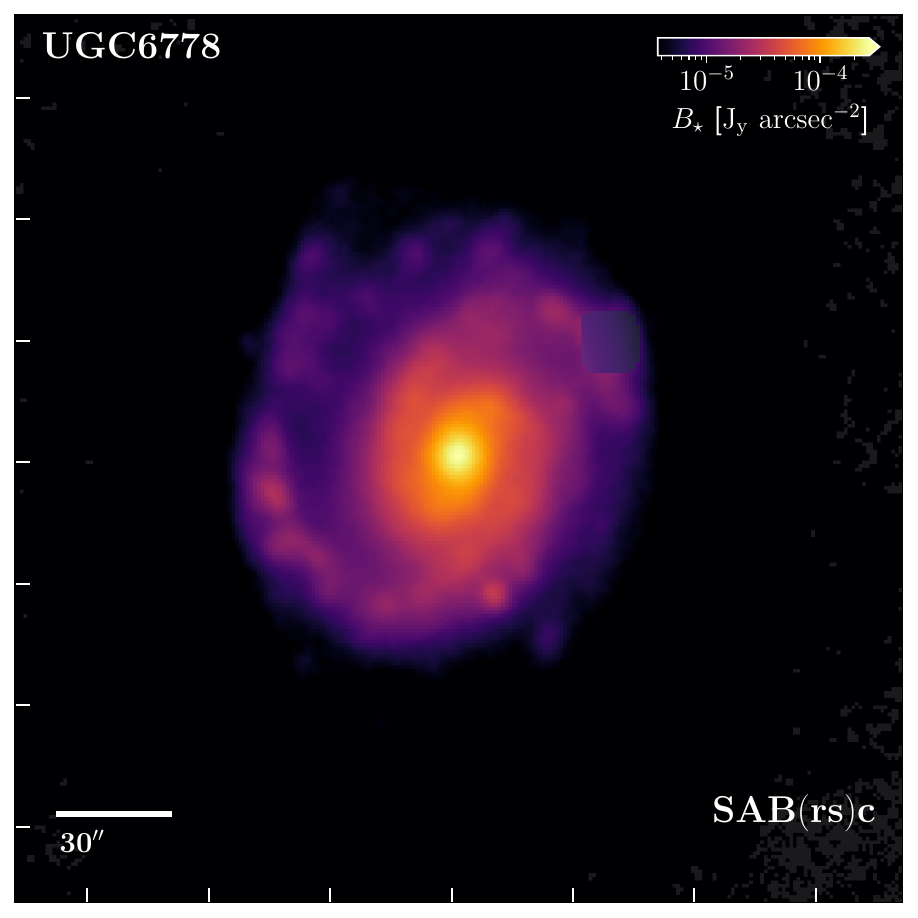}
    \end{subfigure}
    \begin{subfigure}[b]{0.245\textwidth}
        \includegraphics[width=\textwidth]{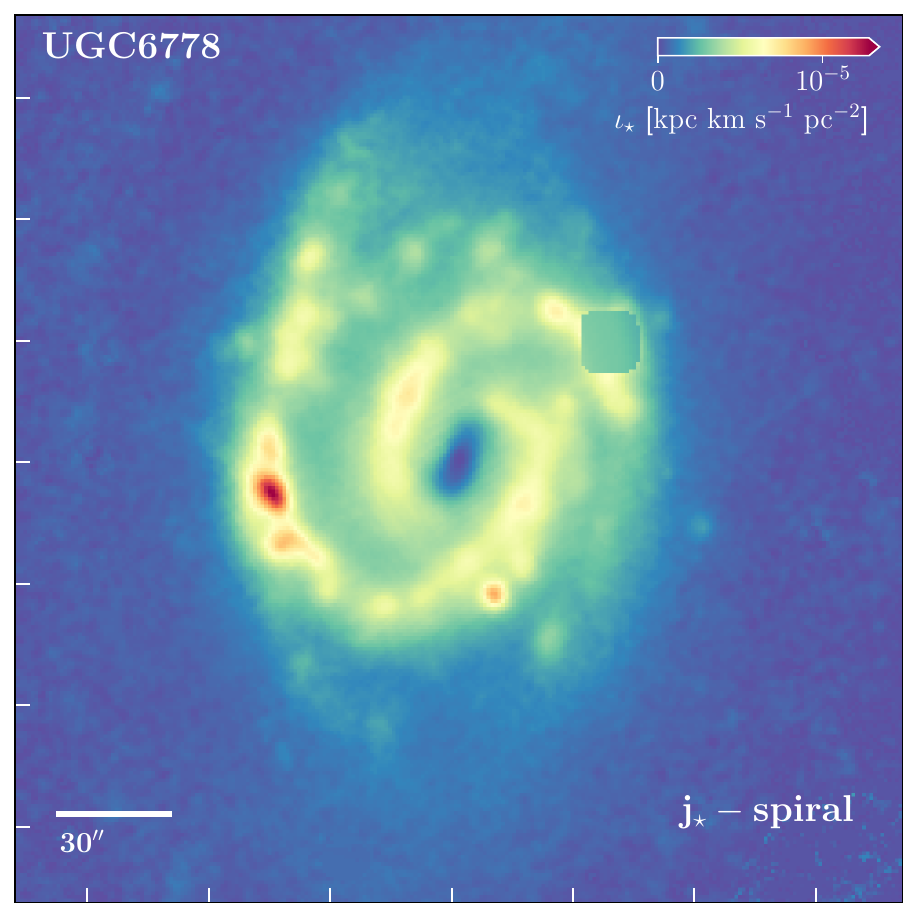}
    \end{subfigure}
    \caption{Maps of $B_\star$ and stellar sAMSD for UGC11852, UGC10075, UGC8334, UGC3734, UGC11012, UGC5253, UGC2855, UGC6778, UGC11670, UGC9649, UGC12754, UGC5414, UGC4325, UGC4499, UGC1913, UGC9179, UGC11597, UGC4284, UGC7323, UGC9858, UGC2800, UGC5251, UGC7766, UGC3574, and UGC11914.}
    \label{fig: app_j_maps}
\end{figure*}

\begin{figure*}[ht!]\ContinuedFloat
    \centering

    \begin{subfigure}[b]{0.245\textwidth}
        \includegraphics[width=\textwidth]{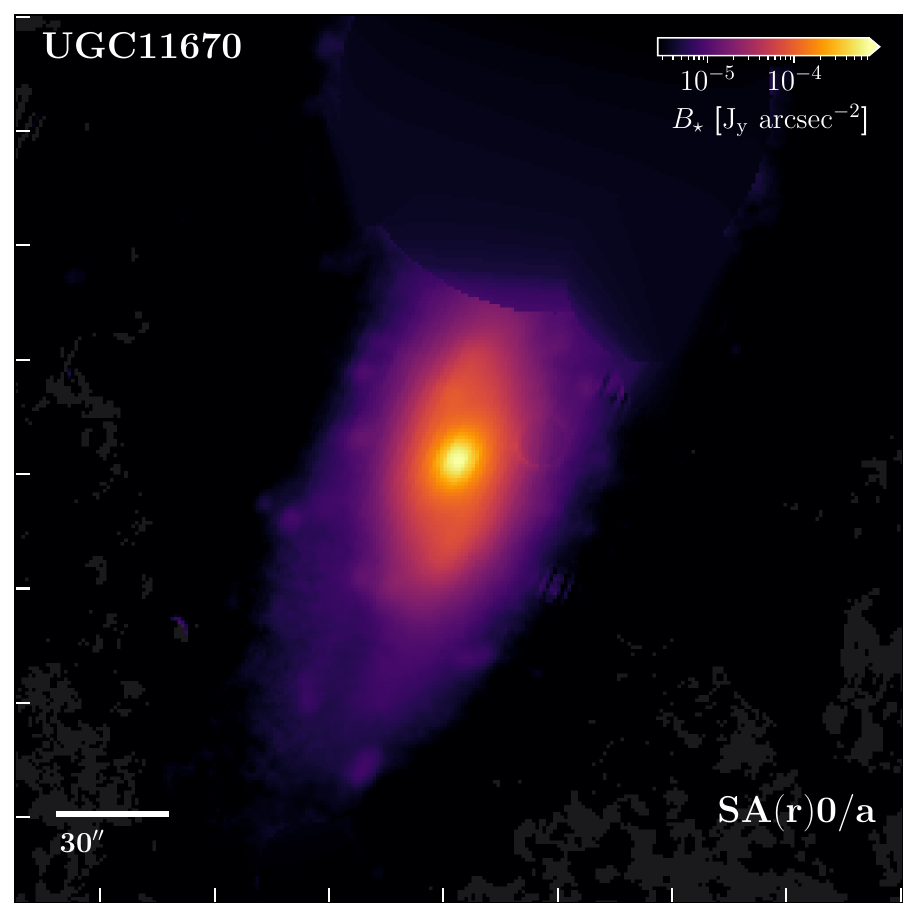}
    \end{subfigure}
    \begin{subfigure}[b]{0.245\textwidth}
        \includegraphics[width=\textwidth]{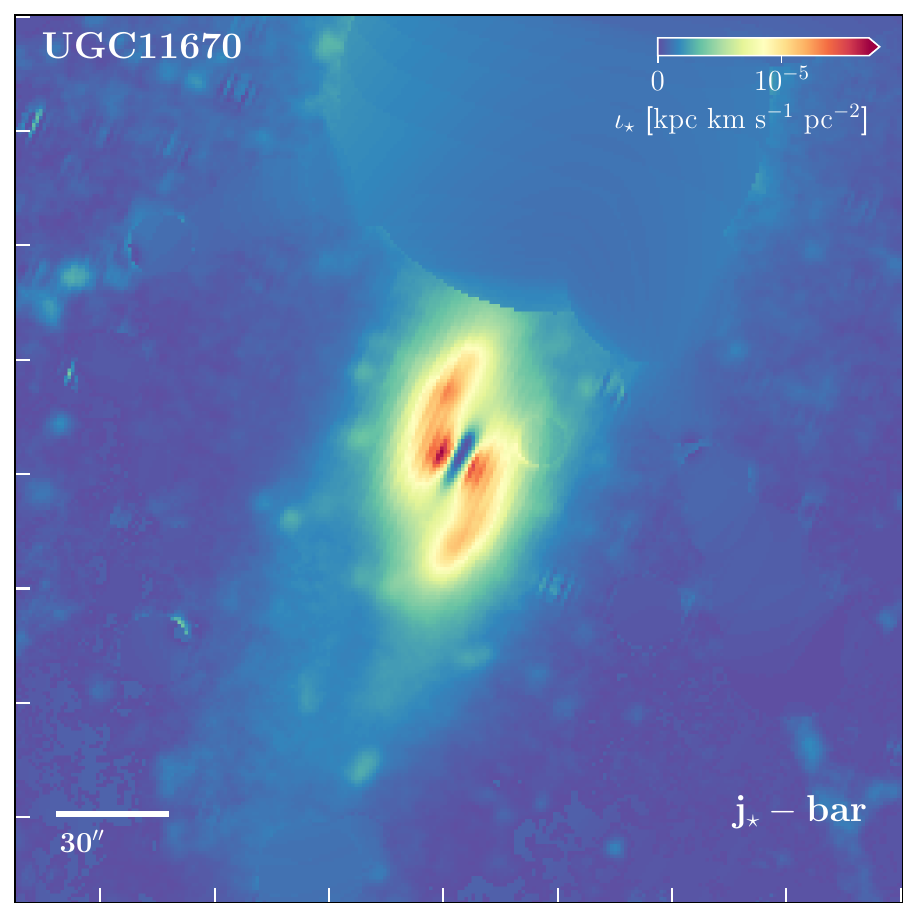}
    \end{subfigure}
    \begin{subfigure}[b]{0.245\textwidth}
        \includegraphics[width=\textwidth]{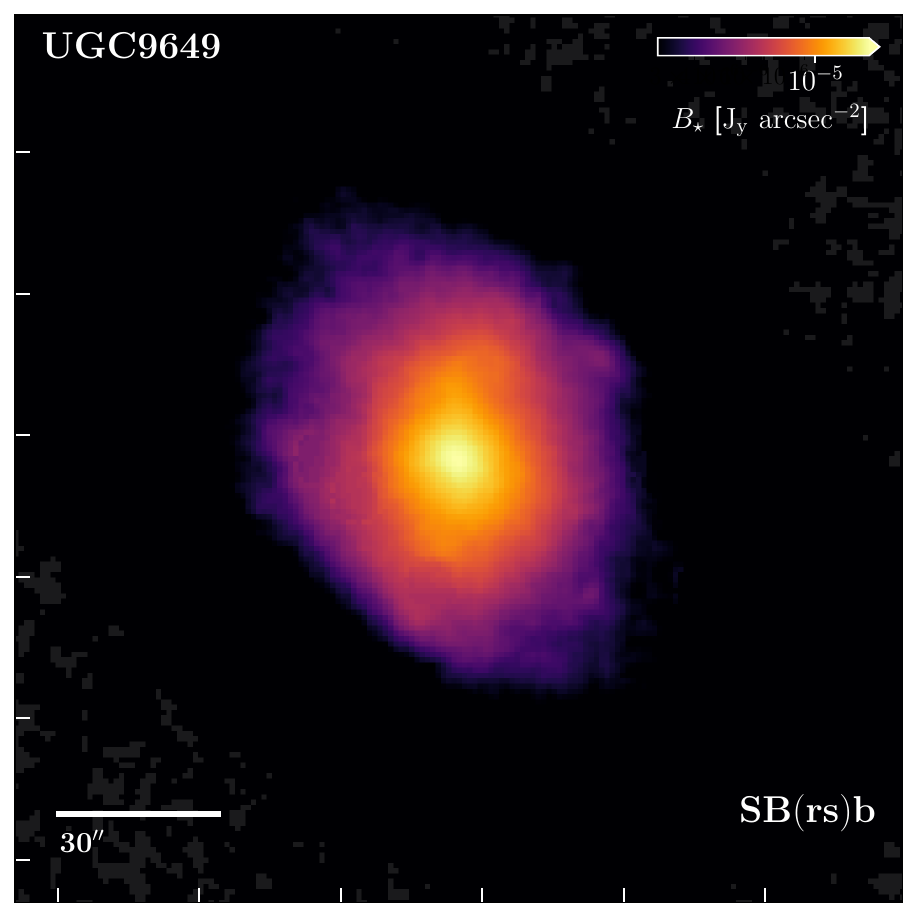}
    \end{subfigure}
    \begin{subfigure}[b]{0.245\textwidth}
        \includegraphics[width=\textwidth]{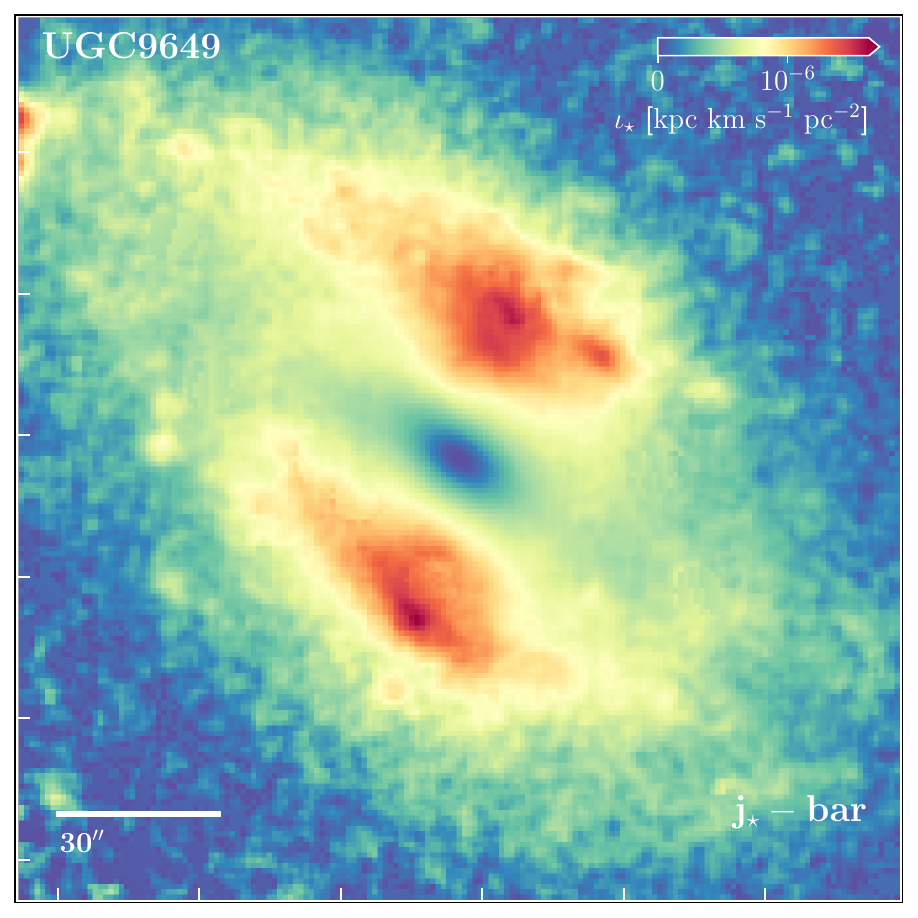}
    \end{subfigure}
    \\

    \begin{subfigure}[b]{0.245\textwidth}
        \includegraphics[width=\textwidth]{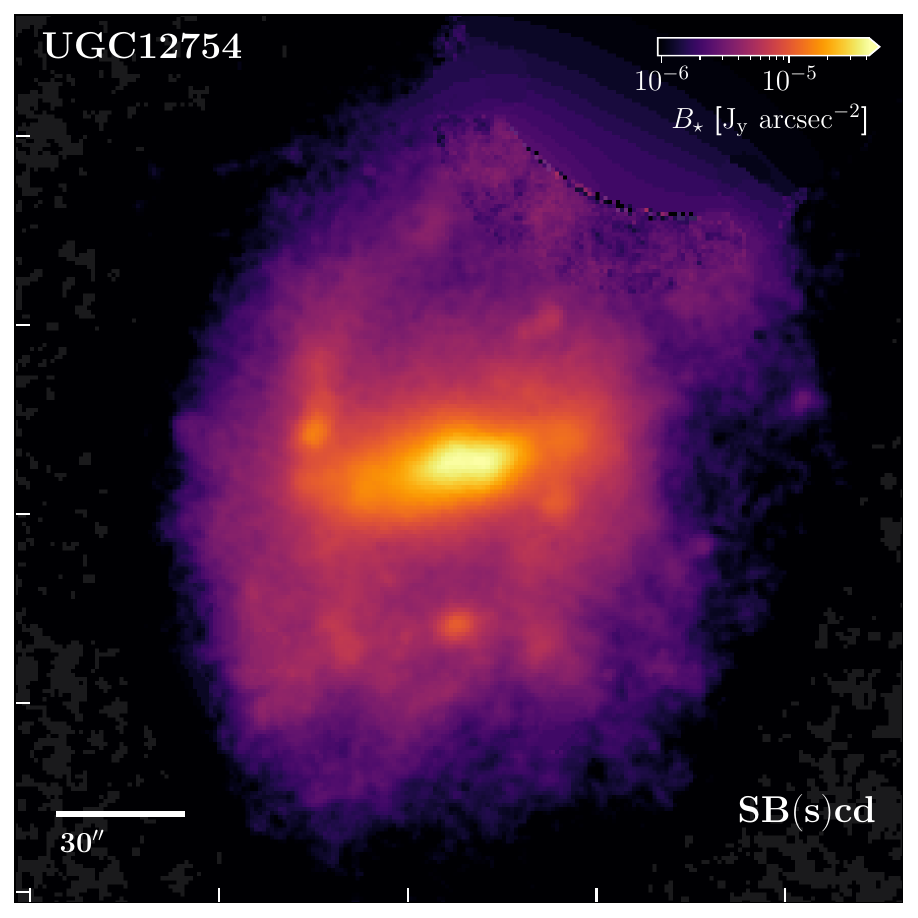}
    \end{subfigure}
    \begin{subfigure}[b]{0.245\textwidth}
        \includegraphics[width=\textwidth]{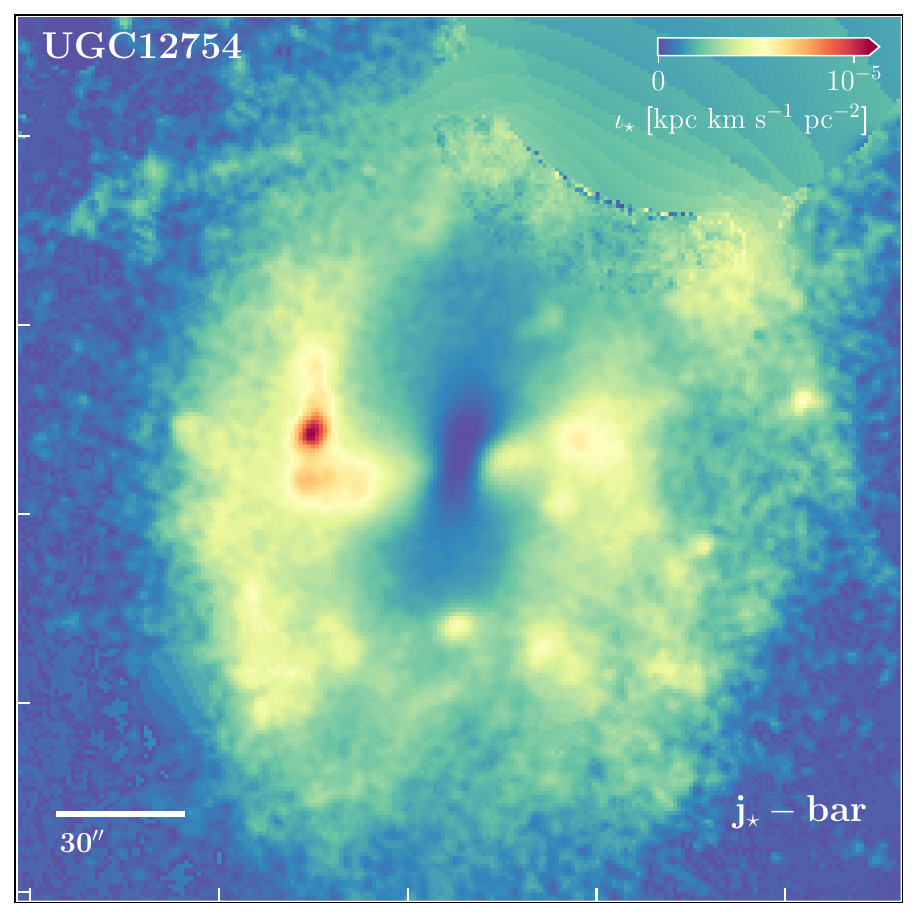}
    \end{subfigure}
    \begin{subfigure}[b]{0.245\textwidth}
        \includegraphics[width=\textwidth]{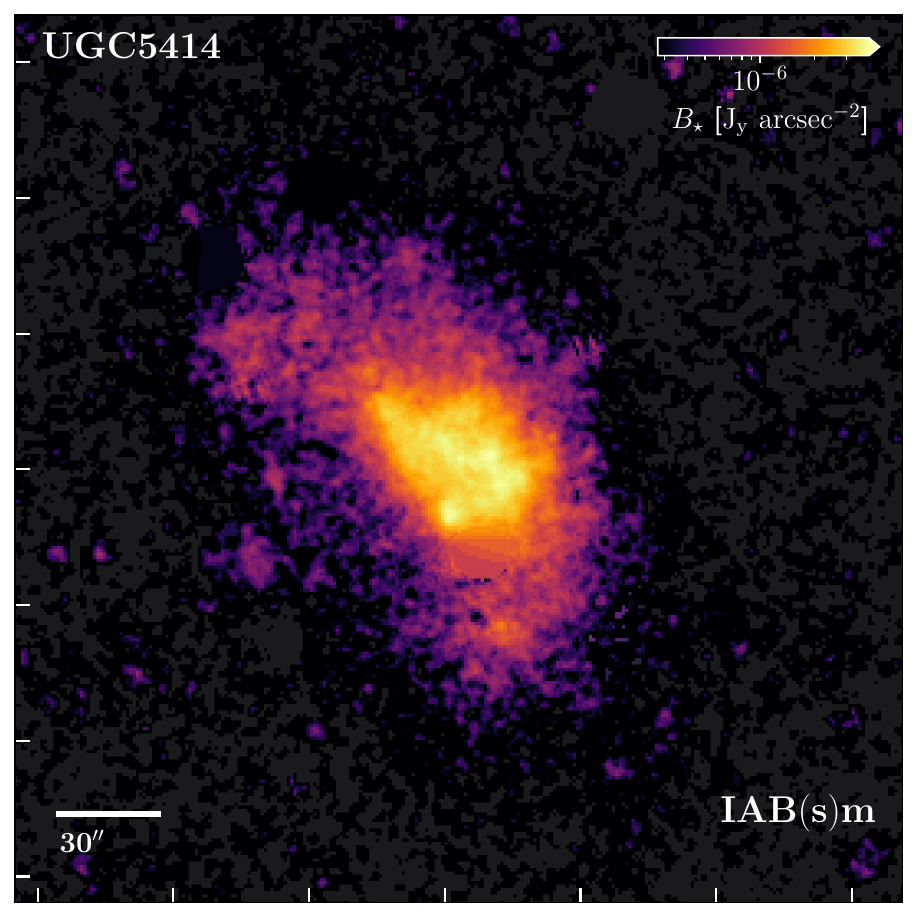}
    \end{subfigure}
    \begin{subfigure}[b]{0.245\textwidth}
        \includegraphics[width=\textwidth]{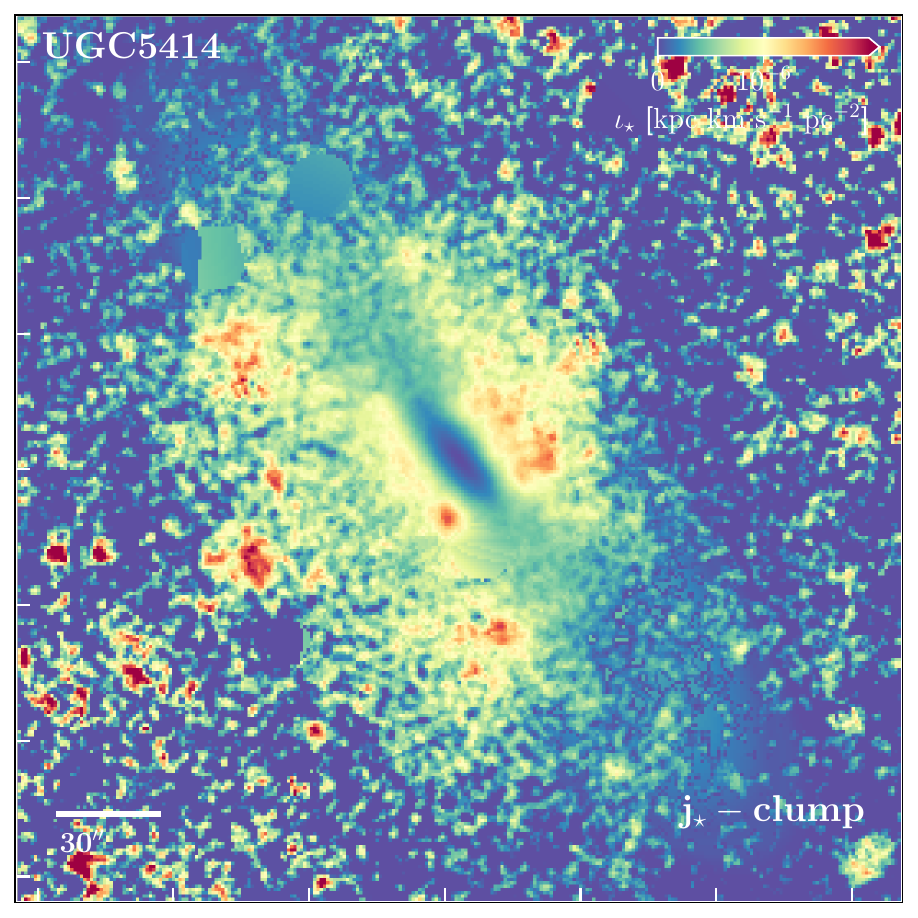}
    \end{subfigure}
    \\
    \begin{subfigure}[b]{0.245\textwidth}
        \includegraphics[width=\textwidth]{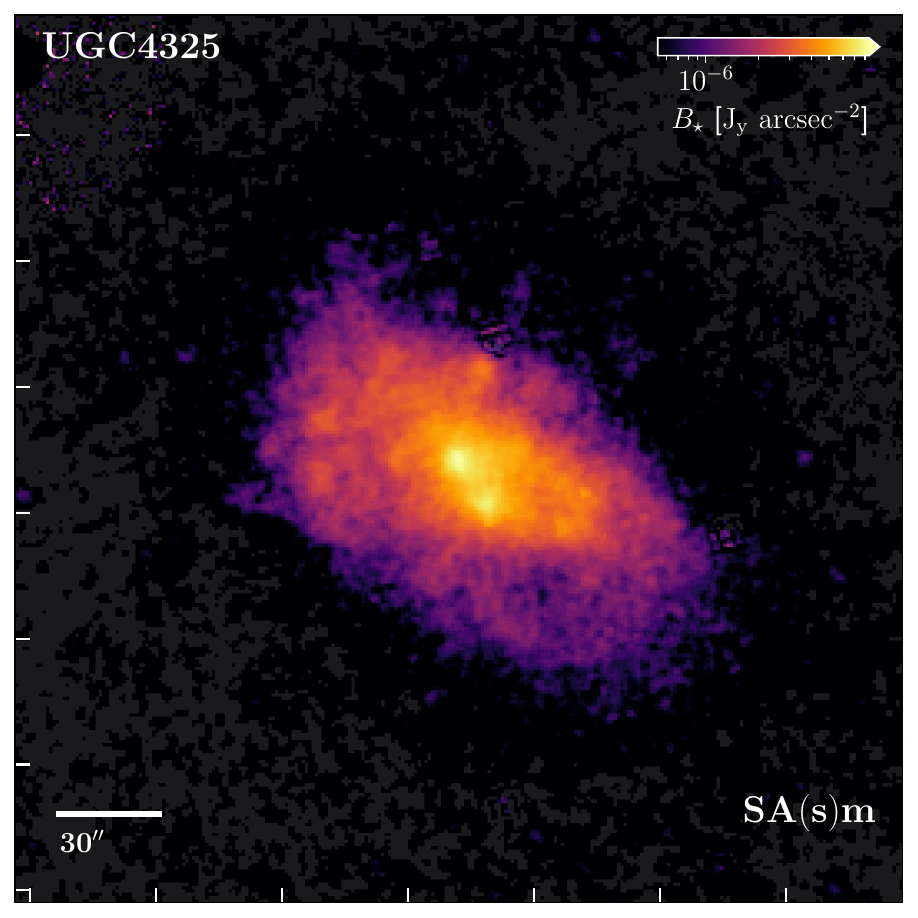}
    \end{subfigure}
    \begin{subfigure}[b]{0.245\textwidth}
        \includegraphics[width=\textwidth]{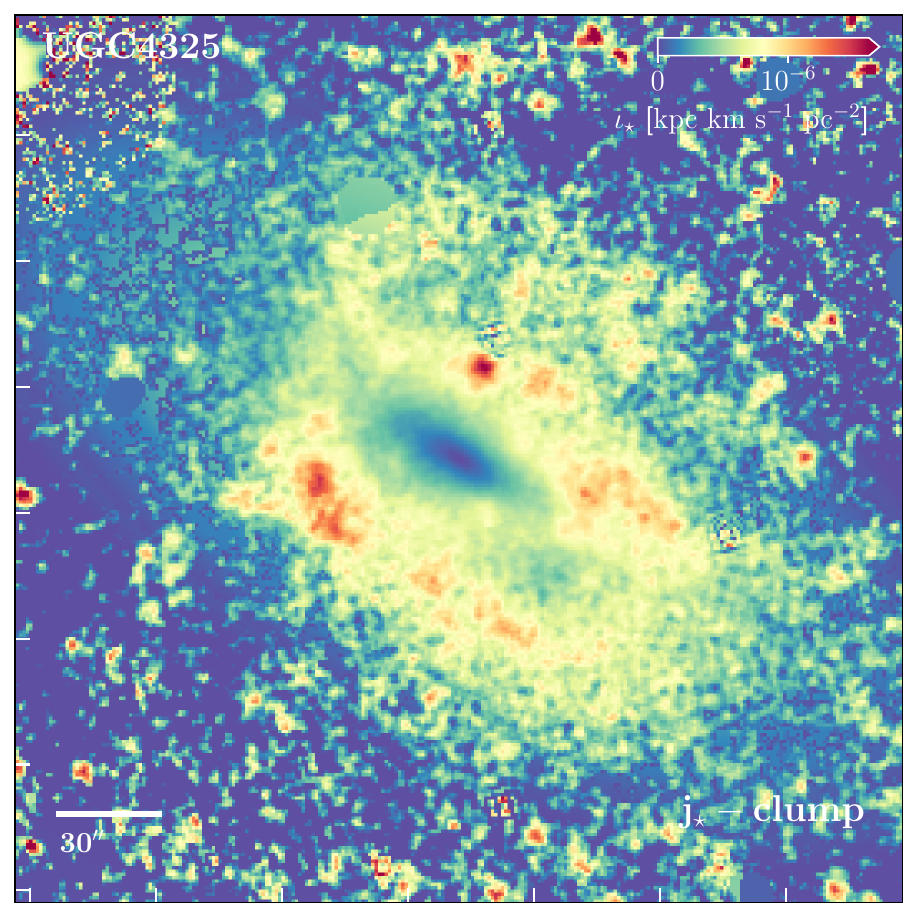}
    \end{subfigure}
    \begin{subfigure}[b]{0.245\textwidth}
        \includegraphics[width=\textwidth]{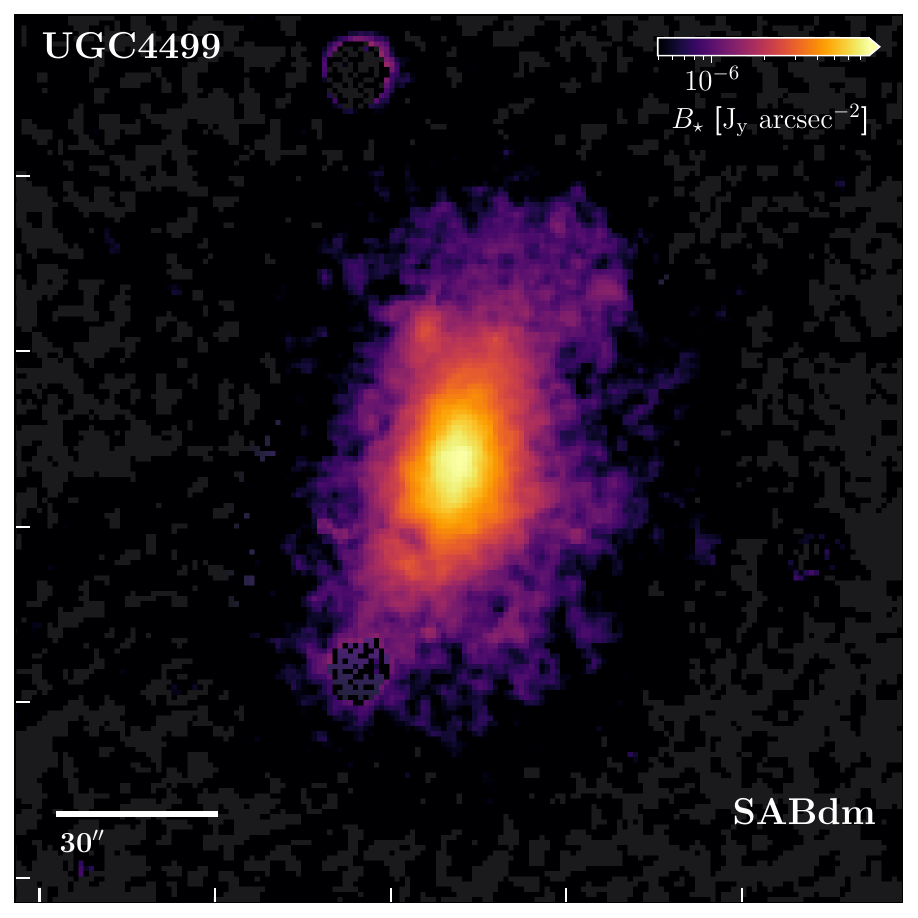}
    \end{subfigure}
    \begin{subfigure}[b]{0.245\textwidth}
        \includegraphics[width=\textwidth]{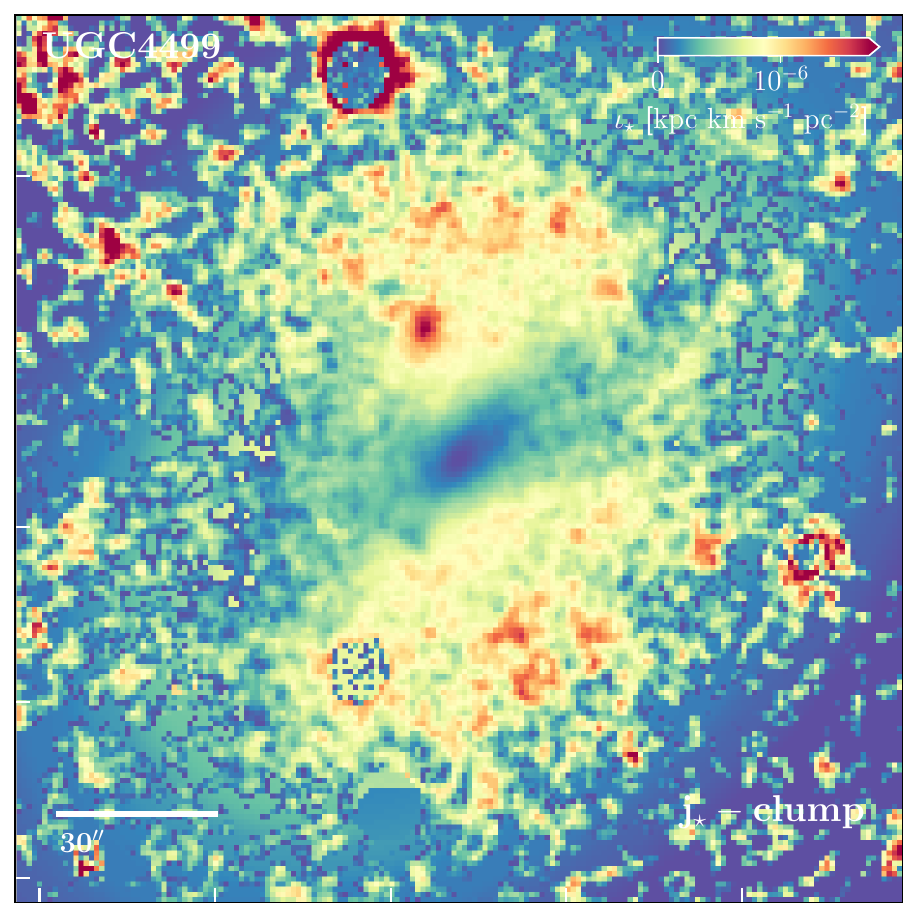}
    \end{subfigure}
    \\
    \begin{subfigure}[b]{0.245\textwidth}
        \includegraphics[width=\textwidth]{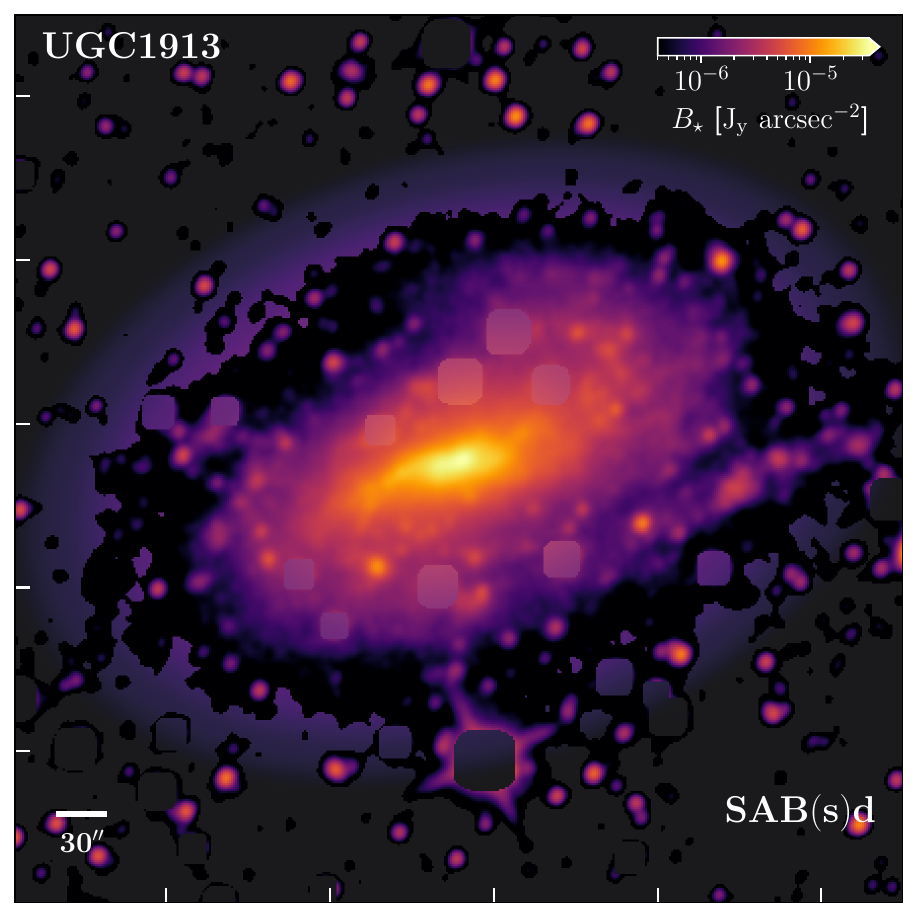}
    \end{subfigure}
    \begin{subfigure}[b]{0.245\textwidth}
        \includegraphics[width=\textwidth]{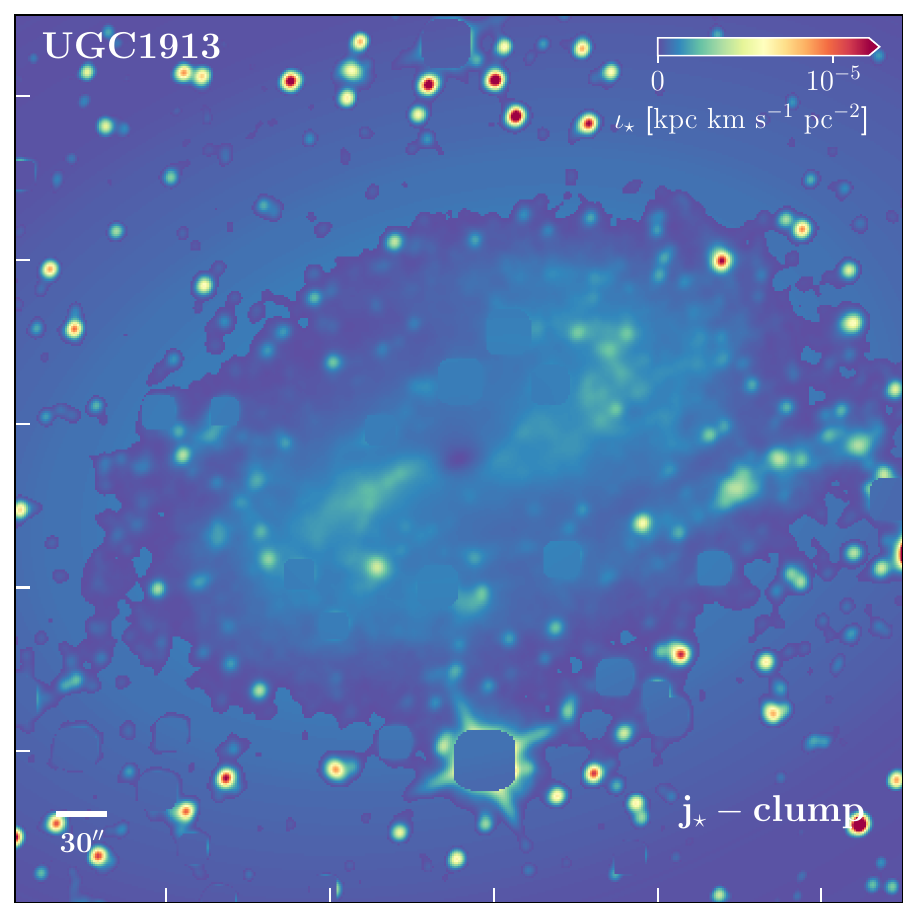}
    \end{subfigure}
    \begin{subfigure}[b]{0.245\textwidth}
        \includegraphics[width=\textwidth]{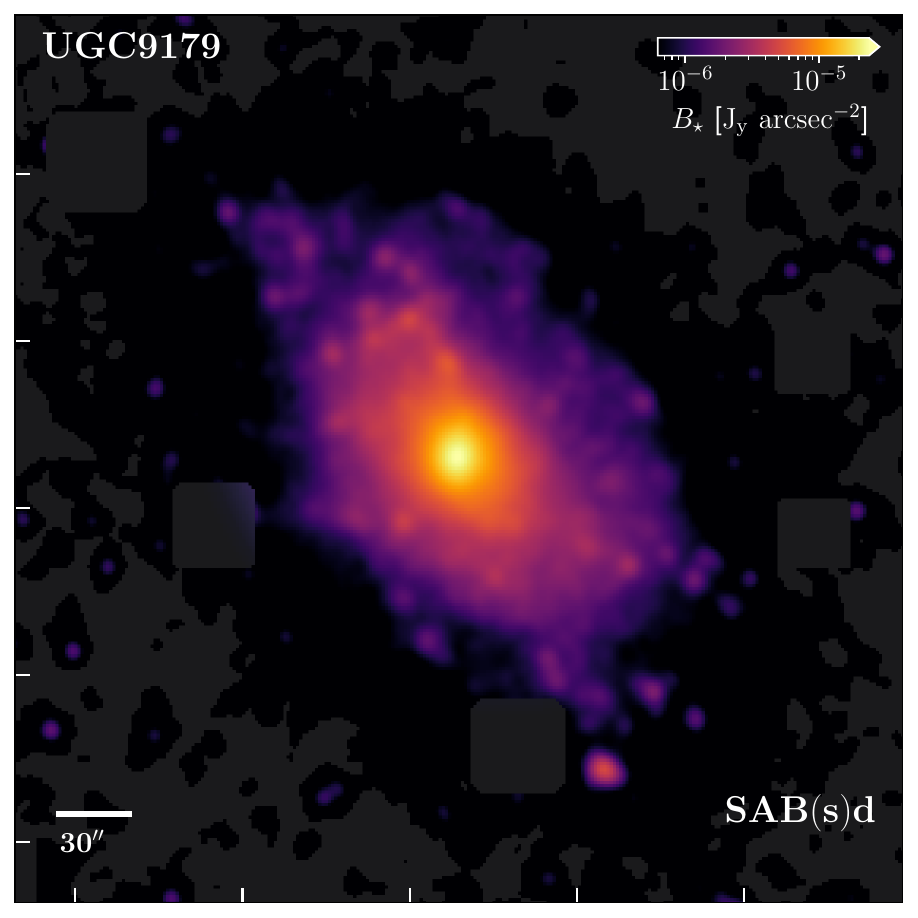}
    \end{subfigure}
    \begin{subfigure}[b]{0.245\textwidth}
        \includegraphics[width=\textwidth]{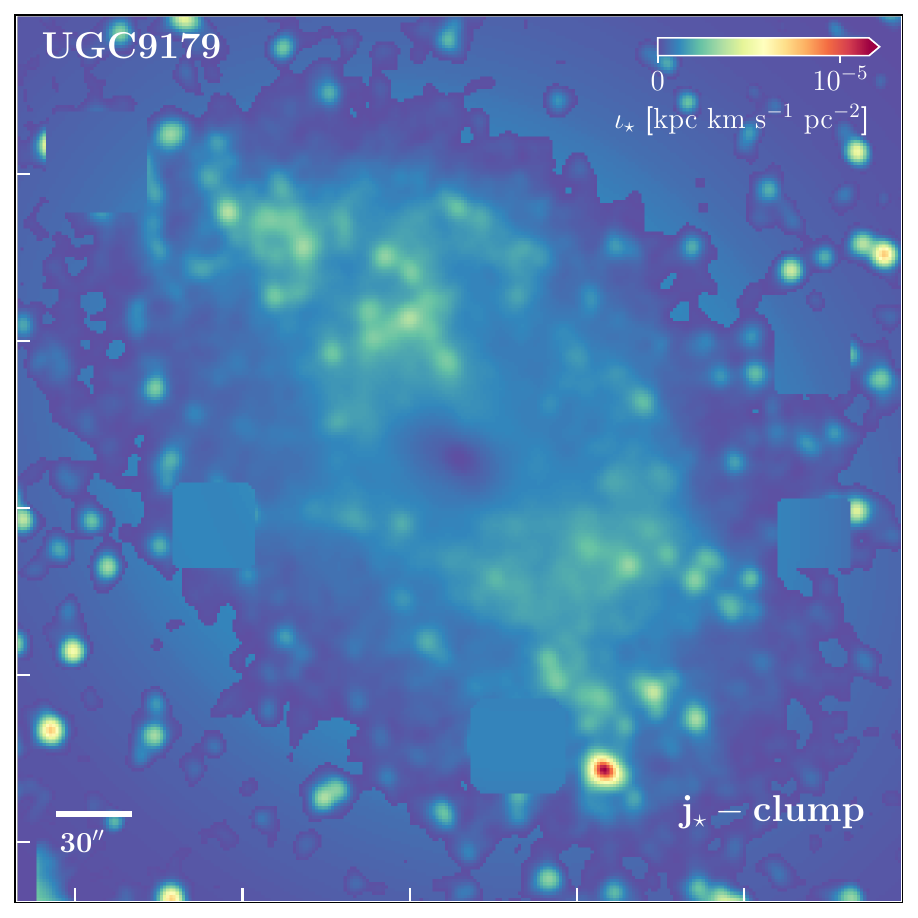}
    \end{subfigure}
    \\
    \begin{subfigure}[b]{0.245\textwidth}
        \includegraphics[width=\textwidth]{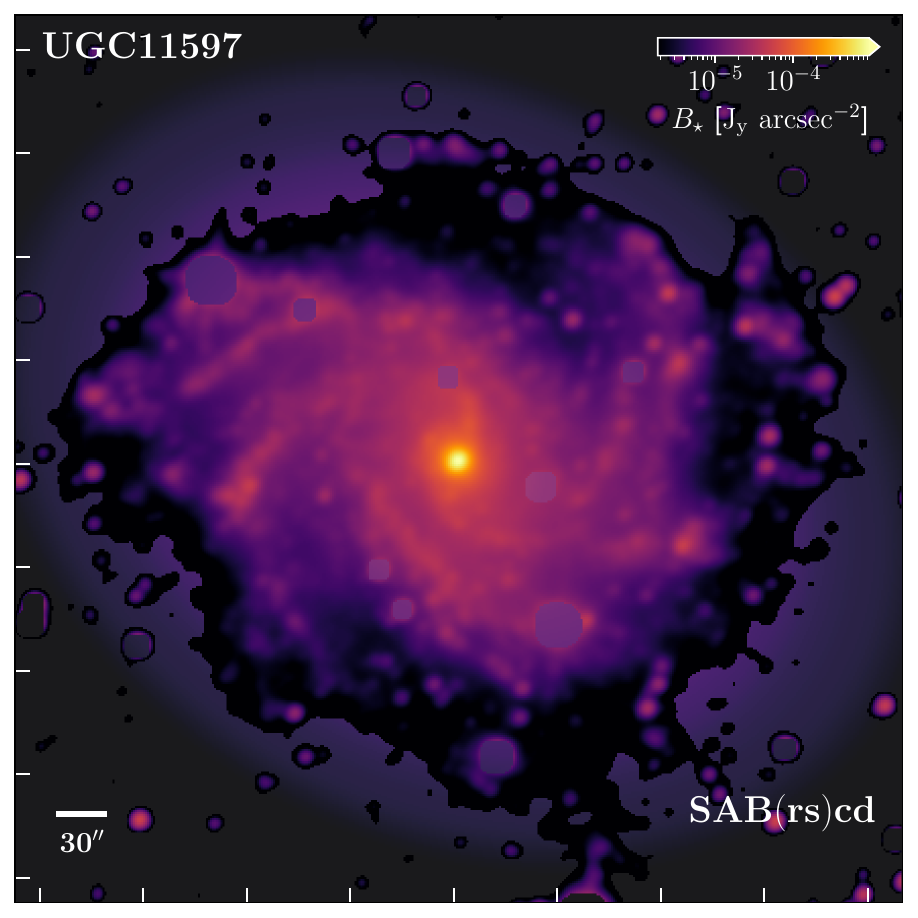}
    \end{subfigure}
    \begin{subfigure}[b]{0.245\textwidth}
        \includegraphics[width=\textwidth]{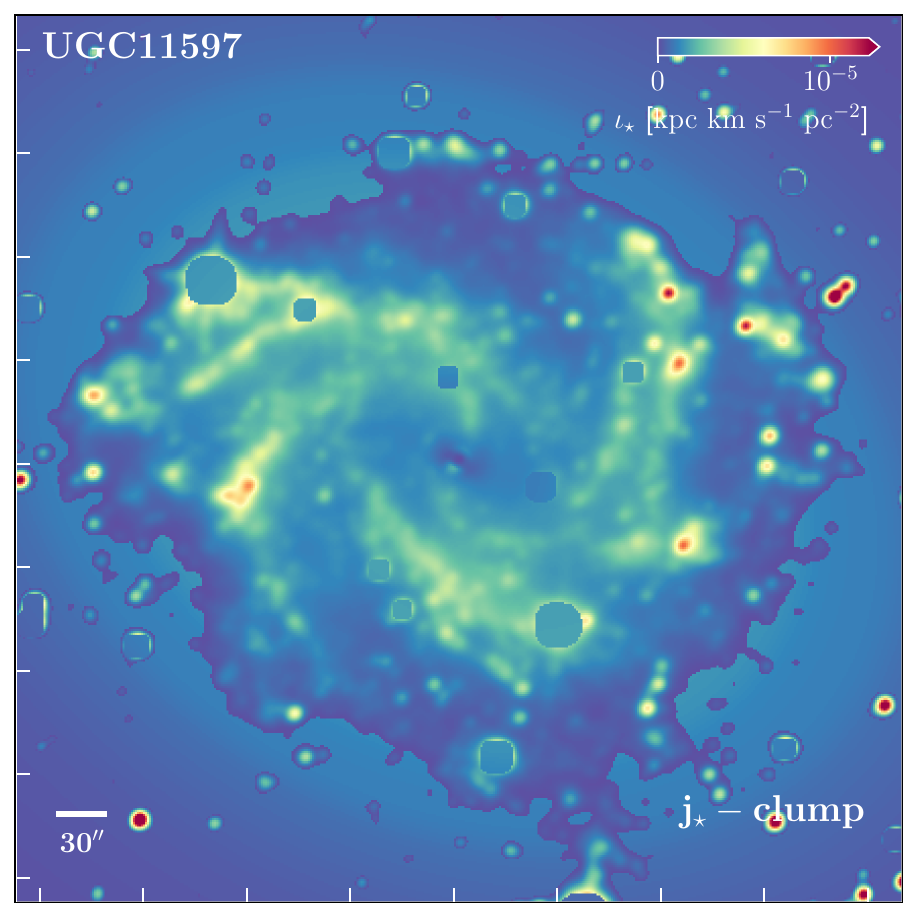}
    \end{subfigure}
    \begin{subfigure}[b]{0.245\textwidth}
        \includegraphics[width=\textwidth]{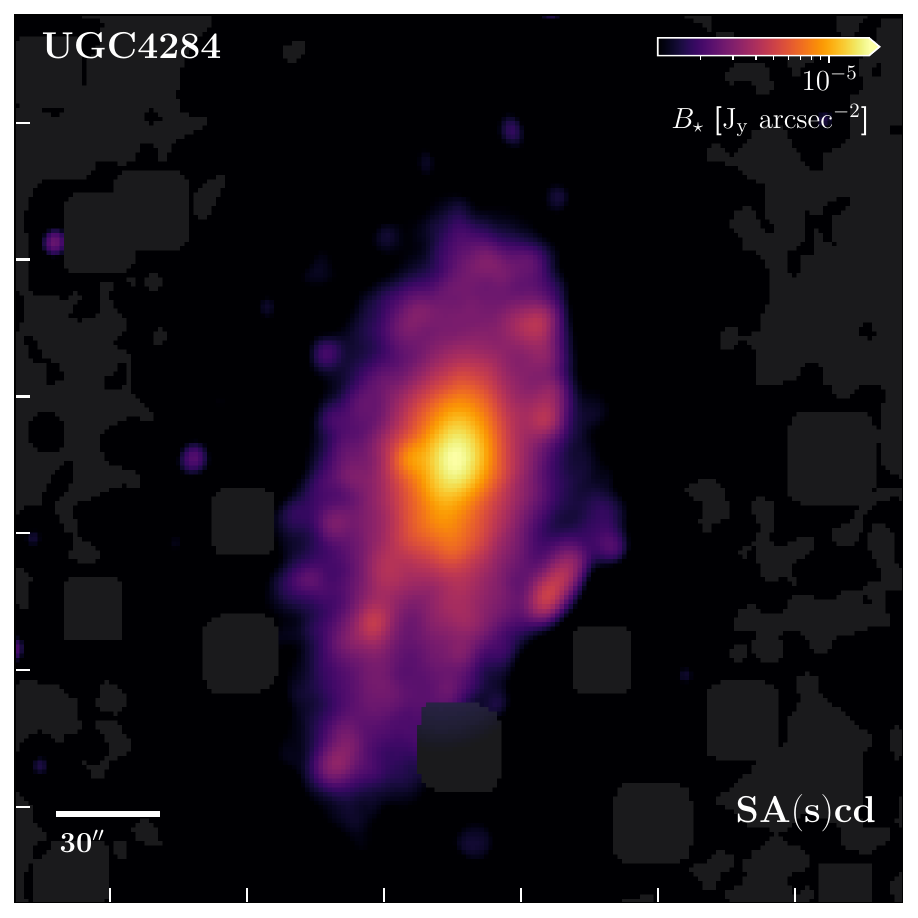}
    \end{subfigure}
    \begin{subfigure}[b]{0.245\textwidth}
        \includegraphics[width=\textwidth]{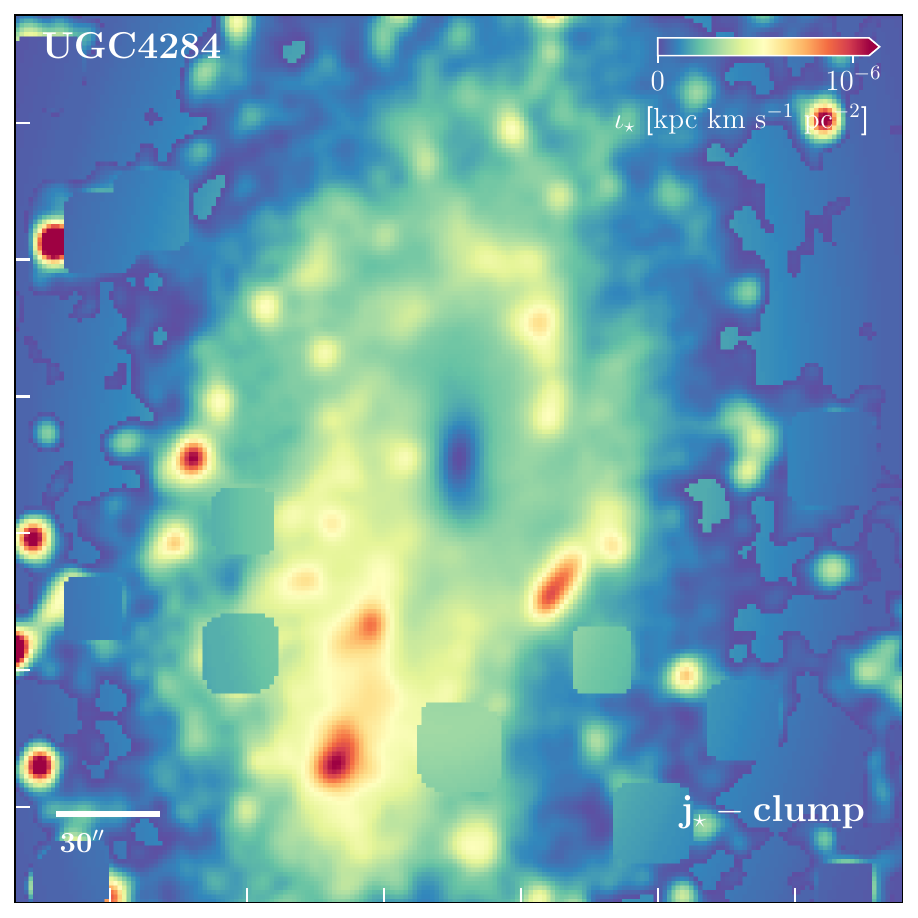}
    \end{subfigure}
    \caption{Continued.}
\end{figure*}

\begin{figure*}[ht!]\ContinuedFloat
    \centering
    \begin{subfigure}[b]{0.245\textwidth}
        \includegraphics[width=\textwidth]{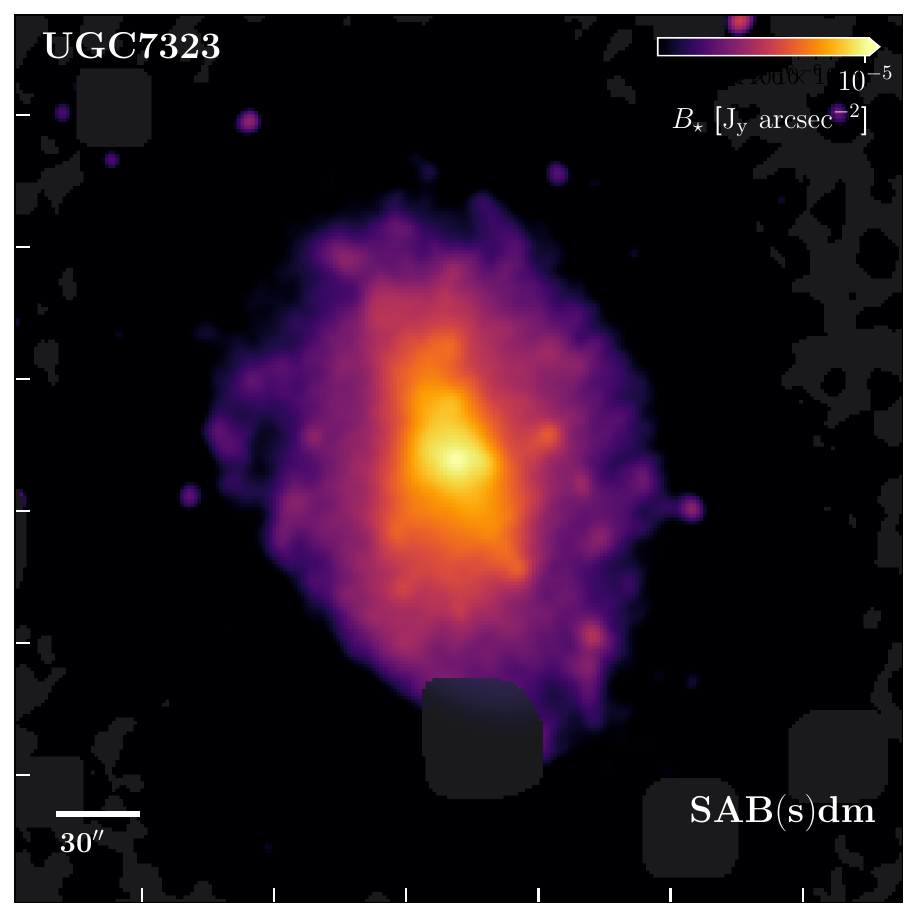}
    \end{subfigure}
    \begin{subfigure}[b]{0.245\textwidth}
        \includegraphics[width=\textwidth]{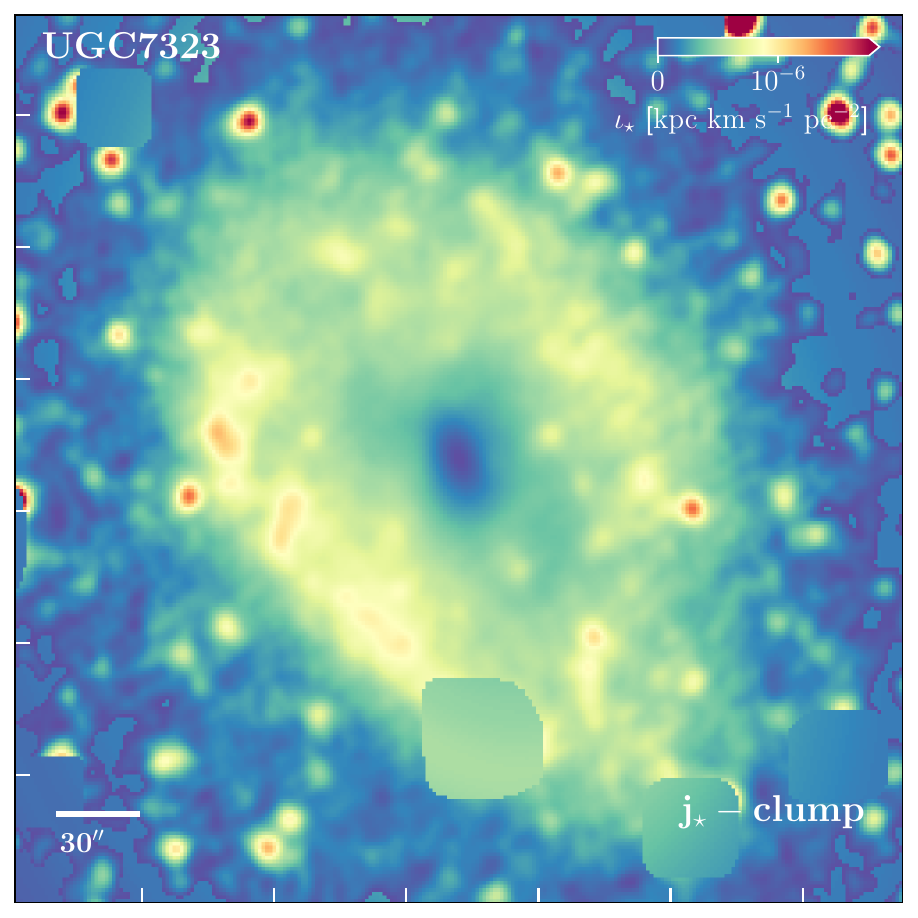}
    \end{subfigure}
    \begin{subfigure}[b]{0.245\textwidth}
        \includegraphics[width=\textwidth]{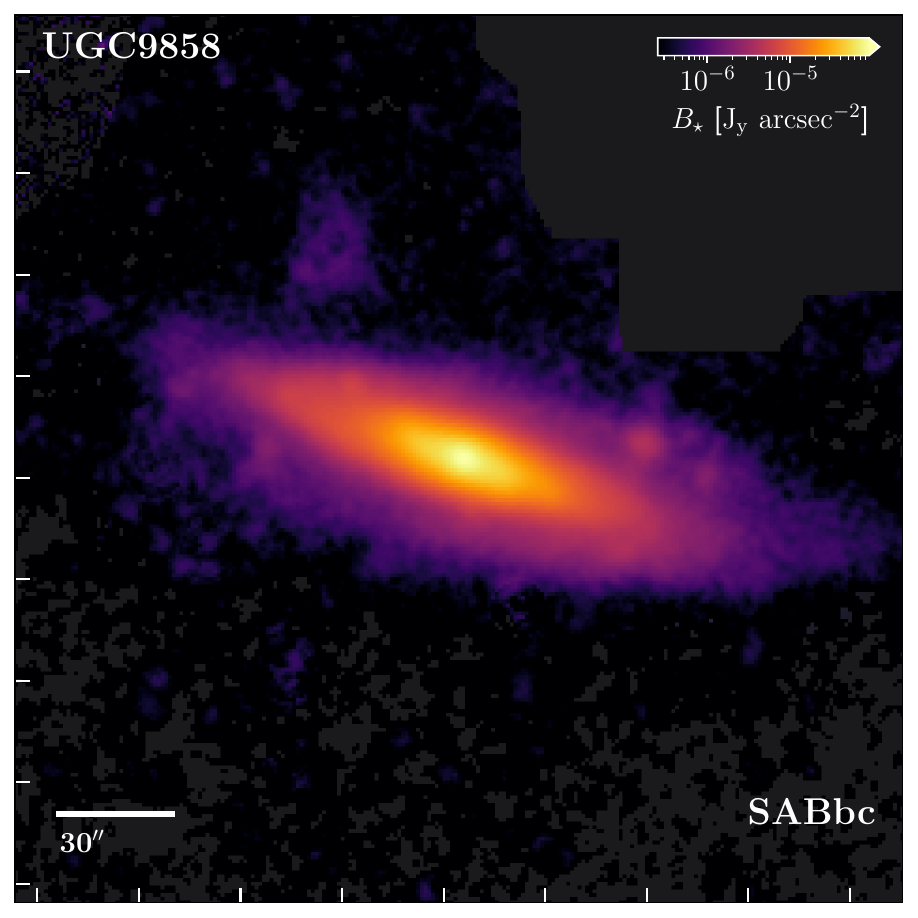}
    \end{subfigure}
    \begin{subfigure}[b]{0.245\textwidth}
        \includegraphics[width=\textwidth]{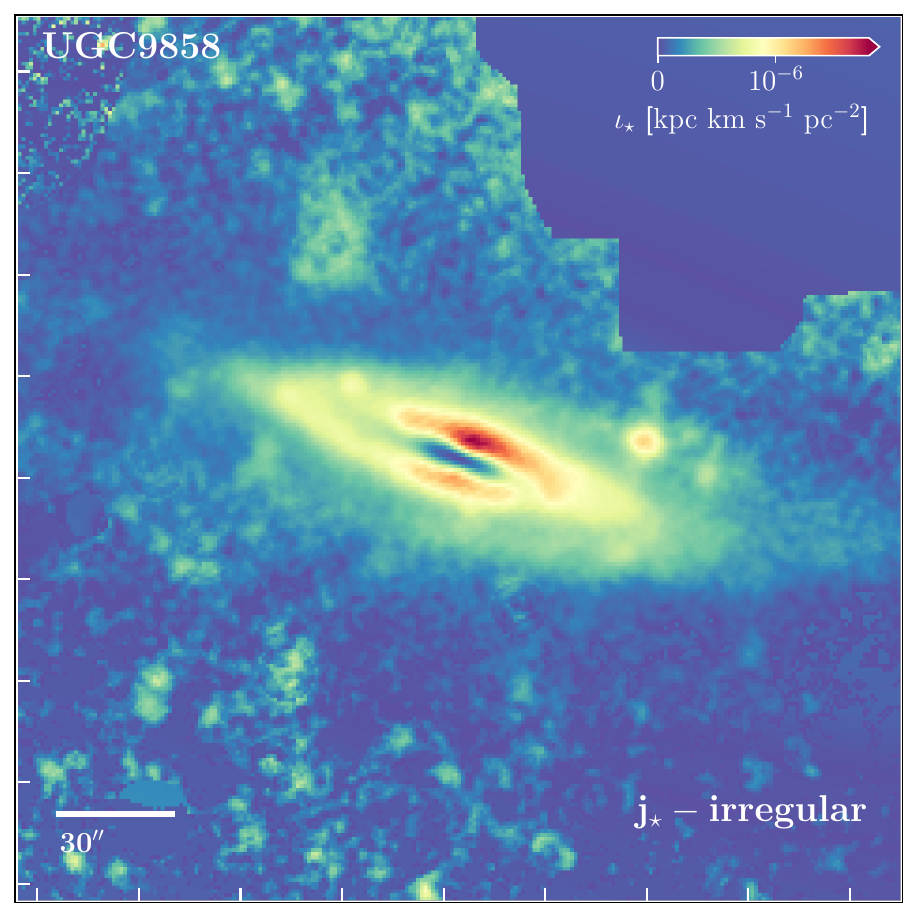}
    \end{subfigure}
    \\
    \begin{subfigure}[b]{0.245\textwidth}
        \includegraphics[width=\textwidth]{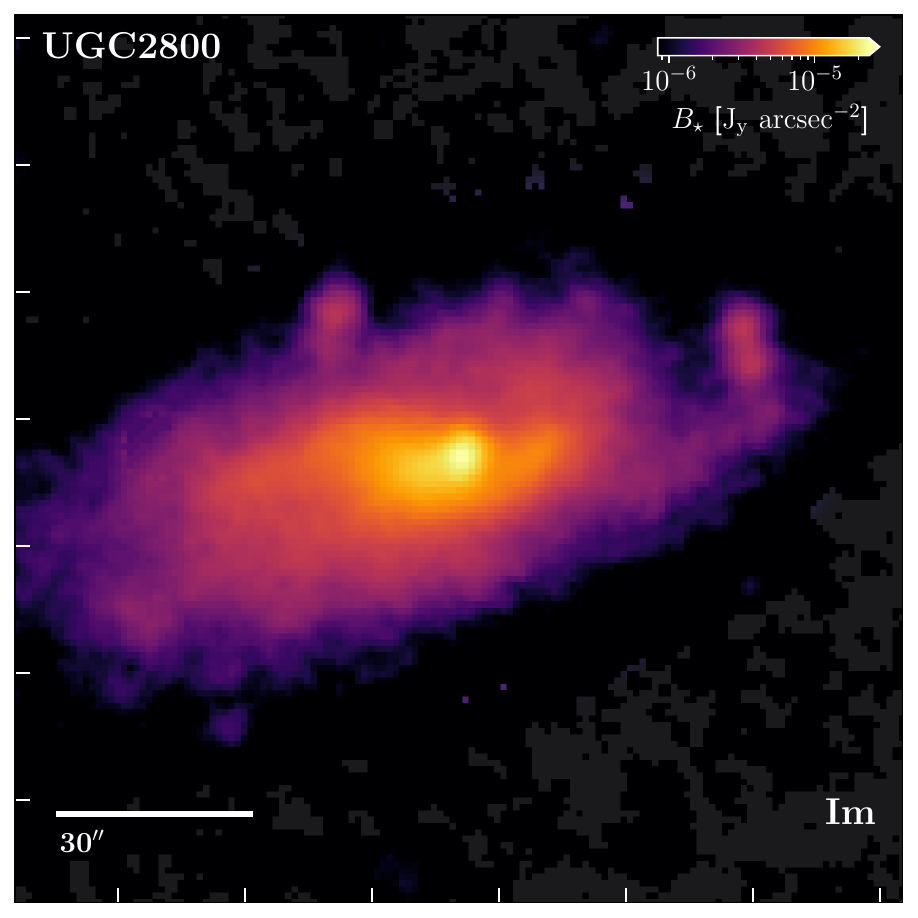}
    \end{subfigure}
    \begin{subfigure}[b]{0.245\textwidth}
        \includegraphics[width=\textwidth]{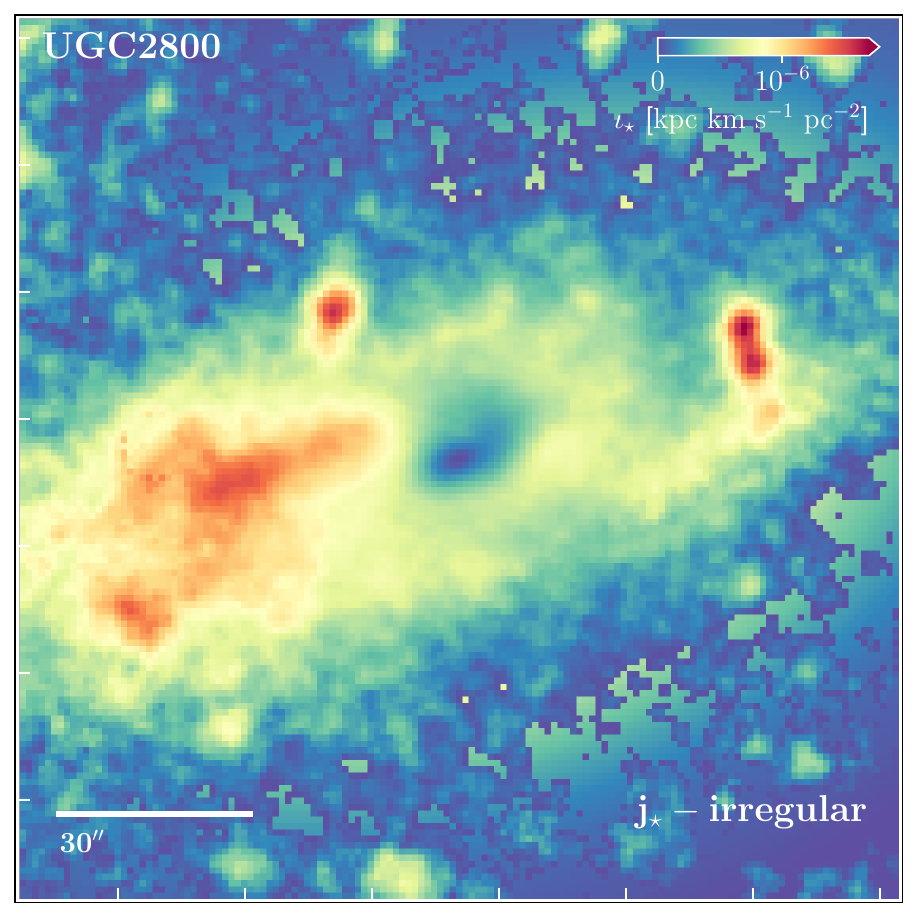}
    \end{subfigure}
    \begin{subfigure}[b]{0.245\textwidth}
        \includegraphics[width=\textwidth]{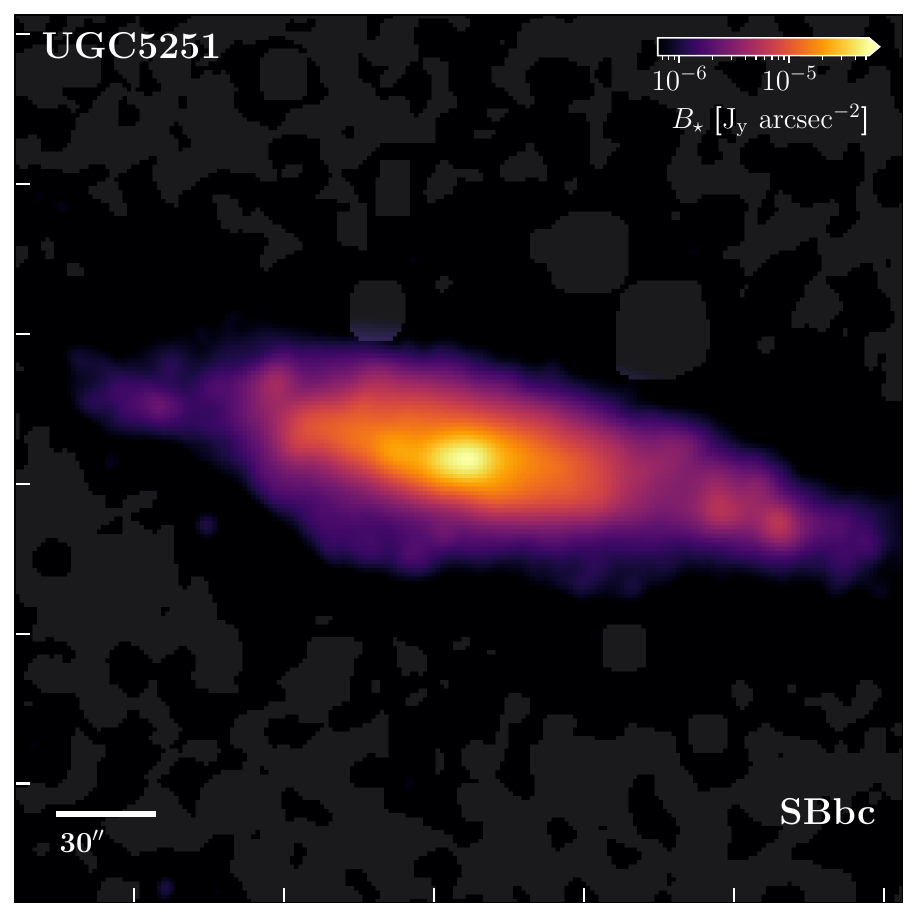}
    \end{subfigure}
    \begin{subfigure}[b]{0.245\textwidth}
        \includegraphics[width=\textwidth]{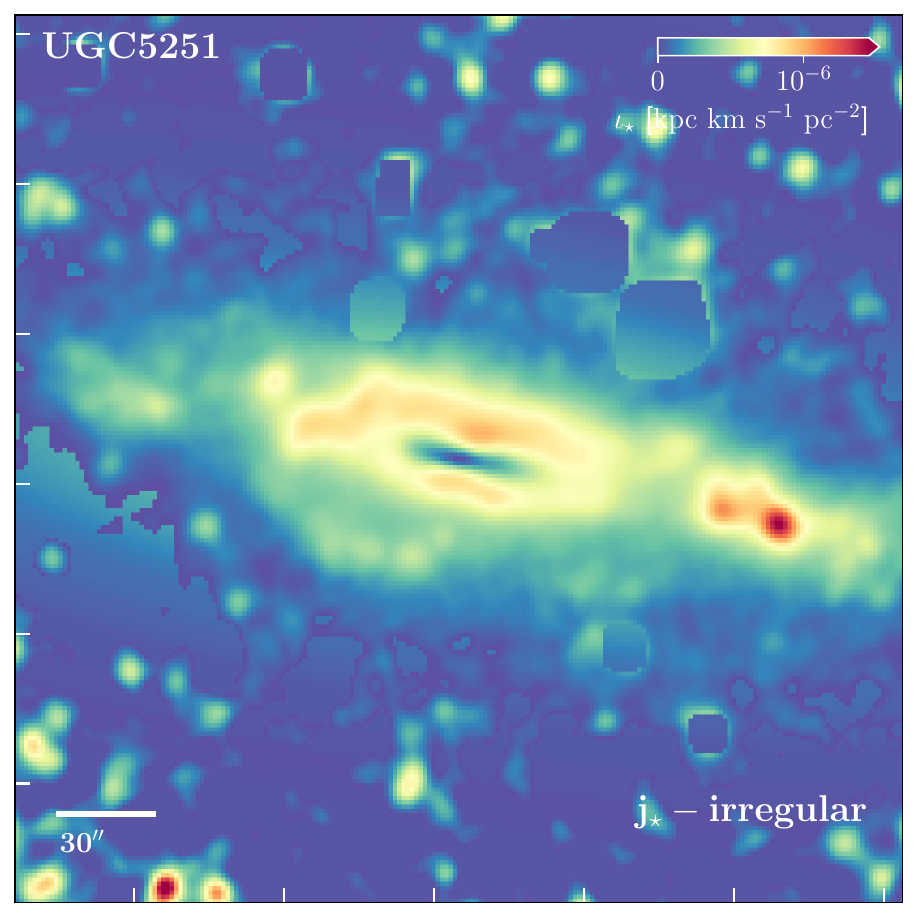}
    \end{subfigure}
    \\
    \begin{subfigure}[b]{0.245\textwidth}
        \includegraphics[width=\textwidth]{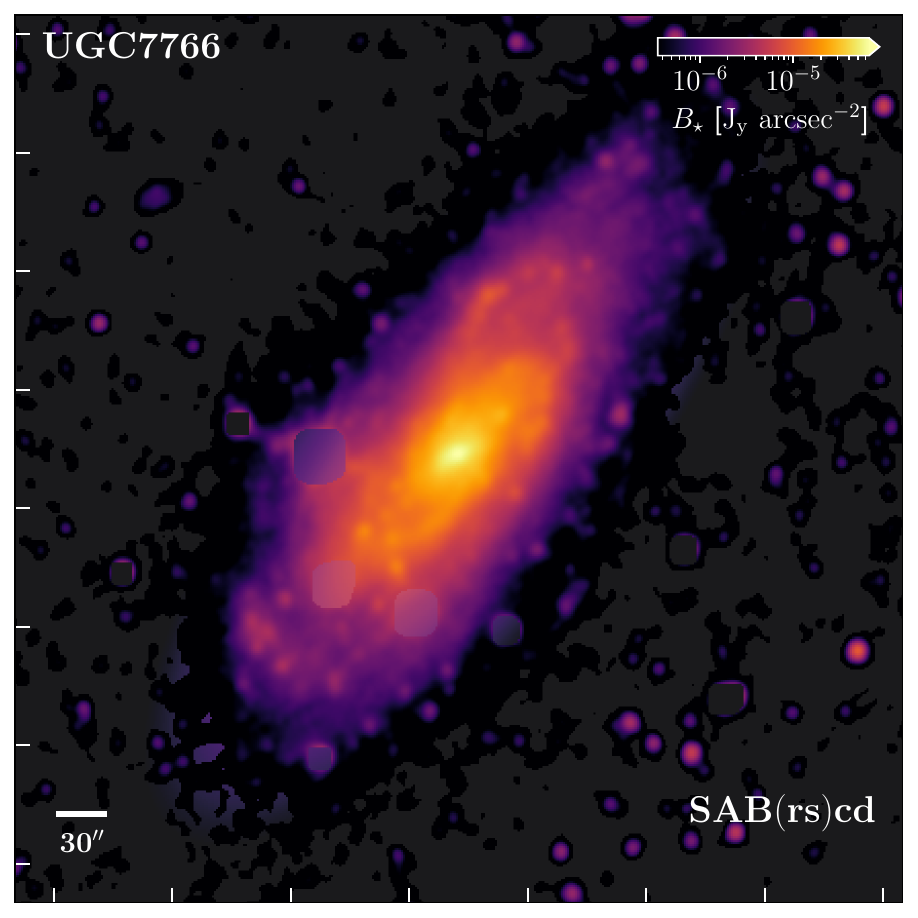}
    \end{subfigure}
    \begin{subfigure}[b]{0.245\textwidth}
        \includegraphics[width=\textwidth]{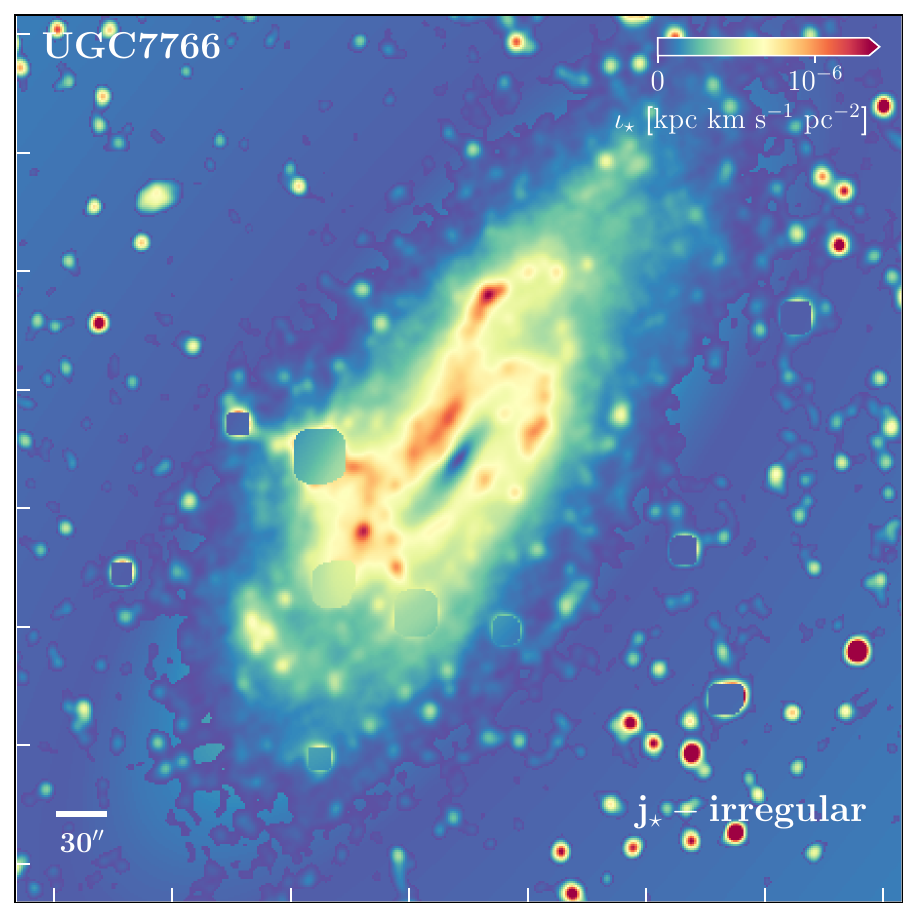}
    \end{subfigure}
    \begin{subfigure}[b]{0.245\textwidth}
        \includegraphics[width=\textwidth]{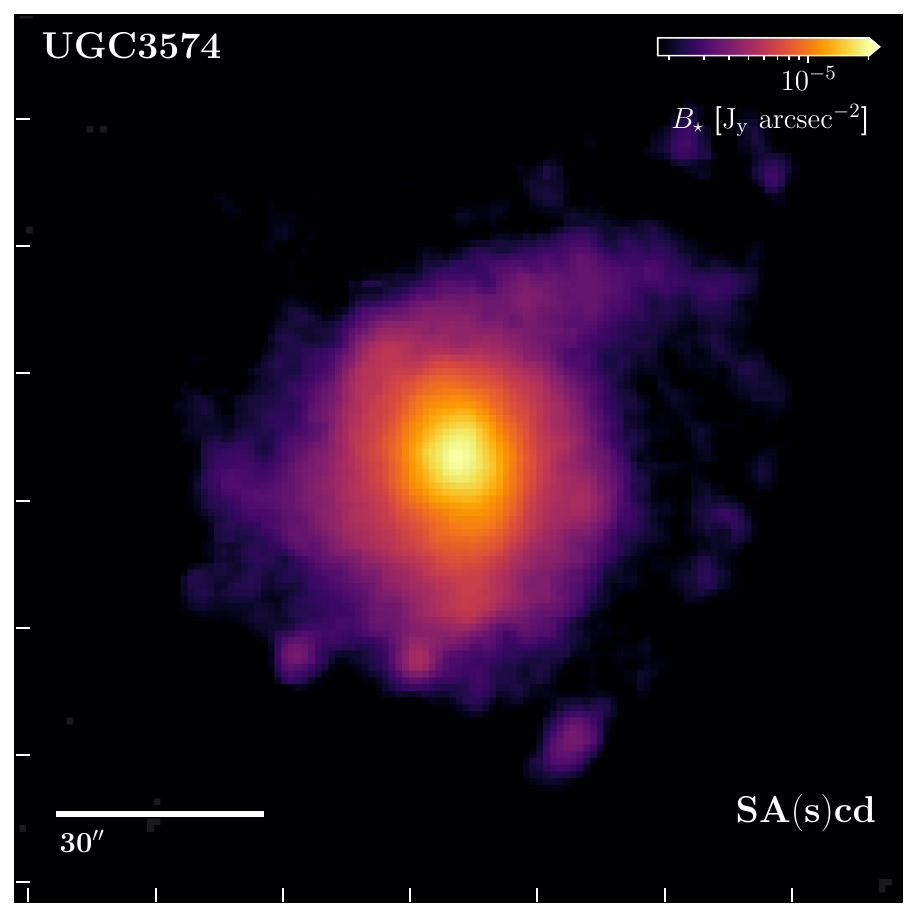}
    \end{subfigure}
    \begin{subfigure}[b]{0.245\textwidth}
        \includegraphics[width=\textwidth]{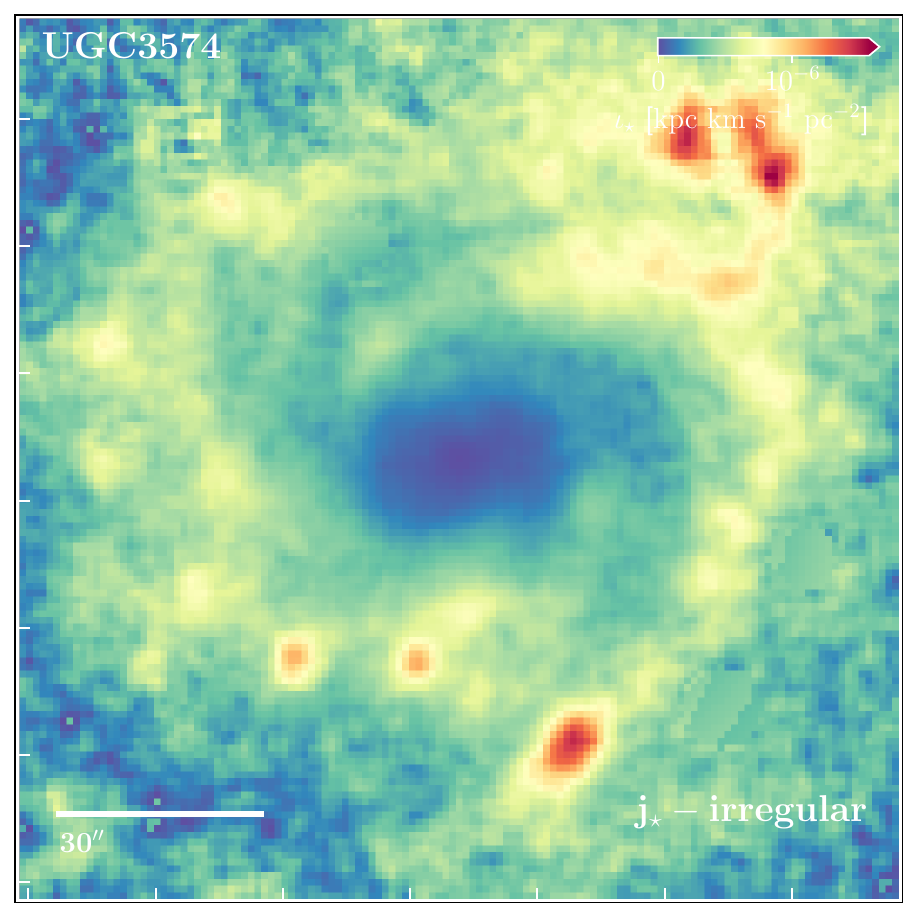}
    \end{subfigure}
    \\
    \begin{subfigure}[b]{0.245\textwidth}
        \includegraphics[width=\textwidth]{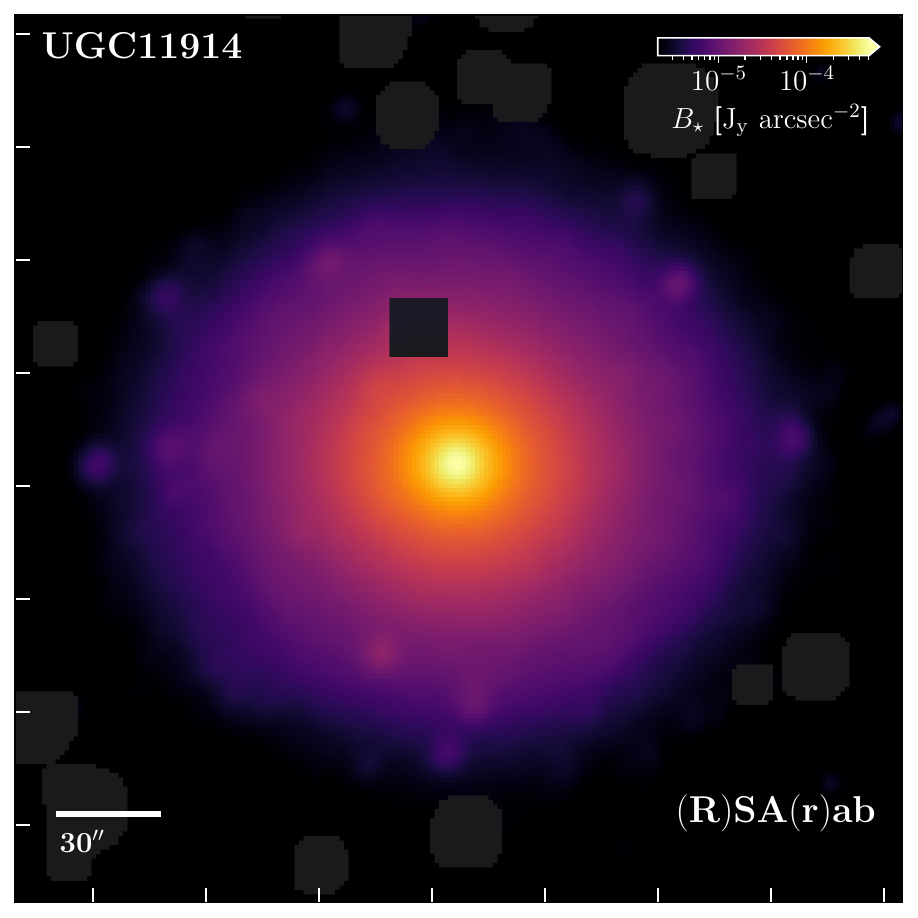}
    \end{subfigure}
    \begin{subfigure}[b]{0.245\textwidth}
        \includegraphics[width=\textwidth]{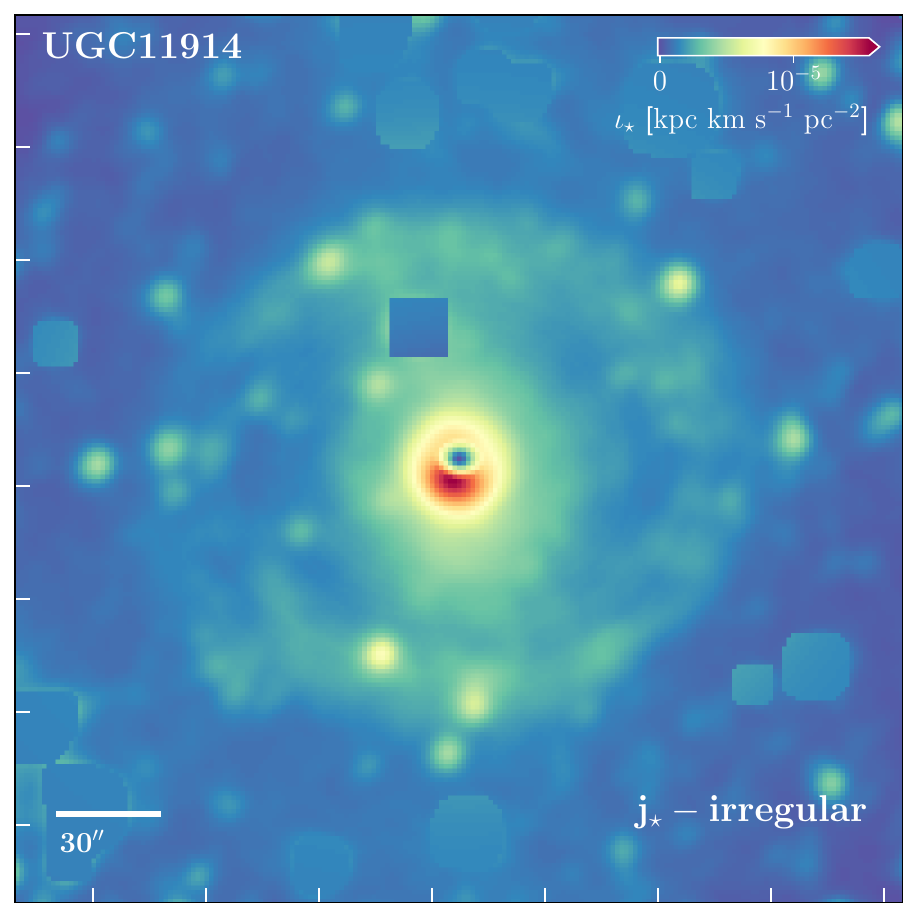}
    \end{subfigure}
    \caption{Continued.}
\end{figure*}

\FloatBarrier
\clearpage
\twocolumn

\section{Freeman disc stellar sAMSD map}\label{app: Freeman disc}

\begin{figure}[h]
\centering
\includegraphics[width=\hsize]{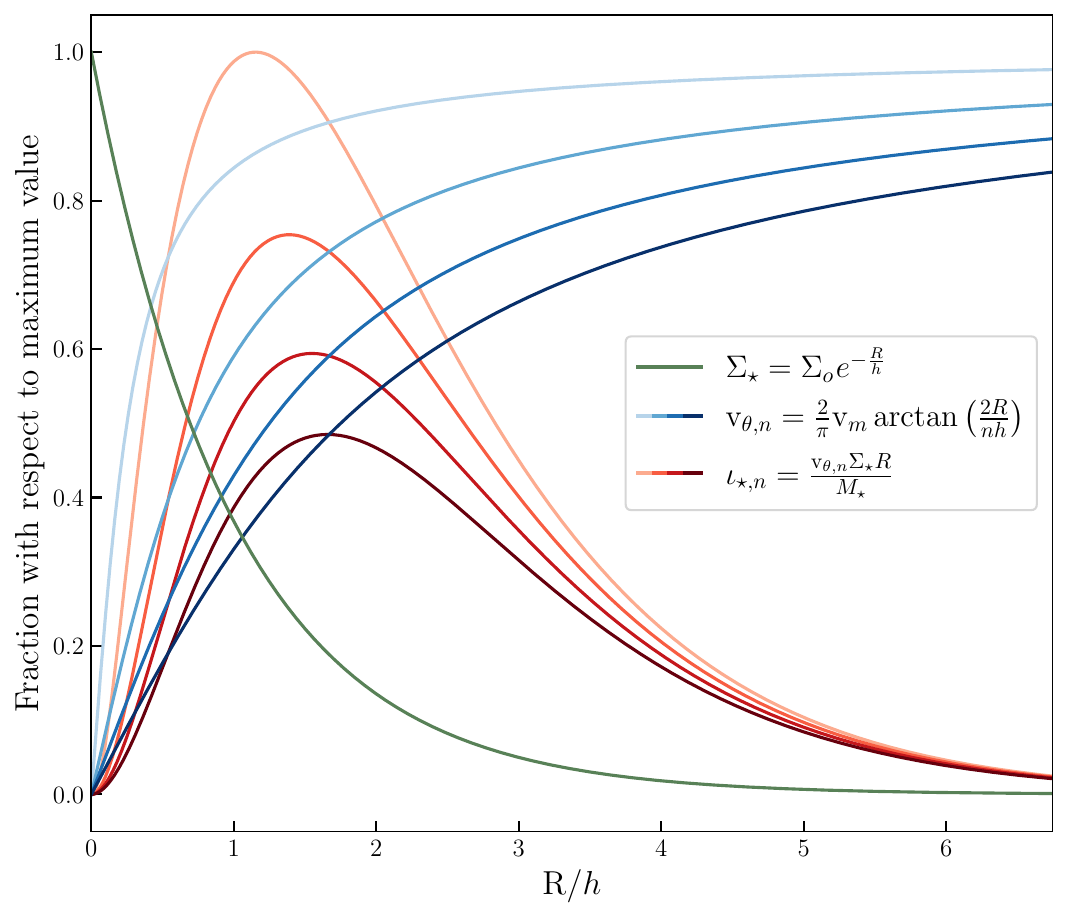}
\caption{Radial stellar sAMSD profiles for a Freeman disc mass distribution convolved with a set of varying arctan RCs. The x-axis represents the radius normalised by the disc scale length $h$, while the y-axis shows each one-dimensional profile normalised with respect to its maximum value, achieving a consistent comparison of their shapes. The solid green line corresponds to the radial stellar surface mass density for a Freeman disc. The solid blue lines represent the set of tangential RCs modelled using a parametric arc-tangent RC with a constant maximum velocity and with various turnover radii, whereas the group of red lines represents the radial profiles of stellar sAMSD corresponding to these RC models (see Eq. \ref{eq: iota}). The turnover radius is set to $n\times h$, with $n$ equal to 0.5, 1.5, 2.5, and 3.5. The line darkness increases as $n$ increases.}
\label{fig: Freeman disc 1D}
\end{figure}

Figure \ref{fig: Freeman disc 1D} shows the $\iota_{\star,n}$ radial profiles (red lines) obtained for an exponential disc surface mass density (green line) and a set of arc-tangent RC models (blue lines). We obtained these profiles by including the axisymmetric hypothesis in Eq. (\ref{eq: iota}). As expected, the bell-shaped $\iota_{\star,n}$ profiles show that the galactic centre of a Freeman disc has a stellar sAMSD equal to zero. This is due to the null rotation of the stars at this point, coupled with the fact that angular momentum is measured with respect to the perpendicular axis that originates at the centre of the galactic plane. As $R$ increases, the density of stars begins to decline, but the tangential velocity gains predominance, causing a rapid rise in angular momentum that reaches its maximum at the radius where $\Sigma_\star$ and $v_\theta \times R$ are balanced. From this point onwards, the exponential decay of the stellar distribution forces the sAMSD to decline, falling back to zero where there are no stars left that can store a significant amount of angular momentum. This behaviour arises naturally when assuming continuous axisymmetric mass distributions, being independent of the RC model. Variations in the turnover radius (different blue lines depending on $n$), maximum velocity, or $v_{\theta}$ shape can change the position and width of the stellar sAMSD peak, but not its intrinsic form (see Fig. \ref{fig: Freeman disc 1D}).

\begin{figure}
\centering
\includegraphics[width=.9\hsize]{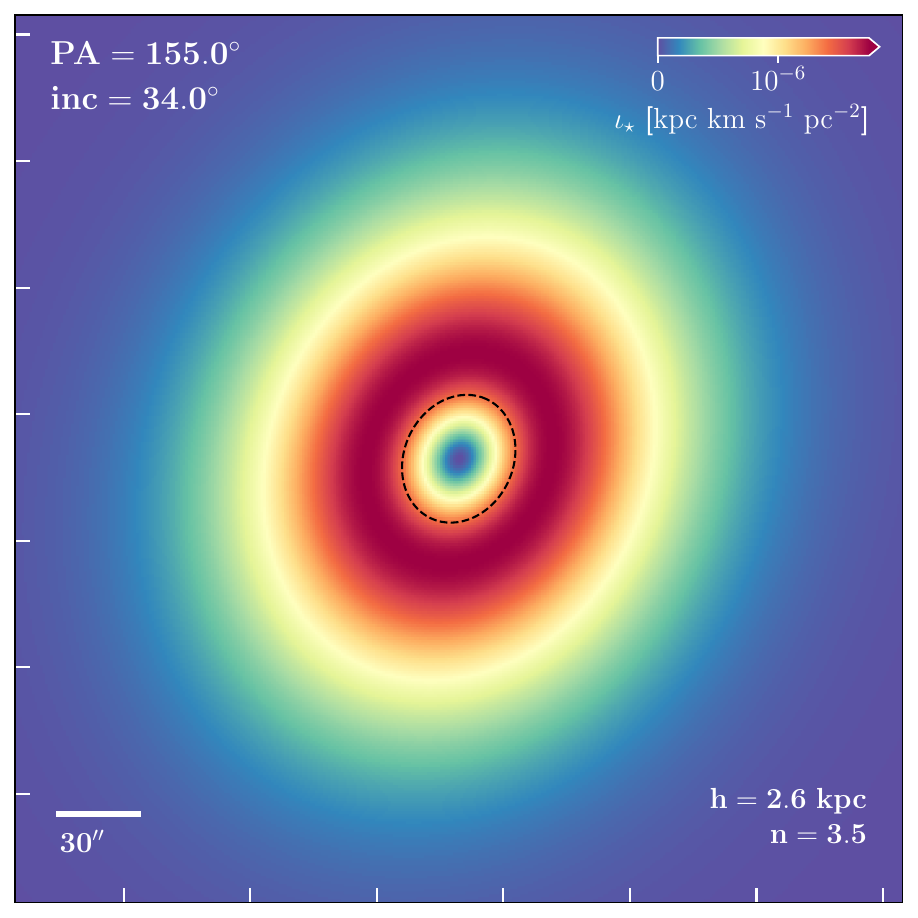}
\caption{Stellar sAMSD two-dimensional distribution of the $n=3.5$ profile presented in Fig. \ref{fig: Freeman disc 1D} with $h = 2.6$ kpc at a distance of 21.2 Mpc. The geometrical parameters used for the projection of this profile in the sky are shown in the upper-left corner of the map, while the units for $\iota_{\star}$ are shown in the upper-right corner. The dotted black line indicates the disc scale length. The spatial resolution and pixel scale used here correspond to the values of the $W_1$ band of WISE.}
\label{fig: Freeman disc 2D}
\end{figure}

The two-dimensional stellar sAMSD distribution for a Freeman disc can be obtained by directly projecting its one-dimensional radial profile onto the sky, assuming a certain inclination and PA. This is exactly what can be seen in Fig. \ref{fig: Freeman disc 2D}, where we present the stellar sAMSD map corresponding to the $n = 3.5$ profile (the darker red curve in Fig. \ref{fig: Freeman disc 1D}). Figure \ref{fig: Freeman disc 2D} confirms what we stated in Sect. \ref{sec: sAM maps results}: the expected morpho-kinematics for an unperturbed galactic disc is a $j_{\star}$- ring. The bell-shaped radial profile naturally corresponds to a torus in two dimensions.  The width and position of this structure in the galactic plane are determined by the balance between $h$ and the parameters of the RC.

\section{Morpho-kinematic metrics}\label{app: metrics}

\begin{table}
\caption{Morpho-kinematic metrics for our sample.}
\label{table: morphokinematic_params}
\centering
\setlength{\tabcolsep}{3pt}
\setlength\extrarowheight{1.5mm}
\begin{tabular}{c c c c c c c}
\hline\hline\\[-2ex]
\makecell{N$^{\circ}$ \\ UGC \\ (1)} & \makecell{Morpho-kinematic \\ type \\ (2)} & \makecell{$\rm R^2$ \\ \\ (3)} & \makecell{$\rm A_2$ \\ \\ (4)} & \makecell{C \\ \\ (5)} & \makecell{A \\ \\ (6)} & \makecell{S \\ \\ (7)}\\
\hline
   11852 & $j_{\star}$- ring & 0.87 & 0.06 & 2.35 & 0.27 & 0.08\\
   9969 & $j_{\star}$- ring & 0.85 & 0.03 & 1.86 & 0.24 & 0.07\\
   10075 & $j_{\star}$- ring & 0.80 & 0.04 & 2.80 &     0.45 & 0.18\\
   8334 & $j_{\star}$- ring & 0.79 & 0.16 & 2.62 & 0.18 & 0.09\\
   3734 & $j_{\star}$- ring & 0.75 & 0.08 &     2.01 & 0.12 & 0.06\\
   11012 & $j_{\star}$- ring & 0.74 & 0.04 & 3.02 & 0.36 & 0.15\\
   5253 & $j_{\star}$- ring & 0.32 & 0.05 & 2.43 & 0.10 & 0.05\\[.5ex]
   \hline
   6537 & $j_{\star}$- spiral & 0.68 & 0.09 & 1.68 & 0.39 & 0.10\\
   2855 & $j_{\star}$- spiral & 0.63 & 0.04 & 1.98 & 0.39 & 0.11\\
   6778 & $j_{\star}$- spiral & 0.59 & 0.08 & 2.05 & 0.34 & 0.11\\[.5ex]
   \hline
   11670 & $j_{\star}$- bar & 0.40 & 0.26 & 3.28 & 0.47 & 0.11\\
   10470 & $j_{\star}$- bar & 0.44 & 0.21 & 1.45 & 0.23 & 0.06\\
   9649 & $j_{\star}$- bar & 0.66 & 0.21 & 1.89 & 0.22 & 0.09\\
   12754 & $j_{\star}$- bar & 0.56 & 0.18 & 1.55 & 0.41 & 0.10\\[.5ex]
   \hline
   5414 & $j_{\star}$- clump & -0.25 & 0.24 & 2.22 & 0.84 & 0.39\\
   4325 & $j_{\star}$- clump & 0.07 & 0.08 & 2.13 & 0.62 & 0.29\\
   4499 & $j_{\star}$- clump &  -0.10 & 0.20 & 1.76 & 0.61 & 0.27\\
   1913 & $j_{\star}$- clump & -0.15 & 0.19 & 1.83 & 0.59 & 0.26\\
   9179 & $j_{\star}$- clump & 0.09 & 0.13 & 1.92 & 0.53 & 0.21\\
   11597 & $j_{\star}$- clump & 0.14 & 0.07 & 1.88 & 0.49 & 0.20\\
   8490 & $j_{\star}$- clump & 0.42 & 0.13 & 2.12 & 0.51 & 0.18\\
   4284 & $j_{\star}$- clump & 0.29 & 0.08 & 2.03 & 0.51 & 0.18\\
   7323 & $j_{\star}$- clump & 0.43 & 0.07 & 1.75 & 0.37 & 0.15\\[.5ex]
   \hline
   9858 & $j_{\star}$- irregular & 0.31 & 0.02 & 2.77 & 0.63 & 0.23 \\
   2800 & $j_{\star}$- irregular & 0.42 & 0.10 & 2.31 & 0.54 & 0.17 \\
   5251 & $j_{\star}$- irregular & 0.54 & 0.03 & 2.50 & 0.53 & 0.27 \\
   7766 & $j_{\star}$- irregular & 0.42 & 0.07 & 2.96 & 0.51 & 0.22 \\
   3574 & $j_{\star}$- irregular & 0.21 & 0.03 & 1.26 & 0.41 & 0.12 \\
   10359 & $j_{\star}$- irregular & 0.29 & 0.15 & 1.67 & 0.31 & 0.10 \\
   11914 & $j_{\star}$- irregular & -1.11 & 0.15 & 2.44 & 0.28 & 0.13 \\[1ex]
\hline
\end{tabular}
\tablefoot{(1) Name of the galaxy in the UGC catalogue, (2) morpho-kinematic type defined in Sect. \ref{sec: sAM maps results}, (3) ringness of the stellar sAMSD map, (4) bar predominance of the stellar sAMSD map, (5), (6) and (7) are the concentration, asymmetry and smoothness of the stellar sAMSD map. The detailed definition of parameters (3), (4), (5), (6), and (7) can be found in Sect. \ref{sec: sAM maps results}.}
\end{table}

Table \ref{table: morphokinematic_params} shows the values of all the morpho-kinematic metrics presented in Sect. \ref{sec: sAM maps results} for each galaxy in our sample, grouped by $j_{\star}$ type. We describe in the following the $\rm R^2$ and $\rm A_2$ metrics proposed in this study as part of the $j_{\star}$ classification.

\subsection{$\rm R^2$ definition and implementation}\label{app: R^2}

The $\rm R^2$ factor is a statistical measure of how similar the stellar sAMSD map of a galaxy is to that of an unperturbed stellar disc. We therefore computed it as

\begin{equation}
    \mathrm{R}^2 = 1 - \frac{\sum_{x^{\prime},y^{\prime}} \left[\iota_\star (x^{\prime},y^{\prime}) - \iota_{\rm Freeman}(x^{\prime},y^{\prime})\right]^2}{\sum_{x^{\prime},y^{\prime}} \left[\iota_\star (x^{\prime},y^{\prime}) - \overline{\iota_{\star}}\right]^2}
    \label{eq: R2}
\end{equation}

\noindent using the general definition of the coefficient of determination. We estimated $\iota_\star (x^{\prime},y^{\prime})$ following the methodology explained in Sect. \ref{sec: stellar sAM maps}, while we obtained the modelled stellar sAMSD distribution $\iota_{\rm Freeman}(x^{\prime},y^{\prime})$ by applying the procedure described in Appendix \ref{app: Freeman disc} assuming the Freeman disc parameters presented in \cite{korsaga2018ghasp} and the H$\alpha$ + \ion{H}{i} RC of each galaxy. The distance, pixel scale, and geometrical parameters are the same in both maps, ensuring a proper pixel-by-pixel comparison. $\overline{\iota_{\star}}$ is the mean stellar sAMSD.

Under this definition, galaxies with an $\rm R^2$ close to 1 have a stellar sAMSD map with a $j_\star$- ring morpho-kinematic typical of an unperturbed galactic disc (see Appendix \ref{app: Freeman disc}), while galaxies with an $\rm R^2$ close to 0 show no trace of a ring structure in sAMSD space. Negative values indicate that the stellar sAMSD model for a Freeman disc fits the stellar sAMSD map of the galaxy worse than a constant $\overline{\iota_{\star}}$ map.

\subsection{$\rm A_2$ definition and implementation}\label{app: A_2}

The $\rm A_2$ parameter quantify the predominance of bars in the stellar sAMSD map by applying the same technique traditionally used in galaxy images and simulations \citep{ohta1990surface,athanassoula2002bar,diaz2016stellar}. The idea is to express the azimuthal variation of the stellar sAMSD for a given radius as a Fourier series:

\begin{equation}
    \iota_\star(R, \theta) = \mathrm{A}_0(R) + \sum_{m=1}^{\infty} \mathrm{A}_m(R) \cos\!\big(m\theta - \phi_m(R)\big),
\end{equation}

\noindent where $\iota_\star(R, \theta)$ is the stellar sAMSD map, and $\mathrm{A}_m(R)$ and $\phi_m(R)$ are the amplitude and phase of the $m$-th Fourier mode. The bars are bi-symmetric substructures, so their relevance in the azimuthal variation of $\iota_\star$ is captured in mode $m=2$, whose amplitude is defined as

\begin{equation}
    \mathrm{A}_2(R) = \frac{\left|\sum_{\theta} \iota_\star(R, \theta)\, e^{-2i\theta}\right|}{\sum_\theta \iota_\star(R,\theta)}.
    \label{eq: A2}
\end{equation}

\noindent We identified that the accumulation of angular momentum in the bars occurs mainly at their outer edges. These edges are located in the region where these galaxies should have a $j_\star$- ring. Therefore, we decided to construct a single value of $\rm A_2$ for each galaxy by averaging $\mathrm{A}_2(R)$ between the borders of its expected Freeman $j_\star$- ring. We define these borders for each galaxy as the radii at which its radial profile $\iota_{\rm Freeman}$ (computed as described in Appendix \ref{app: R^2}) has 75\% of its maximum value, establishing a uniform criterion for our entire sample.

\section{Decision tree for morpho-kinematic classification}\label{app: decision_tree}

\FloatBarrier

\begin{figure*}[ht!]
    \centering
    \includegraphics[width=1\textwidth]{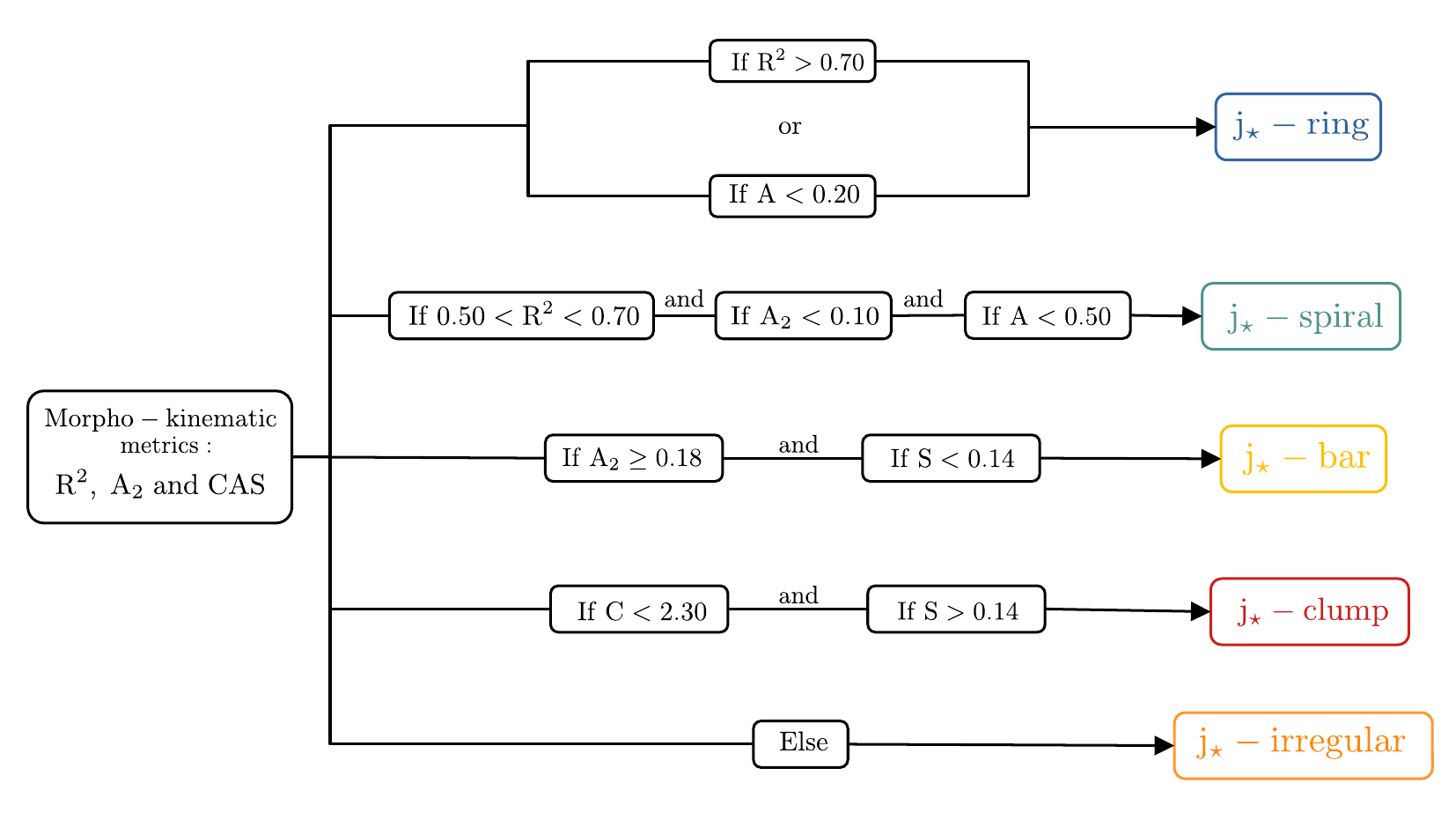}
    \caption{Decision tree describing the numeric conditions used on each parameter to determine the morpho-kinematic classification.}
\label{fig: decision_tree}
\end{figure*}

\end{appendix}
\end{document}